\documentclass[12pt,english]{article}

\renewcommand{\Re}{{\rm Re \, }}

\newcommand{\dS}{{\rm dS }}

\newcommand{\SO}{{\rm {SO}}}

\newcommand{\SU}{{\rm {SU}}}
\newcommand{\U}{{\rm {U}}}

\newcommand{\ket}[1]{\bigl| #1 \bigr\rangle}
\newcommand{\bra}[1]{\bigl\langle #1 \bigr|}

\usepackage[latin9]{}
\usepackage{geometry}
\usepackage{tikz}
\usetikzlibrary{shapes.geometric}
\usepackage{youngtab}

\usepackage{babel}
\usepackage{cite}
\usepackage{verbatim,enumitem}
\usepackage{prettyref}
\usepackage{amsmath}
\usepackage{amssymb}
\usepackage{mathabx}
\usepackage{mathrsfs}
\usepackage{graphicx}
\usepackage{esint}
\usepackage{physics}
\usepackage{slashed}
\usepackage{subcaption}
\usepackage{dsfont}
\usepackage{xcolor}
\usepackage{color}
\definecolor{darkgreen}{rgb}{0,0.5,0}
\definecolor{darkblue}{rgb}{0,0,0.6}
\definecolor{purple2}{rgb}{0.4,.2,0.7}
\usepackage[colorlinks=true,citecolor=purple,linkcolor=black,urlcolor=black]{hyperref}

\definecolor{darkpastelgreen}{rgb}{0.01, 0.75, 0.24}

\numberwithin{equation}{section}
\numberwithin{figure}{section}
\numberwithin{table}{section}

\def\IC{{\mathbb C}}
\def\one{{\mathds 1}}

\def\tr{\,{\rm tr}\,}

\def\IR{{\mathbb R}}
\def\rP{{\mathsf P}}

\newcommand{\rme}{\mathrm e}
\newcommand{\rmd}{\mathrm d}
\newcommand{\rmi}{\mathrm i}
\addtocounter{page}{0}

\newcommand{\TFD}{\text{TFD}}

\newcommand{\bluecheck}{}%
\DeclareRobustCommand{\bluecheck}{%
  \tikz\fill[scale=0.4, color=blue]
  (0,.35) -- (.25,0) -- (1,.7) -- (.25,.15) -- cycle;%
}
\newcommand{\redcross}{}%
\DeclareRobustCommand{\redcross}{%
  \textcolor{red}{$\times$}%
}
\begin{document}

\begin{center}
{\LARGE \bf Quasinormal modes and complexity in saddle-dominated \texorpdfstring{$\SU(N)$}{} spin systems}  
\vskip3mm
$\textbf{Sergio E. Aguilar-Gutierrez}^a\textbf{, Yichao Fu}^b\textbf{, Kuntal Pal}^b$ $\textbf{and Klaas Parmentier}^c$\\
${}^a\,$Qubits and Spacetime Unit, Okinawa Institute of Science and Technology\\
1919-1 Tancha, Onna, Okinawa 904 0495, Japan\\
${}^b\,$ Department of Physics and Photon Science, Gwangju Institute of Science and Technology\\
123 Cheomdan-gwagiro, Gwangju 61005, Republic of Korea\\
${}^c\,$Department of Physics, Columbia University, New York, NY 10027, USA\\
\textit{E-mail}: \hyperlink{sergio.ernesto.aguilar@gmail.com}{sergio.ernesto.aguilar@gmail.com}, \hyperlink{yichao.fu@gm.gist.ac.kr}{yichao.fu@gm.gist.ac.kr}, \hyperlink{kuntalpal@gist.ac.kr}{kuntalpal@gist.ac.kr}, \hyperlink{k.parmentier@columbia.edu}{k.parmentier@columbia.edu}
\end{center}

\vskip5mm

\noindent  {
\textbf{Abstract}. We study SU($N$) spin systems that mimic the behavior of particles in $N$-dimensional de Sitter space for $N=2,3$. Their Hamiltonians describe a dynamical system with hyperbolic fixed points, leading to emergent quasinormal modes at the quantum level. These manifest as quasiparticle peaks in the density of states. For a particle in 2-dimensional de Sitter, we find both principal and complementary series densities of states from a PT-symmetric version of the Lipkin-Meshkov-Glick model, having two hyperbolic fixed points in the classical phase space. We then study different spectral and dynamical properties of this class of models, including level spacing statistics, two-point functions, squared commutators, spectral form factor, Krylov operator and state complexity. We find that, even though the early-time properties of these quantities are governed by the saddle points -- thereby in some cases mimicking corresponding properties of chaotic systems, a close look at the late-time behavior reveals the integrable nature of the system.

\thispagestyle{empty}

{\setlength{\baselineskip}{16pt}
\newpage
\thispagestyle{empty}
\setcounter{tocdepth}{3}
\tableofcontents}

\section{Introduction}\label{sec:intro}
Certain integrable models can reproduce features that are considered to be typical of chaotic systems, such as early-time exponential behavior of out-of-time-ordered correlators (OTOC) or a peak in spread complexity \cite{Rozenbaum:2019nwn, Xu:2019lhc, Nandy:2024zcd,Erdmenger:2023wjg,Balasubramanian:2022dnj,Balasubramanian:2022tpr,Balasubramanian:2023kwd, Baggioli:2024wbz,Alishahiha:2024vbf, Jha:2024nbl, Camargo:2024deu,  Bhattacharjee:2022vlt, Huh:2023jxt}. In
saddle-dominated scrambling, an unstable fixed point -- classically of measure zero -- gets smeared over a region in phase space due to the finite size of wave packets (see e.g. \cite{Xu:2019lhc, pilatowsky, kidd2021saddle} and references therein). This can lead to exponential growth of squared commutators \cite{Rozenbaum:2019nwn, Xu:2019lhc}. Although hyperbolic saddles are not chaotic by themselves, it is interesting to note in this context that they are typically responsible for the onset of chaos under small perturbations.

One famous system, known to feature saddle-dominated scrambling, is the Lipkin-Meshkov-Glick (LMG) model \cite{Lipkin:1964yk,Meshkov:1965btx,Glick:2002fef}, which in general is defined as \cite{Lipkin:1964yk}
\begin{eqnarray}\label{eq:more general LMG}
    H_{\rm LMG}=A\qty(J_+^2+J_-^2)+B\qty(J_+J_-+J_-J_+)+CJ_z~,
\end{eqnarray}
where $A,~B$ and $C$ are constants and $J_{\pm}$, $J_z$ are the angular momentum operators acting on the spin-$j$ irrep of $\SU(2)$.
This family of models is classically integrable but also displays some characteristics of chaotic quantum systems, as measured by OTOCs \cite{Xu:2019lhc} or Krylov complexity \cite{Huh:2023jxt, Bhattacharjee:2022vlt}. These latter works focused on one particular type of LMG system where $A=B$ in (\ref{eq:more general LMG}).
In this work we consider $B=0$, $A=\tfrac{1}{4j}$, $C=\frac{\nu}{j}$ in (\ref{eq:more general LMG}):
\begin{equation}\label{eq:hamintro}
    H_j = \frac{1}{4j}(J_+^2 + J_-^2) + \frac{\nu}{j} J_z\,.
\end{equation}
Our interest in this particular system stems from the fact it mimics the behavior of a massive particle, of scaling dimension $\Delta = \frac12 + \rmi \nu$, in 2-dimensional de Sitter (dS) space \cite{Parmentier:2023axg}.\footnote{{Note however, that this microscopic model in and of itself is not holographically dual to dS$_2$ space. As mentioned above, this integrable model (in contrast with dS space being maximally or hyperfast chaotic \cite{Aalsma:2020aib,Susskind:2021esx}) captures some properties of a massive quantum particle in dS space. Still, we hope this model can teach us lessons about how to incorporate matter in dS holography.}} In particular, it reproduces at large-$j$ the corresponding density of states, whose poles are the $\dS$ quasinormal modes (QNMs). In this sense, the QNMs emerge from the spin system.

The similarity between a particle in $\dS_2$ and the spin system \eqref{eq:hamintro} becomes clear when noting that at large spin, \eqref{eq:hamintro} has a classical description in terms of coherent spin states \cite{Berezin:1974du, Perelomov_1977}. As we discuss in sec.\,\ref{sec:SU(N)}, this classical phase space has two saddle points. The same is true for the static patch Hamiltonian in $\dS_{2}$, which acts as a dilatation on the future conformal boundary circle. This can also be appreciated from the static patch perspective, since a particle in $\dS_2$ experiences an upside-down harmonic oscillator potential. The corresponding towers of resonances are the $\dS$ QNMs.

\paragraph{Purpose of this work.}

The aim of this work is two-fold. 

Since the simple SU(2) spin model with Hermitian Hamiltonian \eqref{eq:hamintro} reproduces the density of states for particles in the $\dS_2$ principal series, it is natural to wonder:
\begin{quote}
Can this system be modified to reproduce the density of states and QNM frequencies of other types of fields, corresponding for instance to the complementary and discrete series unitary irreps of $\SO(1,2)$? Are there similar generalizations to higher-dimensional $\dS$ space, for instance by increasing the dimension of the group on which the spin system is based?    
\end{quote}
Our first purpose is to try and answer these questions. We will find that by allowing a complexified but still PT-symmetric version of the SU$(2)$ Hamiltonian, we can also recover the complementary series density of states. Moreover, by considering an SU$(3)$ extension of \eqref{eq:hamintro}, we obtain a density of states related to that of a massive particle in dS$_3$. This generalization to higher dimensions is also natural because systems with one degree of freedom are somewhat special as far as their level spacing statistics are considered. This generalization will make the analysis more clear.\\

Secondly, given the lack of a clear consensus in the literature about which specific measures are suitable to distinguish saddle-dominated scrambling and quantum chaos, we ask:

\begin{quote}
    What specific measures of quantum chaos can distinguish it from saddle-dominated scrambling in the toy model of our consideration?
\end{quote}

Our results show that many commonly used measures of quantum chaos can indeed serve this purpose if one investigates them in sufficient detail. The specific measures we studied and that fulfill this condition include the spectral form factor (SFF) and the level spacing statistics from the spectral side, while dynamical measures, such as the square commutators, Krylov operator complexity, as well as spread complexity, also encode characteristic signals of the integrability of the system, specifically, when these quantities are computed at late times. 
%which is mainly manifested by an oscillating evolution and modifications in the range of these functions (which we elaborate more in the main text).
As far as we are aware, \eqref{eq:hamintro} has not been studied in this way. Our investigations, therefore, also help to confirm and solidify earlier claims in the literature. 

%\footnote{\sa{The style of the table summary and Tab. \ref{tab:comparizon_probes} is motivated in part by \cite{Baiguera:2025uss}.}}
\begin{table}[t!]
    \centering
    \begin{tabular}{c|c|c}
         \textbf{Results}&  SU(2) & SU(3)\\\hline
         Hyperbolic fixed points& Sec.\,\ref{sec:SU(N)}, App.\,\ref{app:saddle_derivation}& Sec.\,\ref{sec:SU(N)}\\
         Spectral properties&Sec.\,\ref{sec:spectra} & Sec.\,\ref{sec:spectra}\\
         Dynamical probes of chaos& Sec.\,\ref{sec:spread complexity}, App.\,\ref{SC_lowest}& \\
         Complementary series extension& Sec.\,\ref{sec:complementary series} &
    \end{tabular}
    \caption{{An overview of the main topics studied in this work, as elaborated on in the Outline below. We indicate in which section the results can be found and to which specific system, $\SU(2)$ or $\SU(3)$, these pertain.}}
    \label{tab:results SUN}
\end{table}

\paragraph{Outline.} The structure of this work -- {see also Tab.\ref{tab:results SUN} for a quick overview of the treated topics --} is as follows. In sec.\,\ref{sec:SU(N)} we explore the classical large-spin dynamics of several SU(N) spin systems. We study the appearance of saddle points in phase space, focusing on the cases of SU(2) and SU(3). In sec.\,\ref{sec:spectra} we relate this discussion with the spectrum and QNMs of particles in dS$_2$ and dS$_3$. In sec.\,\ref{sec:spread complexity} we study the two-point functions, OTOCs, Krylov operator and spread complexity of the SU(2) spin model. We show these display features typical of saddle-dominated scrambling, {as summarized in Tab. \ref{tab:comparison_props}.}
Sec.\,\ref{sec:complementary series} deals with a PT-symmetric version of the $\SU(2)$ Hamiltonian and its connection with the complementary series of dS$_2$.
In sec.\,\ref{sec:discussion} we provide a summary of our results as well as ideas for future work and a few more speculative comments. The appendices contain supplementary material.

\section{Spin systems with hyperbolic fixed points}\label{sec:SU(N)}
In this section we study spin systems based on $\SU(N)$ irreps with the highest weight state labeled by (half-)integer spin $j$. In the large-$j$ limit, their Hamiltonian determines a classical dynamical system with phase space $\mathbb{C}\mathsf{P}^{N-1}$. For the reasons outlined in the introduction, we are interested in scenarios where this dynamics has hyperbolic fixed points.

\subsection{\texorpdfstring{$\SU(2)$}{} }\label{sec:classSU2}
We start with the following Hamiltonian acting on the spin-$j$ representation of SU$(2)$: 
\begin{equation}\label{eq:hamds2}
    H_j = \frac{\rmi}{4j}(J_-^2 - J_+^2) + \frac{\nu}{j} J_z\,, 
\end{equation}
where $J_\pm = J_x \pm \rmi J_y$. Note that \eqref{eq:hamds2} is related to \eqref{eq:hamintro} by a unitary conjugation. The above will be the most convenient form in what follows and it is also the one studied in \cite{Parmentier:2023axg}. The operators $J_{\pm}$ and $J_z$ satisfy the following commutation relations 
\begin{equation}\label{eq:su(2) algebra}
    \qty[J_z,~J_\pm]=\pm J_\pm~,\quad \qty[J_+,~J_-]=2J_z~.
\end{equation}
Hamiltonian \eqref{eq:hamds2} is a particular instance of the 
LMG model \cite{Lipkin:1964yk,Meshkov:1965btx,Glick:2002fef}. It stands out among the LMG systems due to its augmented symmetry: a rotation of $\pi$ around the $x$-axis (or $y$-axis) flips the sign of $H_j$. 

As reviewed in app.\,\ref{app:cohstates}, we can describe the system with coherent spin states. These are labeled by points on the 2-sphere, indicating the direction of the spin expectation value. In the large-spin limit, their dynamics becomes classical \cite{Berezin:1974du}. The classical phase space is the 2-sphere, whose volume form (with $j$ acting as $1/\hbar$) serves as symplectic 2-form. The classical dynamics is governed by the coherent state expectation value of $H_j$. In casu:
\begin{equation}\label{H_classical}
H = j XY + \nu Z\,,\qquad  X^2+Y^2+Z^2=1\,, \qquad \Omega = \frac{j}{Z}\,\rmd X \wedge \rmd Y.   
\end{equation}
The classical phase space orbits are depicted in fig.\,\ref{fig:orbitds2}. 
At fixed $\nu$ and large $j$, there are 4 elliptic and 2 hyperbolic fixed points, whose properties are analyzed in app.\,\ref{app:saddle_derivation}. More details on the classical orbits can be found, for instance, in sec.\,5.2 of \cite{Parmentier:2023axg}. 

\begin{figure}
    \centering
  \begin{subfigure}{0.25\textwidth}
            \centering
            \begin{tikzpicture}
              \node[inner sep = 0pt] at (0,0) {\includegraphics[width=\textwidth]{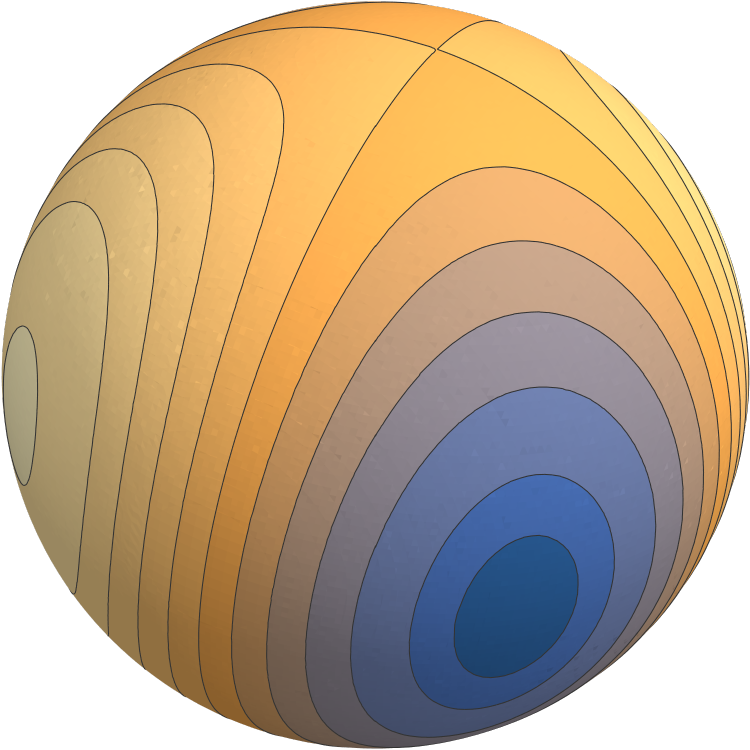}};
            \end{tikzpicture}
            \caption[]
            {\small $\nu/j = 0$}
        \end{subfigure}
        \hspace{0.3cm}
        \begin{subfigure}{0.25\textwidth}  
            \centering 
        \begin{tikzpicture}
              \node[inner sep = 0pt] at (0,0) {\includegraphics[width=\textwidth]{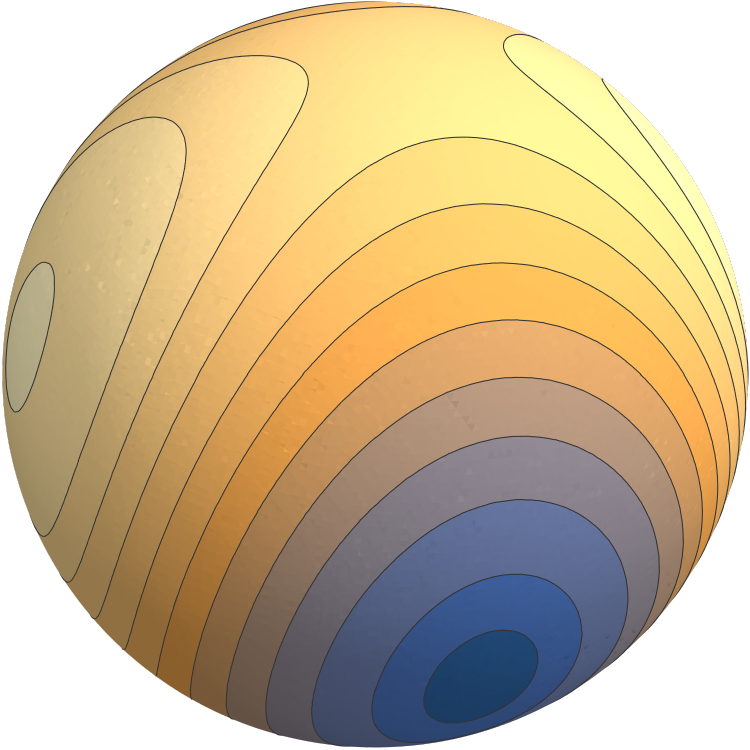}};
            \end{tikzpicture}
            \caption[]
            {\small $\nu/j = 1/4$}
        \end{subfigure}
        \hspace{0.3cm}
        \begin{subfigure}
        {0.25\textwidth}  
            \centering
            \begin{tikzpicture}
              \node[inner sep = 0pt] at (0,0) {\includegraphics[width=\textwidth]{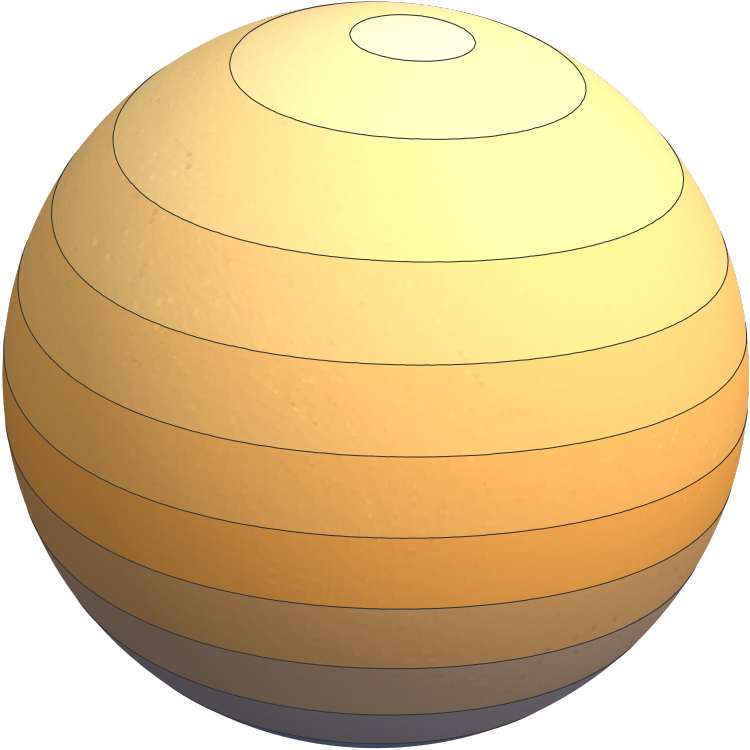}};
            \end{tikzpicture}
            \caption[]
            {\small $\nu/j = 10$}
        \end{subfigure}
        \hspace{0.2cm}
         \begin{subfigure}
        {0.075\textwidth}  
            \centering
            \begin{tikzpicture}
              \node[inner sep = 0pt] at (0,0) {\includegraphics[width=\textwidth]{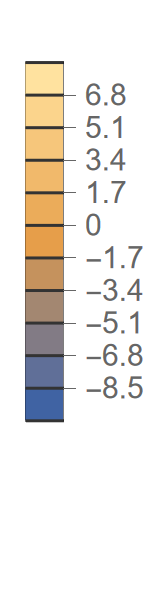}};
            \end{tikzpicture}
        \end{subfigure}
    \caption{$\SU(N)$ coherent states can be identified with points in the phase space $\IC\rP^{N-1}$. In the large-spin limit, their dynamics is classical. Above we show the orbits resulting from the $\SU(2)$ Hamiltonian \eqref{eq:hamds2}, {whose value is indicated by the color scale, ranging from negative energies in blue to positive ones in yellow.} At fixed mass parameter $\nu$ and large spin $j$, the two hyperbolic fixed points give rise to emergent $\dS$ QNMs at the quantum level, while the high-energy elliptic fixed points disappear after coarse-graining. Figure (c) illustrates the opposite limit, where the $J_z$ term dominates, resulting in simple spin precession.}
    \label{fig:orbitds2}
\end{figure}

The linearized Hamiltonian at the hyperbolic fixed points has eigenvalues $\lambda = \pm 1$ and therefore locally approximates the standard upside-down harmonic oscillator. In sec.\ref{sec:spectra} we will see that this knowledge of the classical dynamics is enough to predict the large-$j$ limit of the quantum density of states, which equals that of a massive particle of scaling dimension $\Delta = \frac12 +\rmi \nu$ in $\dS_2$ \cite{Parmentier:2023axg}. One of our goals is to generalize this system to higher dimensions. 

\subsection{\texorpdfstring{$\SU(3)$}{} }\label{sec:classSU3}
We would like to find a system which mimics for some time the behavior of a massive particle in $\dS_3$. From the static patch point of view, a particle experiences an upside-down potential, rolls down, and then appears to freeze out on the horizon. Can we find a discrete quantum system that mimics this behavior? 

Let us generalize \eqref{eq:hamds2} to the $\SU(3)$ case. We will consider degenerate highest weight irreps labeled by $j$. Coherent states are labeled by points in $\mathbb{C}\mathsf{P}^2$. This is the group $\SU(3)$ divided by the subgroup $\SU(2)\times \U(1)$ that leaves the highest weight state invariant. We can label coherent states by homogeneous coordinates $U,V,W$ on $\mathbb{C}\mathsf{P}^2$. As reviewed in app.\,\ref{app:cohstates}, the states are then represented by homogeneous polynomials $\psi(U,V,W)$ of degree $2j$. Let us now consider
\begin{equation}\label{eq:spds3}
    H_j = \frac{\rmi }{4j}(V^2+W^2)\partial^2_U + \text{c.c.}\, \,
\end{equation}
to act on the states $\psi(U,V,W)$. This is the generalization to $\SU(3)$ of \eqref{eq:hamds2} (at $\nu=0$). Note that, written in this way, the further generalization to $\SU(N)$ is also straightforward. We can think of \eqref{eq:spds3} as having even hopping terms on the $2$-dimensional lattice formed by the $\SU(3)$ states, as shown in fig.\,\ref{fig:SU3}.
\begin{figure}
    \centering
\includegraphics[width=0.5\linewidth]{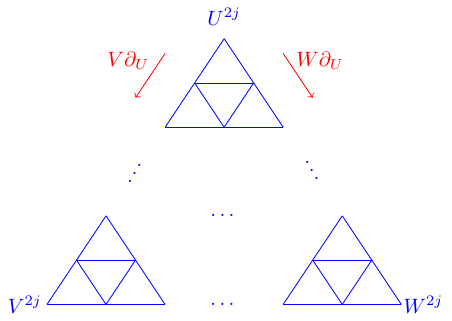}
    \caption{Lattice representation of SU(3) states illustrating the role of $V\partial_U$ and $W\partial_U$ as hopping terms in the Hamiltonian \eqref{eq:spds3}.}
    \label{fig:SU3}
\end{figure}

Since $H_j$ commutes with the angular momentum operator
\begin{equation}
    L = \rmi (V\partial_W - W\partial_V)\,
\end{equation}
a convenient basis of states is given by the following eigenbasis of $L$:
\begin{equation}\label{L_eigenbasis}
    \psi_{n,m}(U,V,W) = \mathcal{N}_{n,m}\,U^{2j-n-m}(V-\rmi W)^{n}(V+\rmi W)^m\,, \quad L\psi_{n,m}=(n-m) \psi_{n,m}\,,
\end{equation}
whose normalization, using \eqref{eq:innerCoh}, is given by
\begin{equation}
    \mathcal{N}^2_{n,m} = \frac{(2j)!}{2^{n+m}(2j-n-m)!n!m!}\,.
\end{equation}
We will use this basis to numerically diagonalize $H_j$ in sec.\,\ref{sec:spectra}. Below, we discuss the classical dynamics in the large-$j$ limit. Knowledge of the fixed points will suffice to determine the leading large-$j$ quantum density of states.

\paragraph{Large-spin dynamics}
The large-spin dynamics is governed by the coherent state expectation value of the Hamiltonian\footnote{{This is computed by acting with \eqref{eq:spds3} on the polynomial $(U\bar U + V \bar V + W \bar W)^{2j}$, and dividing the result by the latter. Some useful properties of coherent states are reviewed in app.\,\ref{app:cohstates}. For the general theory, see \cite{Perelomov_1977}. Useful formulas relating to $\SU(N)$ coherent state expectation values can be found in \cite{Gitman_1993}.}}, whose leading term takes the form
\begin{equation}\label{eq:hamhom}
    H(U,V,W) = \rmi j\,\frac{(V^2+W^2)\bar U^2- (\bar V^2+\bar W^2) U^2 }{(U\bar U + V\bar V + W \bar W)^2}\,.
\end{equation}
In local coordinates $(z_1,z_2)= (V,W)/U$ this becomes 
\begin{equation}\label{eq:leadingH}
    H = \rmi j \, \frac{z_1^2+z_2^2 - \bar{z}^2_1 - \bar{z}^2_2}{(1+|z_1|^2+|z_2|^2)^2}\,.
\end{equation}
In the same limit, the classical angular momentum takes the form
\begin{equation}
    L = 2\,\rmi j\, \frac{ z_1 \bar z_2 - \bar z_1 z_2}{1+|z_1|^2+|z_2|^2}\,.
\end{equation}
The symplectic form \eqref{eq:symp} can be inverted to
\begin{equation}
    \Omega^{\bar{i}j} = \frac{1+|z_1|^2+|z_2|^2}{2\,\rmi j}\begin{pmatrix}
       1+|z_1|^2  & z_1 \bar z_2\\ \bar z_1 z_2 & 1+ |z_2|^2
    \end{pmatrix}_{ij}\, ,
\end{equation}
allowing to verify that indeed $\{H,L\}=0$. Since we have 2 conserved quantities and a 4-dimensional phase space, the system is integrable. 
The classical equations of motion are
\begin{equation}\label{eq:zeom}
    \dot z_1 = \{H, z_1\} = (1+|z_1|^2+|z_2|^2)\qty(\frac{1+|z_1|^2}{2\rmi j}\partial_{\bar z_1} H + \frac{z_1\bar z_2 }{2\rmi j}\partial_{\bar z_2}H) \, ,
\end{equation}
which simplifies to
\begin{equation}\label{eq:zeomsimple}
    \dot z_1 =  - \frac{\bar{z}_1+z_1(z^2_1+z^2_2)}{1+|z_1|^2+|z_2|^2} \, ,
\end{equation}
together with the ones where $1\leftrightarrow 2$ (under which $H$ is invariant but $L$ flips sign), as well as their complex conjugates. Using the notation $j(2l,\,h)=(L,\,H)$ this can also be written as:
\begin{equation}
    \dot z_1 =   -\bar{z}_1 + \rmi l \bar{z}_2 + \rmi (1+|z_1|^2+|z_2|^2)h z_1 \, ,
\end{equation}
indicating that in the large-$j$ limit, states with fixed $H, L$ will behave as if in an inverted harmonic oscillator (IHO) potential, for as long as $z_1, z_2$ remain small. To restrict to this sector in the quantum system, we can coarse-grain in both the time and the angular directions. Before doing so, we study the structure of the classical orbits, as determined by \eqref{eq:leadingH}.

\paragraph{Hyperbolic fixed point at the origin}
Near $(z_1,z_2)\approx (0,0)$ the classical system becomes
\begin{equation}
  H \approx \rmi j\,(z_1^2+z_2^2 - \bar z_1^2 - \bar z_2^2)\,,\qquad \Omega \approx 2\rmi j\, \rmd z \wedge \rmd \bar z \, .
\end{equation}
This is the standard 2d IHO. The origin is a hyperbolic fixed point with energy $H=0$. In sec.\,\ref{sec:spectra}, we will see that it is responsible for a distinct quasiparticle peak in the density of states. Some orbits close to the origin are visualized in fig.\,\ref{fig:orbits}.

\begin{figure}[!ht]
    \centering
    \centering
  \begin{subfigure}{0.47\textwidth}
            \centering
            \begin{tikzpicture}
              \node[inner sep = 0pt] at (0,0) {\includegraphics[width=\textwidth]{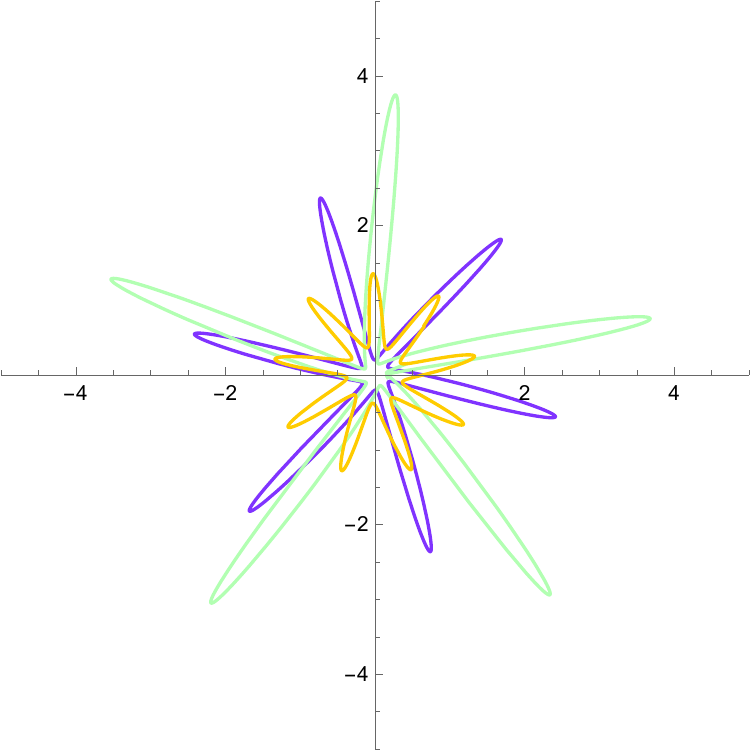}};
              \node at (0.9,3.4) {\text{$\Re(z_2)$}};
              \node at (3.5,-0.7) {\text{$\Re(z_1)$}};
            \end{tikzpicture}
            \caption[]
            {\small }
        \end{subfigure}
        \hspace{0.4cm}
        \begin{subfigure}{0.47\textwidth}  
            \centering 
        \begin{tikzpicture}
              \node[inner sep = 0pt] at (0,0) {\includegraphics[width=\textwidth]{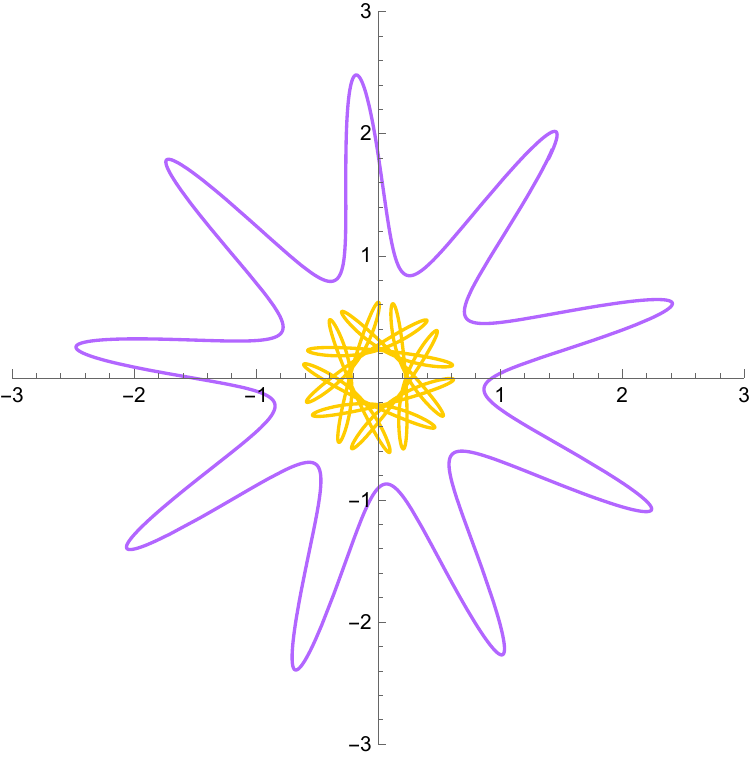}};
            \end{tikzpicture}
            \caption[]
            {\small }
        \end{subfigure}
    \caption{In (a) we show classical orbits in the $\Re(z_1)\Re(z_2)$ plane at a particular value $l = 0.135$ of the angular momentum. The three orbits have energies $h = 0.08, 0.17, 0.36$. The ones with lower energy reach larger radii. For aesthetic purposes, we chose particular parameters in order for the orbits to close. In (b), we show the highest energy orbit of (a) in the $(x_1 x_2)$ (yellow) and $p_1p_2$ (purple) planes. These are the coordinates $\Re(z_i)=x_i+p_i$ in which the Hamiltonian becomes an upside-down oscillator $p^2-x^2$ close to the origin. }
    \label{fig:orbits}
\end{figure}

\paragraph{Circles of elliptic fixed points}
The fixed points are determined from \eqref{eq:zeomsimple} by requiring
\begin{equation}
    \bar{z}_1+z_1(z^2_1+z^2_2) = \bar{z}_2 + z_2(z^2_1+z^2_2) =0\,.
\end{equation}
If $z^2_1+z^2_2 = 0$, then we are led to $(z_1,z_2)=(0,0)$, which we already found. If not, then we must have $(z_1,z_2) = (\cos\phi,\, \sin\phi)\, \rme^{\rmi \theta}$, with $\theta \in \{\frac{\pi}{4}, \frac{3\pi}{4}, \frac{5\pi}{4}, \frac{7\pi}{4}\}$. These constitute 4 circles of fixed points, each with angular momentum $L = 0$ and energy $H= -\frac{j}{2}\sin2\theta = \pm \frac{j}{2}$. These are therefore high-energy fixed points, which are not expected to contribute to coarse-grained quantities in the large-$j$ limit. The only fixed points we may have missed in the above considerations are the ones at infinity. We turn to these next.

\paragraph{Sphere of hyperbolic fixed points at infinity}

The 2-sphere at infinity corresponds to  $z_1, z_2 \to \infty$, or equivalently, taking the homogeneous coordinate $U \to 0$. From \eqref{eq:hamhom} it is clear that $H \to 0$ in this limit. Defining new inhomogeneous coordinates $(w_1,w_2)= (U,V)/W$, the sphere at infinity consists of the points $(0,w_2)$. Close to it, we find 
\begin{equation}
    H_j \approx j\,\frac{\bar w_1^2(1+w_2^2) -  w_1^2(1+\bar w_2^2)}{(1+ w_2 \bar w_2)^2} \,,\qquad \Omega \approx 2\rmi j\, \rmd w \wedge \rmd \bar w \, .
\end{equation}
It is then easy to check that $(0,w_2)$ is a fixed point for any $w_2$, with angular momentum
\begin{equation}
    L = 2\,\rmi j \, \frac{w_2 -\bar w_2}{1+|w_2|^2}\,.
\end{equation}
Close to the fixed points, we find
\begin{equation}
   \dot w_1 \approx -\frac{\bar{w}_1 (1+w_2^2)}{1+|w_2|^2} \, , \qquad \dot w_2 \approx 0\,.
\end{equation}
Applying this twice, we get
\begin{equation}\label{eq:lyapunov}
    \ddot{w}_1 = \lambda^2 w_1\, , \qquad \lambda^2 = \frac{(1+w^2_2)(1+\bar{w}^2_2)}{(1+|w_2|^2)^2} \,.
\end{equation}
We therefore have a collection of 1d IHOs, whose frequency is modified as a function of $w_2$. Note however that $\lambda^2 = 1 \Leftrightarrow w_2 \in \IR \Leftrightarrow L/j=0$. In other words, coarse-graining over $L$ only retains states with the standard IHO spectrum.  

% Due to the IHO behavior, orbits take an infinite amount of time to reach the fixed points. To each value of $-j \leq L\leq j$ corresponds a circle at infinity. Since there are no $L\neq 0$ fixed points away from infinity, the 
% $L\neq 0$ orbits must begin and end at infinity. We can then conclude that all $H=0$ orbits take an infinite amount of time as $j\to \infty$ and that only the $L=0$ ones can connect the origin to infinity, i.e. the observer to the horizon.  

\paragraph{Adding a mass term}
Since our aim will be to mimic the spectrum of a massive particle in $\dS_3$, with scaling dimension $\Delta= 1 + \rmi \nu$, we can add a mass term of the form
\begin{equation}
    \frac{\nu}{j} (U\partial_U-j) \approx \nu \,\frac{U^2-V^2-W^2}{U^2+V^2+W^2}\,,
\end{equation}
{where as before $\approx$ indicates the classical limit or coherent state expectation value.} 
Just like the $J_z$ term in \eqref{eq:hamds2}, it interpolates between $\nu$ at the origin and $-\nu$ at infinity. Besides this additive constant at the hyperbolic fixed points, it only has a subleading effect on the large-$j$ dynamics.

\section{Spectral properties of the quantum Hamiltonian}\label{sec:spectra}
A hyperbolic fixed point in the classical phase space yields a tower of resonances at the quantum level. This allows us to write down an analytic expression for the large-spin density of states. We verify this prediction by comparing it with results obtained by numerical diagonalization. We also analyze the level spacing statistics and spectral form factor of the spin Hamiltonians introduced in the previous section. 

\subsection{\texorpdfstring{$\SU(2)$}{}}
If we label the states by their distance $0\leq n\leq 2j$ to the lowest-spin state, we have the standard relations 
\begin{align}
    J_z&\ket{n}=(n-j)\ket{n}~,\\
    J_+&\ket{n}=\sqrt{(n+1)(2j-n)}\ket{n+1}~,\\
    J_-&\ket{n}=\sqrt{n(2j-n+1)}\ket{n-1}~.
\end{align}
The Hamiltonian  \eqref{eq:hamds2} then acts on these states as
\begin{equation}\label{eq:Hamiltonian spin model}
    H_j\ket{n}=\rmi c_{n}\ket{n-2}- \rmi c_{n+2}\ket{n+2}+ \qty(\frac{n}{j}-1)\nu\ket{n}~,
\end{equation}
%\begin{equation}\label{eq:Hamiltonian spin model}    H_j\ket{j,j-n}=\nu\qty(1-\frac{n}{j})\ket{j,j-n}+\rmi\qty[c_{n}\ket{j,j-n-2}-c_{n+2}\ket{j,j-n+2}]~,\end{equation}
where
\begin{equation}
    c_{n}=\frac{1}{4j}\sqrt{n(n-1)(2j-n+1)(2j-n+2)}~.
\end{equation}
Even and odd subspaces are clearly invariant. We can use the above expressions to numerically diagonalize $H_j$, and analyze its spectrum.

\subsubsection{Character and density of states}
It was shown in \cite{Parmentier:2023axg} that the inverse level spacing of the Hamiltonian \eqref{eq:hamds2} converges to the density of states \eqref{eq:exactrho} for a massive particle in $\dS_2$ with scaling dimension $\Delta=\frac12+\rmi \nu$. The numerical result is shown in fig.\,4.3 in \cite{Parmentier:2023axg}, reproduced here in fig.\,\ref{fig:avmatch}. See also \cite{PhysRevE.78.021106} for a related large-$j$ analysis of the various phases of the LMG model using different methods.  

\begin{figure}
    \centering
  \begin{subfigure}{0.45\textwidth}
            \centering
            \begin{tikzpicture}
              \node[inner sep = 0pt] at (7,3) {\includegraphics[width=\textwidth]{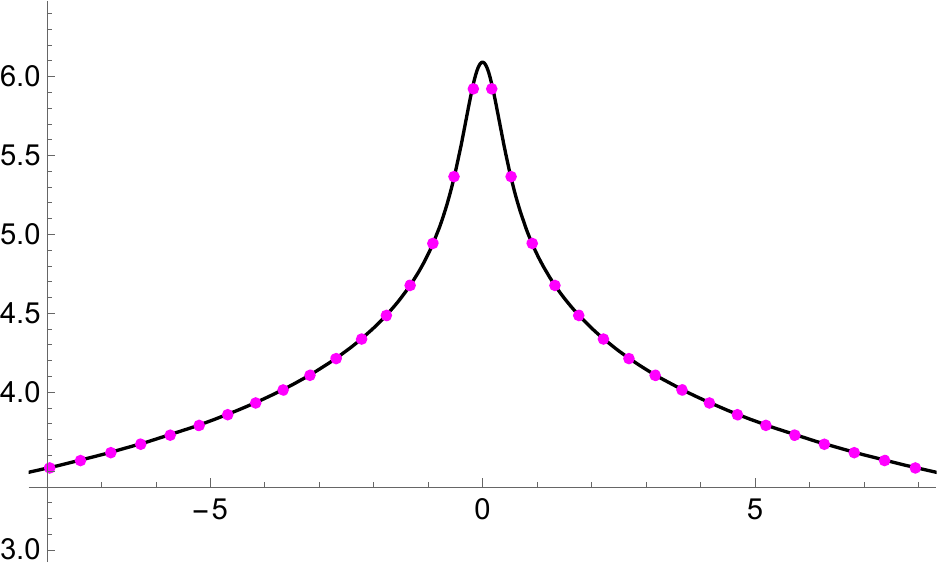}};
               \node (x2) at (4.3,5.05) {\text{$\rho$
        }};\node (jplus) at (10.5, 0.9) {\text{$\omega$
        }};
        \end{tikzpicture}
            \caption[]%
            {{\small  $\nu=0$}} 
        \end{subfigure}
        \hspace{0.5cm}
        \begin{subfigure}{0.45\textwidth}  
            \centering 
            \begin{tikzpicture}
              \node[inner sep = 0pt] at (7,3) {\includegraphics[width=\textwidth]{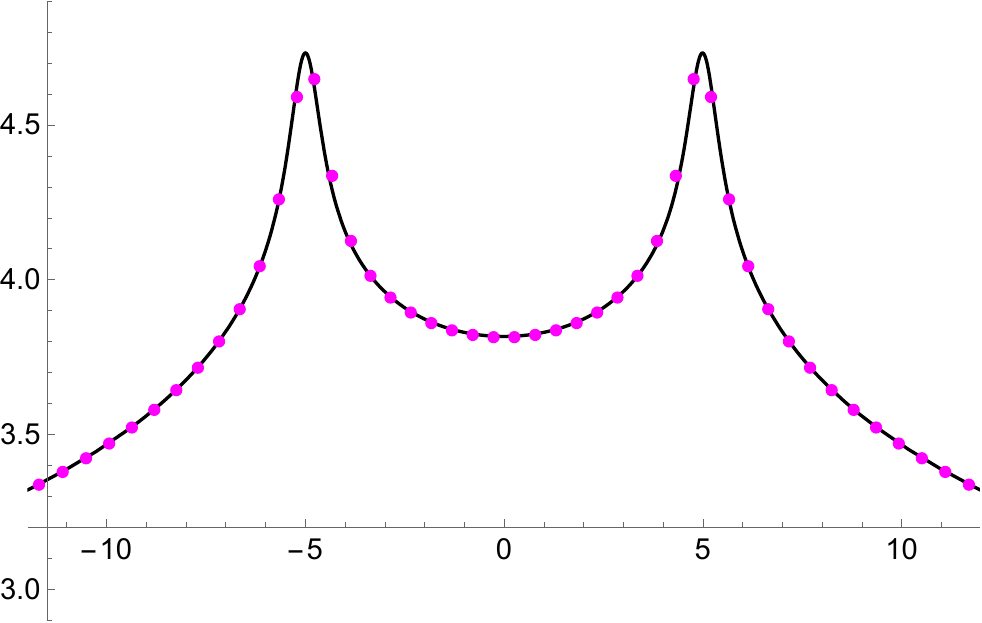}};
               \node (x2) at (4.3,5.05) {\text{$\rho$
        }};\node (jplus) at (10.5, 0.9) {\text{$\omega$
        }};
        \end{tikzpicture}
            \caption[]%
            {{\small  $\nu=5$}}
        \end{subfigure}
    \caption{The total density of states found by numerically diagonalizing the model in \eqref{eq:hamds2} with 1001 states is in excellent agreement with the analytic result \eqref{eq:exactrho} shown by the black curve for a massive particle in $\dS_2$ with scaling dimension $\Delta=\frac12+\rmi \nu$.}\label{fig:avmatch}
\end{figure}
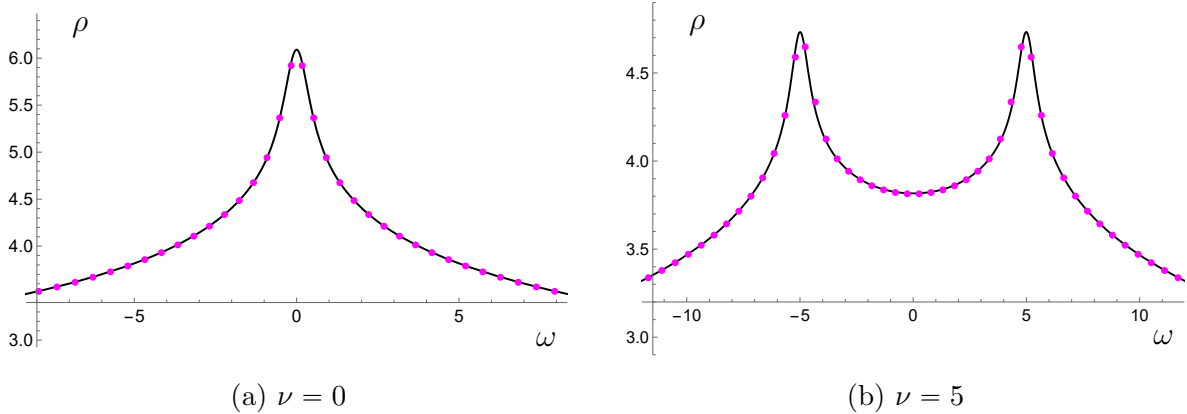

Demonstrating the convergence is simplest at the level of the coarse-grained character 
\begin{equation}
    \chi_{j,\epsilon}(t) = \sum_n \rme^{-\rmi t \omega_n - \epsilon^2 \omega^2_n}\,,
\end{equation}
where the coarse-graining parameter $\epsilon$ represents a Gaussian averaging over a time window of size $\epsilon$, {and $\omega_n$ is the frequency of the $n$th energy level}. Knowledge of the classical dynamics, discussed in sec.\,\ref{sec:classSU2}, is sufficient to understand the large-$j$ behavior of this quantity. One can use coherent spin states to show that in the large-$j$ limit, $\chi_{j,\epsilon}$ receives contributions from the two hyperbolic fixed points\footnote{The elliptic fixed points have energies that grow with spin $j$ and will not contribute to coarse-grained observables. The same holds for periodic orbits, which only contribute after a time $\sim \log \epsilon j$ \cite{Parmentier:2023axg}. The reason is that the zero-energy orbits have infinite periods in the limit $j\to \infty$.}, each of which contributes to the character as an IHO \cite{Parmentier:2023axg}:
\begin{equation}
     \chi_{j,\epsilon}(t) \to \frac{\rme^{-\Delta t}+\rme^{-\bar \Delta t}}{|1-\rme^{-t}|}\,.
\end{equation}
The RHS equals the Harish-Chandra character $\chi_{\dS_2}(t)= \tr\rme^{-\rmi t H}$ for a massive particle in $\dS_2$, where $H$ is the static patch Hamiltonian. Expanding $\chi_{\dS_2}(t)$ at late times, it takes the form of a sum over QNMs. Having demonstrated convergence at the level of the character, one then takes a Fourier transform from $\chi(t)$ to $\rho(\omega)$ show that the inverse level spacing of $H_j$ converges to that of a particle in $\dS_2$, given in \eqref{eq:exactrho}. See app.\,\ref{ds_dos} and \cite{Parmentier:2023axg} for more details.

In the next section, we will apply the same method to understand the density of states of the $\SU(3)$ system \eqref{eq:spds3}. This method also shows that certain modifications of $H_j$ in \eqref{eq:hamds2} leave the large-$j$ density unchanged. A simple example is to replace the mass term by $\nu (\frac{J_z}{j})^3$, which does not change the leading classical dynamics near the hyperbolic fixed points.

\subsubsection{Level spacing statistics}\label{sec:spectral_stat}
The $\SU(2)$ spin system \eqref{eq:Hamiltonian spin model} we consider is classically integrable. Indeed, it is a Hamiltonian system with a 2-dimensional phase space. Such systems cannot have chaos \cite{poincare1886courbes,bendixson1901courbes}. 

The Berry-Tabor conjecture \cite{berry77} characterizes the level-spacing statistics of integrable quantum systems. The unfolded spacings $s_i = \epsilon_{i+1}-\epsilon_i$ are expected to obey Poisson statistics 
\begin{equation}\label{eq:Prob distr}
    P_{\rm int}=\rme^{-s}.
\end{equation}
The unfolding procedure is required to have statistically meaningful level spacings, independent of the local model-dependent density of states $\rho(\omega)$. This is done by computing the average number of levels less than a given energy value $\omega_i$,
\begin{eqnarray}\label{eq:unfolded epsilon}
    \varepsilon_i=\int_{-\infty}^{\omega_i}\rho(\omega)\rmd \omega~,
\end{eqnarray}
so that the mean level spacing becomes one.

This Poisson distribution is in contrast with chaotic systems satisfying random matrix universality. For instance, for a Hamiltonian drawn from the  Gaussian Orthogonal Ensemble (GOE), one has instead,
\begin{equation}
\quad P_{\rm GOE}=\frac{s\pi}{2}\rme^{-\frac{\pi}{4}s^2}~.
\end{equation}

The Poisson behavior in the integrable case originates from summing over uncorrelated levels. It only applies to systems with more than one degree of freedom and excludes the case of harmonic oscillators. Deviations from Poisson statistics can also point to extra symmetries responsible for extra degeneracies.

With one degree of freedom, as in our case, the spacing is highly sensitive to the details of the system.  In general, the local density of states $\rho(\omega)$ is found by averaging over several eigenvalues, but from fig.\,\ref{fig:avmatch}, we know that this was not necessary for our $\SU(2)$ system \eqref{eq:Hamiltonian spin model}: the local density of states was essentially equal to the inverse level spacing, implying that the unfolded level spacing will simply be $1$. One can make this $\delta$-like level spacing distribution Poisson by turning certain parameters in the Hamiltonian into random variables \cite{Jeong:2024oao}. To achieve this here, we could consider the parameter $\nu$ in \eqref{eq:hamds2} as drawn from a Gaussian distribution, but this does not have an immediate physical interpretation here. The $\SU(3)$ system \eqref{eq:spds3} is more interesting in terms of the level spacing, as we will see in sec.\,\ref{sssec:level DU3}. 

% %%%%%%%%%%%%%%%%%%%%%%%%%%%%%%%%%%
% \begin{figure}
%     \centering
%     \subfloat[]{
%     \centering 
%     \includegraphics[width=0.31\textwidth]{figures/Level_spacing.pdf}\label{Fig: level_spacing_SU2_delta}}
%     \hfill
%     \subfloat[]{
%     \centering 
%     \includegraphics[width=0.31\textwidth]{figures/Level_Spacing_Ran_sigma2.pdf} \label{Fig: level_spacing_SU2__sigma2}}
%     \hfill
%     \subfloat[]{
%     \centering 
%     \includegraphics[width=0.31\textwidth]{figures/Level_Spacing_Ran_sigma5.pdf}\label{Fig: level_spacing_SU2__sigma5}}
%     \caption{Unfolded level spacing distributions in SU(2) the spin model (\ref{eq:hamds2}) for $j=500$. The blue curve is the Poisson distribution. (a) is the plot obtained from the original Hamiltonian. After introducing a Gaussian distributed random term with mean $\langle \nu \rangle=2$ and standard deviation $\sigma=2$ and $\sigma=5$, the distribution follows Poisson in (b) and (c), respectively.\sa{we might want to move it}}
%     \label{Fig: level_spacing_SU2}
% \end{figure}
%%%%%%%%%%%%%%%%%%%%%%%%%%%%%%%%%%

\subsubsection{Spectral form factor}
The SFF is a valuable tool that has been used to probe the chaotic nature of quantum systems since it displays a characteristic dip-ramp-plateau structure for random matrix theories,  and hence is expected to show a similar structure for more generic chaotic quantum systems \cite{brezin1997spectral,Liu:2018hlr,Gaikwad:2017odv,Cotler:2016fpe}. The original relation between the SFF and spectral rigidity was formulated by Berry in \cite{berry85}. The SFF is defined as follows in terms of the analytically continued partition function (below we will always consider the case of infinite temperature, $\beta=0$)
\begin{eqnarray}\label{eq:SFF_av}
    {\rm SFF}(t)=\expval{\tr(\rme^{-\rmi H t})\tr(\rme^{\rmi H t})}\,.
\end{eqnarray}
The expectation value indicates an ensemble averaging over several realizations in case the Hamiltonian belongs to an ensemble. This is not the case for
the Hamiltonian in \eqref{eq:hamds2}. We can numerically evaluate \eqref{eq:SFF_av} using the eigenvalues of the Hamiltonian for large $j$-spin values. {Although there is no ensemble average for this model, we do include a moving average, which calculates the average of a set of data points over a period of time that is ``moved" forward as new data becomes available,} to smoothen out the SFF data, making it easier to interpret.
The results are displayed in fig.\,\ref{fig:spectral_form_factor}. Compared to the SFF of chaotic systems, there is essentially no ramp. Instead of the slope and dips, there is a rapid change to the plateau. The absence of the ramp is another indication that the model is integrable \cite{berry85}. Fig.\,\ref{fig:spectral_form_factor} (b) also features small oscillations in the dip when $\nu\neq0$. It would be interesting to understand this feature in terms of (dis)connected correlations in the SFF (\ref{eq:SFF_av}).\footnote{We thank Pratik Nandy for this suggestion.}

\begin{figure}[ht]
    \centering
  \begin{subfigure}{0.45\textwidth}
            \centering
            \begin{tikzpicture}
              \node[inner sep = 0pt] at (0,0) {\includegraphics[width=\textwidth]{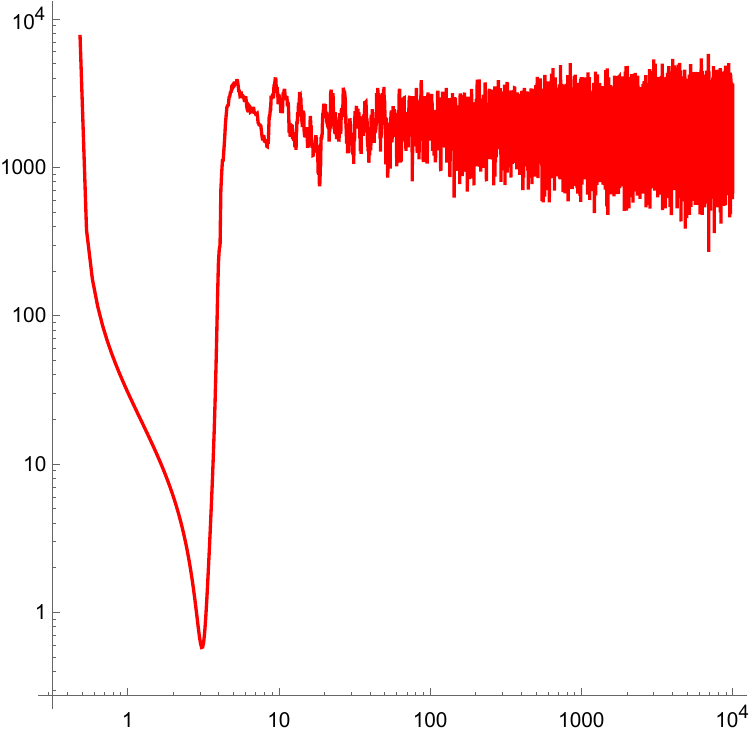}};
              \node at (3.4,-2.6) {$t$};
              \node at (-2.2,3.2) {SFF};
            \end{tikzpicture}
            \caption[]
            {\small $\nu=0$}
        \end{subfigure}
        \hspace{0.6cm}
        \begin{subfigure}{0.45\textwidth}  
            \centering 
        \begin{tikzpicture}
              \node[inner sep = 0pt] at (0,0) {\includegraphics[width=\textwidth]{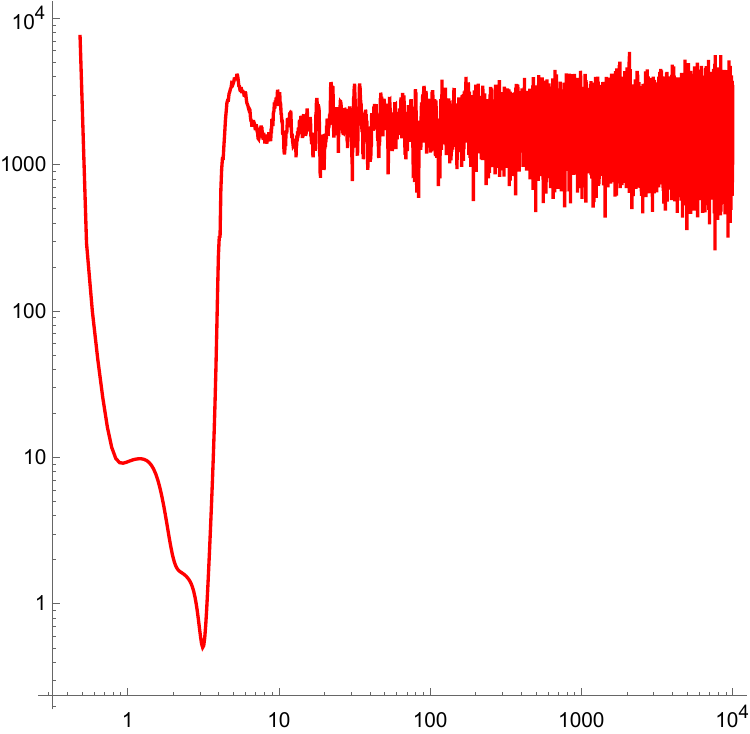}};
              \node at (3.4,-2.6) {$t$};
              \node at (-2.,3.2) {SFF};
            \end{tikzpicture}
            \caption[]
            {\small $\nu=5$}
        \end{subfigure}
    \caption{Evolution of the SFF  for the spin model \eqref{eq:Hamiltonian spin model} with $j=500$, and (a) $\nu=0$ or (b) $\nu=5$. We took samples at time intervals of 0.05, and then performed a moving average over 20 such points, corresponding to $\Delta t = 1$. The effect of $\nu$ is to slightly modify the oscillation pattern. Notably, there is no linear ramp in the log-log scale plot; the plateau emerges immediately after the dip, which occurs at a time $t\approx 2$ independent of $\nu$ and $j$ in the large-$j$ limit.}
    \label{fig:spectral_form_factor}
\end{figure}
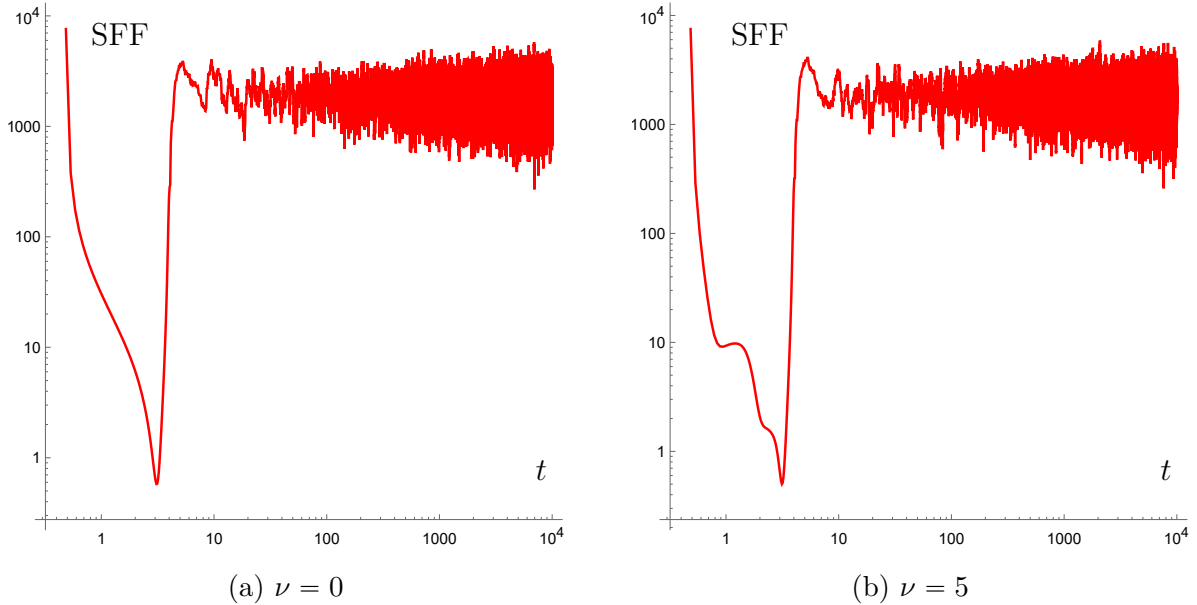

% \begin{figure}
%     \centering
%     \subfloat[]{\includegraphics[width=0.475\textwidth]{figures/spectral2.pdf}}\hfill\subfloat[]{\includegraphics[width=0.475\textwidth]{figures/spectral3.pdf}}
%     \caption{Evolution of the SFF with for the $SU(2)_j$ spin model with: (a) $j=1000$ (\textit{blue}), $j=2000$ (\textit{red}) and $\nu=5$. (b): $j=1000$ with $\nu=0$ (\textit{blue}) or $\nu=5$ (\textit{red}). The effect of modifying $j$ is to change the shape and amplitude of the plots; while changing $\nu$ modifies the plateau oscillation pattern. Notably, there is no linear ramp, the plateau emerges immediately after the dip, which occurs at a time independent of $j$ in the large-$j$ limit. }
%     \label{fig:spectral_form_factor}
% \end{figure}

% {\cp What is the precise shape of the ramp? In \cite{DeClerck:2023fax} the authors found an exponential ramp. Their classical billiard dynamics is chaotic, but there are a lot of conserved quantities. This leads to what is called arithmetic chaos. They also mention certain integrable variants of SYK where such ramps are observed \cite{liao2020many, winer2020exponential}}

% \kp{Fig. 9b in \cite{Balasubramanian:2023kwd} is perhaps interesting. Looks like a non-random system like ours is as having high Dyson index. They also mention `the lower noise in the Lanczos spectrum around the analytical mean computed from RMT for a given density of states, the higher the coherence, and the larger the oscillations before the system settles onto the plateau'.}

\subsection{\texorpdfstring{$\SU(3)$}{}}
From \eqref{eq:spds3} it follows that $H_j$ is block diagonal in the angular momentum basis \eqref{L_eigenbasis}. We can therefore diagonalize within each block. The numerical results agree with the analytic prediction \eqref{eq:SU3an} at fixed energy and angular momentum in the large-$j$ limit. 
\subsubsection{Character and density of states}
We can figure out the large-$j$ density of states in a sector of fixed angular momentum $L$ by following the same strategy as in the $\SU(2)$ case. Indeed, we understood the classical orbits in sec.\,\ref{sec:classSU3}.
The hyperbolic fixed point at the origin is expected to contribute to the inverse level spacing as a 2d IHO, whose density is determined by its QNMs. In the fixed-$L$ sector, this contribution corresponds to:\footnote{These are the $+\nu$ terms in \eqref{eq:rhods3}. The interested reader can consult app.\,\ref{ds_dos} for the intermediate steps leading to the result given above.}
\begin{equation}
    \rho_{+}(\omega, L) = -\frac{1}{4\pi} \sum_{\pm}\psi\big(\tfrac12 + \tfrac{|L|}{2}\pm \tfrac{\rmi}{2} (\nu-\omega)\big)\,,
\end{equation}
which has the same positive frequency behavior as the density of states of a massive particle in $\dS_3$. However, we saw in sec.\,\ref{sec:classSU3} that there is also a fixed point at infinity in each $L$-sector. Near this fixed point, the states experience a 1d IHO. Counting the IHO resonances with the same parity as $L$, we thus expect a further contribution $\rho_\text{even/odd}$, whose explicit expression can be found in \eqref{eq:even/odd res}. The analytic prediction for the total density of states in the large-$j$ limit is then 
\begin{equation}\label{eq:SU3an}
    \rho(\omega,L) = \rho_+(\omega,L) + \rho_{\text{even/odd}}(\omega)\,.
\end{equation}
As shown in fig.\,\ref{fig:dos}, this agrees very well with the numerical results. The rightmost peak, with $\omega>0$, corresponds to that for a fixed angular momentum sector in $\dS_3$. However, since the left peak is created by a 1D IHO, this part of the spectrum has no clear $\dS_3$ interpretation. One could perhaps imagine taking $\nu$ large enough so that the peaks are quite separated and then focus on the $\omega >0$ part of the spectrum. In the rest of the paper, we will study properties of $H_j$, regardless of a possible de Sitter interpretation.

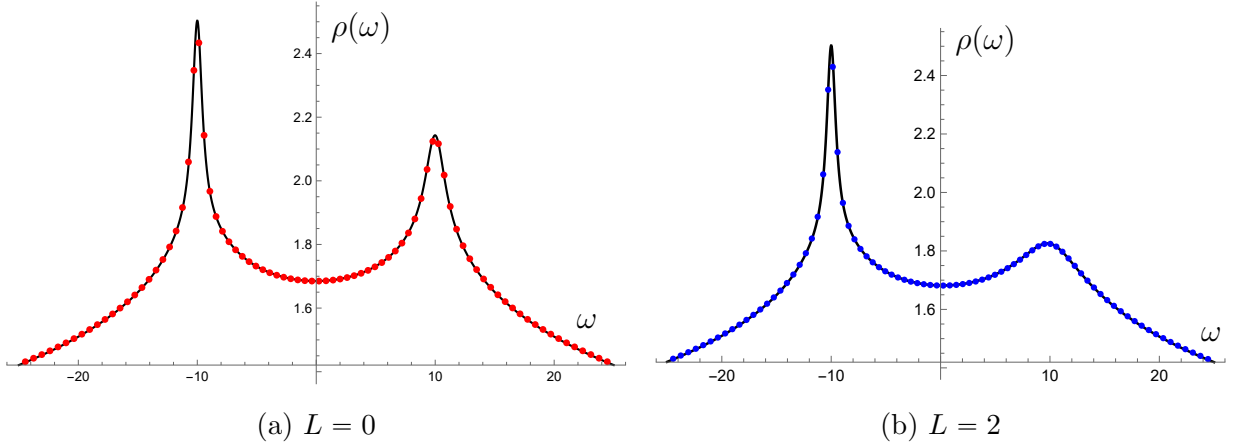
\begin{figure}
    \centering
  \begin{subfigure}{0.5\textwidth}
            \centering
            \begin{tikzpicture}
              \node[inner sep = 0pt] at (0,0) {\includegraphics[width=\textwidth]{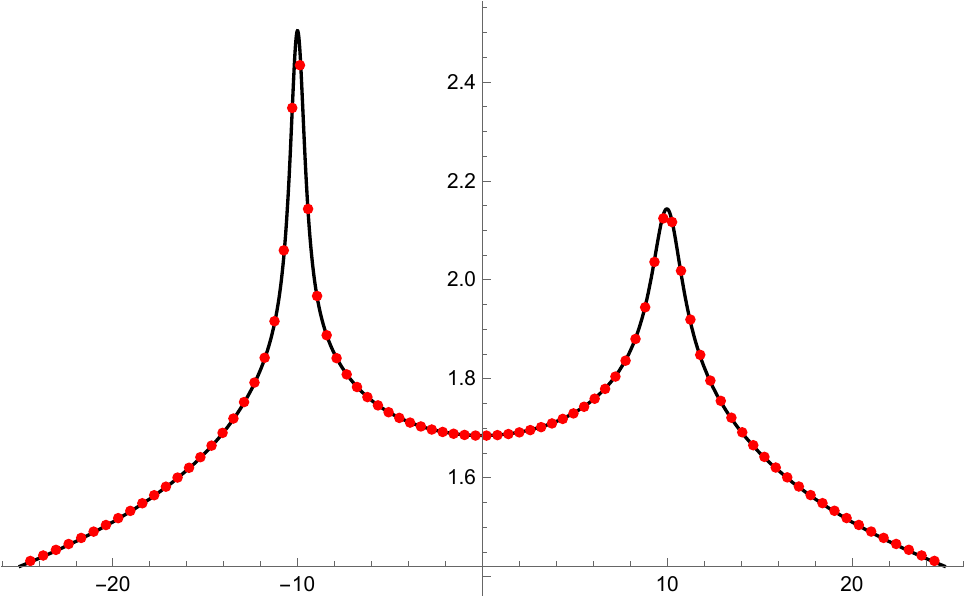}};
             \node at (0.6,2.2) {\text{$\rho(\omega)$}};
             \node at (3.6,-1.7) {\text{$\omega$}};
            \end{tikzpicture}
            \caption[]
            {\small $L=0$}
        \end{subfigure}
        \hspace{0.1cm}
        \begin{subfigure}{0.46\textwidth}  
            \centering 
        \begin{tikzpicture}
              \node[inner sep = 0pt] at (0,0) {\includegraphics[width=\textwidth]{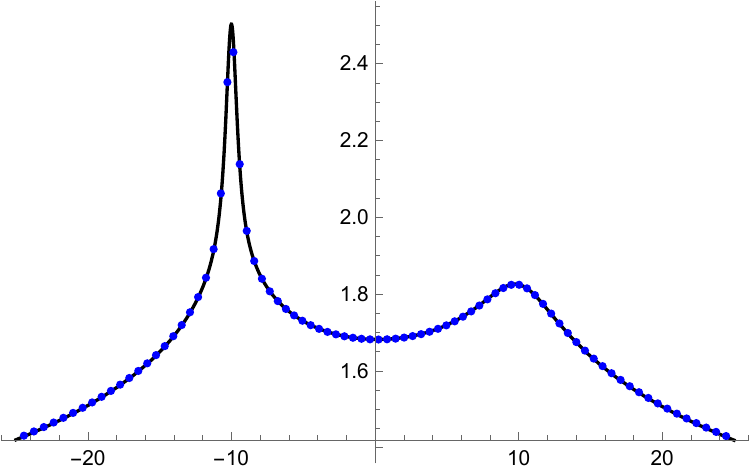}};
              \node at (0.6,2.2) {\text{$\rho(\omega)$}};
              \node at (3.6,-1.7) {\text{$\omega$}};
            \end{tikzpicture}
            \caption[]
            {\small $L=2$}
        \end{subfigure}
    \caption{Inverse level spacings of the $\SU(3)$ system \eqref{eq:spds3} at $j=300$, compared to analytic prediction \eqref{eq:SU3an} for the density of states. The right peak equals that of fixed-$L$ states in $\dS_3$ static patch and is due to the IHO-like behavior near the origin. The left peak instead originates from the behavior at infinity, and is more like that in $\dS_2$ (even/odd resonances for even/odd $L$ resp.). Note how angular momentum smears out the static patch peak.}
    \label{fig:dos}
\end{figure}

\subsubsection{Level spacing statistics}\label{sssec:level DU3}
The $\SU(3)$ system \eqref{eq:spds3} has different sectors labeled by the angular momentum $L$. In each sector, the unfolded level spacings show a $\delta$-peak near $1$, as in the $\SU(2)$ case. However, when summing over different angular momentum sectors, the unfolded level spacings do become Poisson distributed $P(s)=\rme^{-s}$, in line with the general results by Berry and Tabor \cite{berry77}. The level spacing distributions are shown in fig.\,\ref{fig:su3levels}.  

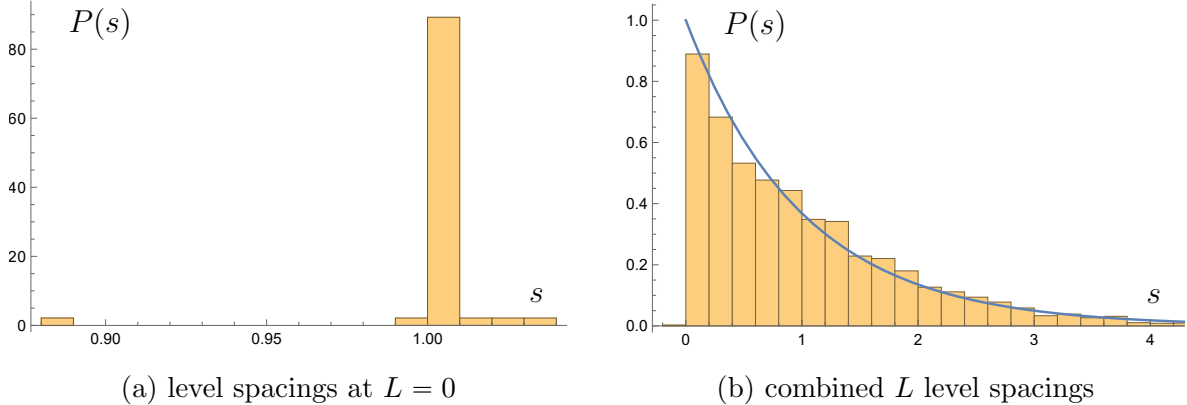
\begin{figure}
    \centering
  \begin{subfigure}{0.45\textwidth}
            \centering
            \begin{tikzpicture}
              \node[inner sep = 0pt] at (0,0) {\includegraphics[width=\textwidth]{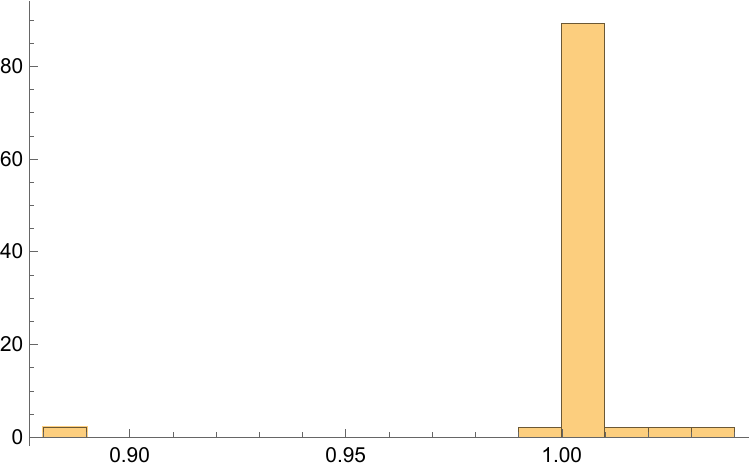}};
               \node at (3.3,-1.6) {$s$};\node at (-2.5,2) {$P(s)$};
            \end{tikzpicture}
            \caption[]
            {\small level spacings at $L=0$}
        \end{subfigure}
        \hspace{0.5cm}
        \begin{subfigure}{0.45\textwidth}  
            \centering 
        \begin{tikzpicture}
              \node[inner sep = 0pt] at (0,0) {\includegraphics[width=\textwidth]{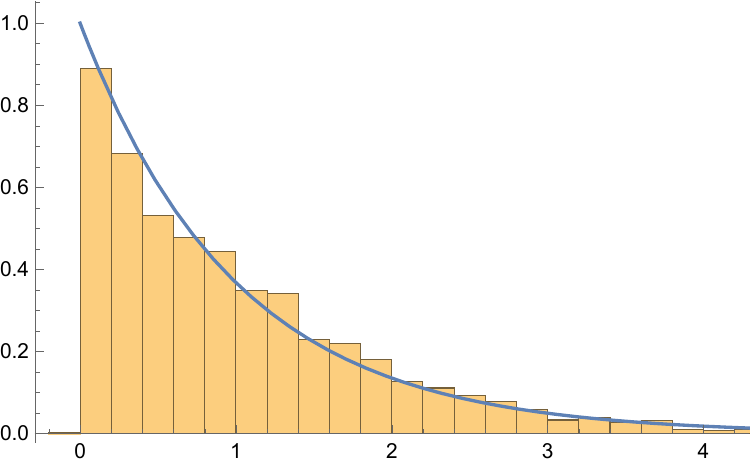}};
              \node at (3.3,-1.6) {$s$};\node at (-2,2) {$P(s)$};
            \end{tikzpicture}
            \caption[]
            {\small combined $L$ level spacings}
        \end{subfigure}
        % \hspace{0.4cm}
        % \begin{subfigure}{0.3\textwidth}  
        %     \centering 
        % \begin{tikzpicture}
        %       \node[inner sep = 0pt] at (0,0) {\includegraphics[width=\textwidth]{figures/unfoldsu3100r.pdf}};
        %     \end{tikzpicture}
        %     \caption[]
        %     {\small $r$-values at combined $l$}
        % \end{subfigure}
    \caption{We consider the $\SU(3)$ Hamiltonian \eqref{eq:spds3} at $j=100$. The PDF histogram plot in (a) shows very regular level spacings in a sector of fixed angular momentum $L=0$. In (b), we see that the level spacings do become Poisson distributed (blue) when considering the combination of all such $L$-sectors, in line with the general results of \cite{berry77}.}
    \label{fig:su3levels}
\end{figure}

\subsubsection{Spectral form factor}
We display the SFF for the $\SU(3)$ system in fig.\,\ref{fig:su3SFF}. From panel (a), we see that there is only a very small ramp when we sum over all modes $\omega_n$ in a fixed angular momentum $L$ sector. This is similar to what we saw in the SU$(2)$ case. In panel (b), we see that by combining the smallest eigenvalue $\omega_i >0$ of different fixed-$L$ sectors, we do get a ramp-like feature in the SFF. This ramp occurs at a much later time and can be seen to be approximately linear in shape, with slope $\approx 1$. Its appearance can be traced back to the fact that the eigenvalues vary quite slowly and in a determinate way as a function of $L$. In random matrix theories, the ramp is known to originate from a similar spectral rigidity \cite{berry85}. This also explains why we only see the ramp-like feature provided that we sum over a fixed number of $L$ states when taking $j$ large. The effect goes away when scaling $L$ with $j$ because then the large $L$ eigenvalues are quite uncorrelated with those at small $L$. See also the comment below \eqref{eq:lyapunov}. {A similar appearance of a ramp after summing over different angular momentum sectors was found in brick-wall systems \cite{Das:2022evy, Jeong:2024oao, Ageev:2024gem}. We comment more on this in the discussion sec.\,\ref{sec:discussion}.}

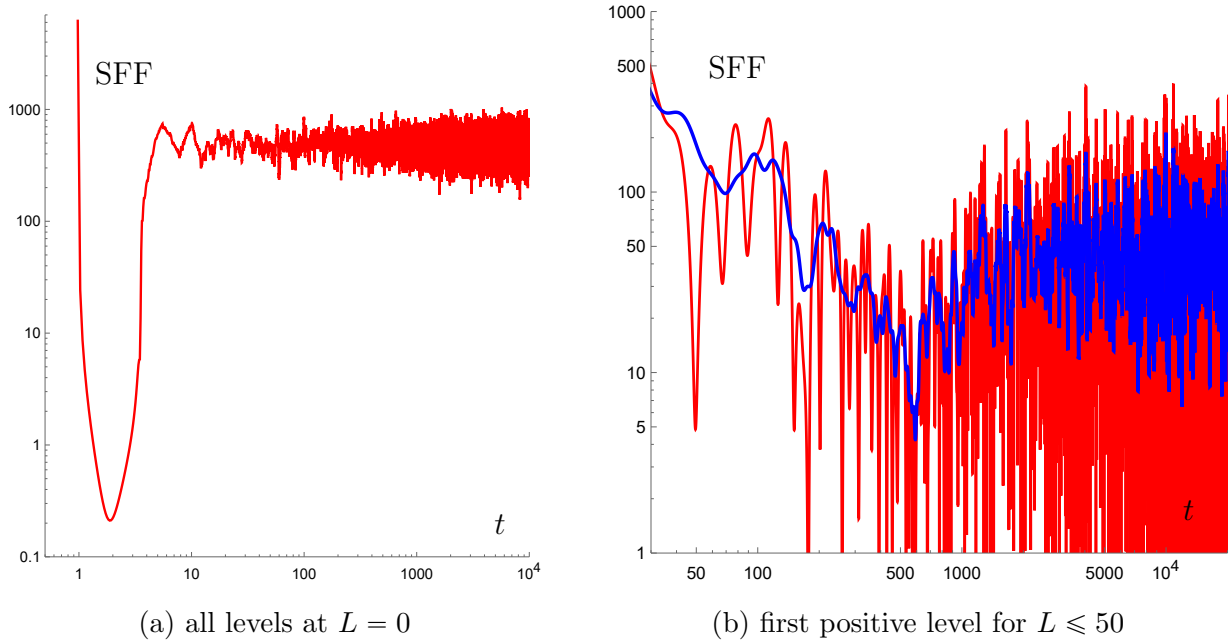
\begin{figure}[ht]
    \centering
  \begin{subfigure}{0.43\textwidth}
            \centering
            \begin{tikzpicture}
              \node[inner sep = 0pt] at (0,0) {\includegraphics[width=\textwidth]{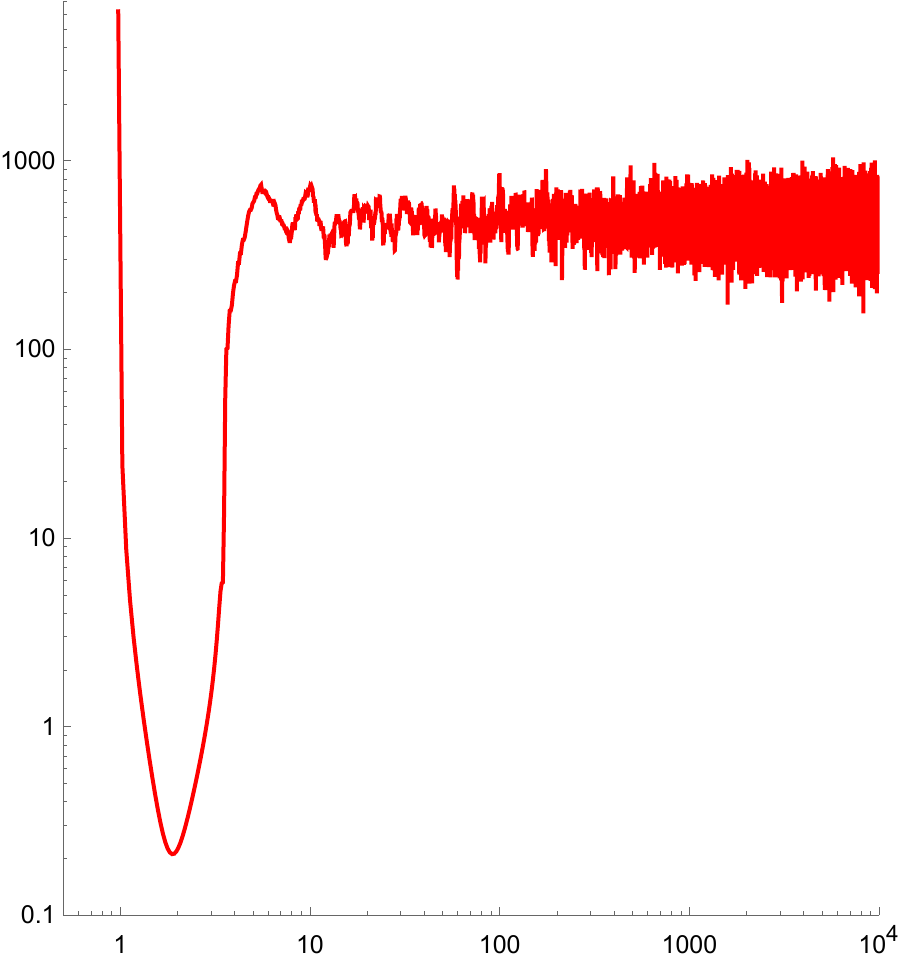}};
              \node at (3,-3) {$t$};
              \node at (-2.,3) {SFF};
            \end{tikzpicture}
            \caption[]
            {\small all levels at $L=0$}
        \end{subfigure}
        \hspace{0.6cm}
        \begin{subfigure}{0.5\textwidth}  
            \centering 
        \begin{tikzpicture}
              \node[inner sep = 0pt] at (0,0) {\includegraphics[width=\textwidth]{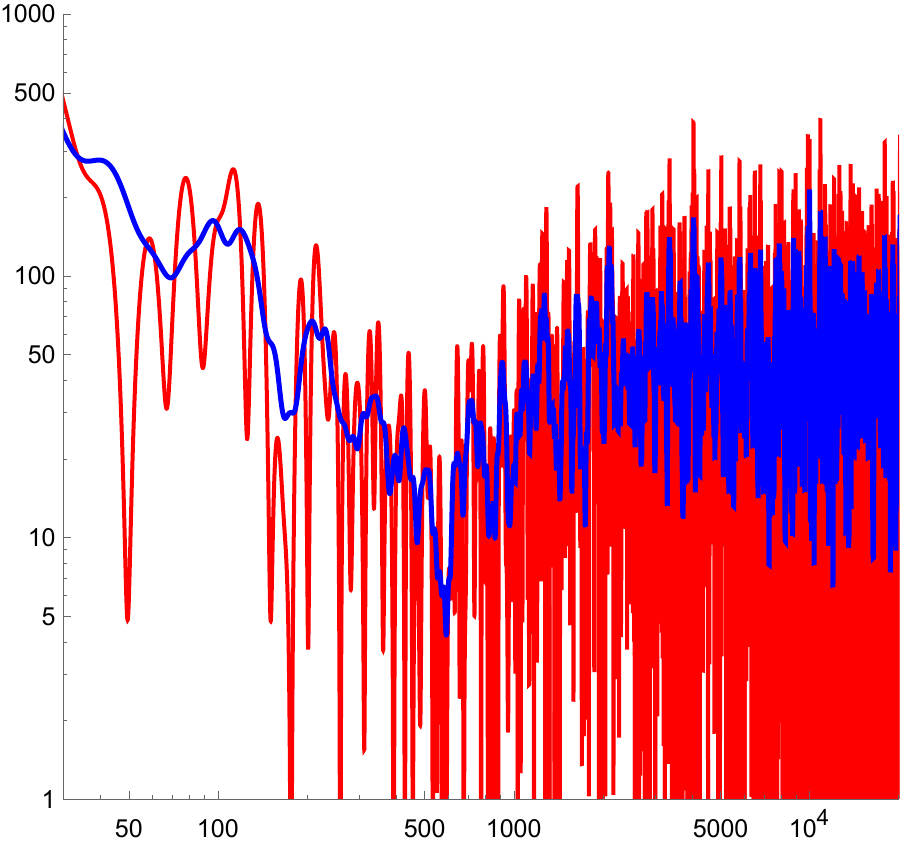}};
              \node at (3.6,-2.9) {$t$};
              \node at (-2.4,3) {SFF};
            \end{tikzpicture}
            \caption[]
            {\small first positive level for $L\leq 50$}
        \end{subfigure}
    \caption{We sample the SFF of the $\SU(3)$ system \eqref{eq:spds3} with $j=500$ and $\nu=0$ at time intervals 0.05. In (a), we consider all energy levels in the sector of fixed angular momentum $L=0$. We show the moving average over 40 data points ($\Delta t=2$). As in the $\SU(2)$ case of fig.\,\ref{fig:spectral_form_factor}, there is only a very small ramp. In (b), we look instead at the first positive energy level and sum over the contributions from $L\leq 50$. This leads to a ramp-like behavior at a larger timescale. In red is the sampled data, in blue their moving average over 1000 points ($\Delta t=50$).}
    \label{fig:su3SFF}
\end{figure}
% \begin{figure}
%     \centering
%     \subfloat[]{\includegraphics[width=0.5\linewidth]{figures/j1000l0.pdf}}\hfill
%     %\subfloat[]{\includegraphics[width=0.5\linewidth]{figures/j1000l02.pdf}}\\
%     \subfloat[]{\includegraphics[width=0.5\linewidth]{figures/j1000l60.pdf}}
%     %\subfloat[]{\includegraphics[width=0.5\linewidth]{figures/j500l1202.pdf}}
%     \caption{Evolution of the SFF (\ref{eq:SFF_av}) for the SU(3)$_j$ spin model (\ref{eq:spds3}) with a moving average to smoothen the data. The parameters include (a) $j=1000$ (blue), $j=2000$ (red) $l=0$ and $\nu=5$. In (b) we evaluate (\ref{eq:SFF_av}) for the lowest positive eigenvalue of (\ref{eq:spds3}) using and $l\in[-l_{\rm max},~l_{\rm max}]$  with $l_{\rm max}=60$, and we take $j=1000$ (blue), $j=500$ (red) with $\nu=0$. We observe than changing $j$ modifies the amplitudes and shapes of the plots (while changing $\nu$ has the effect on the oscillations in the plateau, similar to Fig. \ref{fig:spectral_form_factor}). \sa{Changing $l_{\rm max}$ doesn't make the plot nicer}}
%     \label{fig:SU3 SFF new}
% \end{figure}

%%%%%%%%%%%%%%%%%%%%%%%%%%%%%%%%%%%%%%%%%%%%%%%%%
\section{Saddle-dominated scrambling versus chaos}\label{sec:spread complexity}
%%%%%%%%%%%%%%%%%%%%%%%%%%%%%%%%%%%%%%%%%%%%%%%%%
In this section we study various quantities commonly used in the literature to probe the time evolution of integrable and chaotic systems.  We focus on the two-point correlation functions, squared commutators and Krylov operator complexity for the spin operators, as well as the spread complexity of the TFD state.

% \begin{table}[t!]
%     \centering
%     \begin{tabular}{c|c|c}
%         \textbf{Probe} & \textbf{Success}& \textbf{Late-time features}\\
%         Two-point correlation functions & \bluecheck&Increase and oscillations\\
%         Squared commutators &  \bluecheck&Decay and oscillations\\
%         Krylov operator complexity &  \bluecheck&Decay and oscillations\\
%         Spread complexity & \bluecheck&Decay and oscillations
%     \end{tabular}
%     \caption{{Probes used to discriminate between chaos and saddle-dominated scrambling in the $SU(2)_j$ spin model. All probes succeed in identifying saddle-point dominated scrambling. Even though the \sa{early-time behavior} of these quantities are analogous to those of quantum chaotic systems \sa{\sout{(specifically those refer to early time behavior of the system)}}, \sa{the late-time \sout{there are certain}} properties exhibited by these probes that can be utilized to distinguish underlying integrable dynamics of a saddle-dominated system, such as our spin model, from genuine quantum chaotic systems having level repulsion in the energy spectrum.} }
%     \label{tab:comparizon_probes}
% \end{table}

\begin{table}[t!]
    \centering
    \begin{tabular}{c|c|c}
        \hline
        \textbf{Probe} &  \multicolumn{2}{c}{Capability of distinguishing saddle-dominated}\\ 
         &  \multicolumn{2}{c}{ scrambling from chaos}\\ \cline{1-3}
        & Early time& Late time\\ \cline{2-3}
        Two-point correlation functions & Fast scrambled \redcross& No saturation \bluecheck\\
        Squared commutators & Exponential growth \cite{Xu:2019lhc} \redcross & No saturation \bluecheck\\
        Krylov operator complexity & Exponential growth \cite{ Bhattacharjee:2022vlt} \redcross & Saturation value $> D_{\mathcal{O}}/2$ \bluecheck\\
        Spread complexity (TFD$_{\infty}$) & Peak \cite{Huh:2023jxt}\redcross & Saturates higher than peak \bluecheck\\
        \hline
    \end{tabular}
    \caption{{Probes used to discriminate between chaos and saddle-dominated scrambling in the $\SU(2)_j$ spin model. All probes succeed in identifying saddle-point dominated scrambling. Even though their early-time behavior resembles that of quantum chaotic systems, there are clear late-time properties exhibited by these probes that can be utilized to identify the underlying integrable dynamics of a saddle-dominated system, such as our spin model. These properties can help to distinguish it from genuine quantum chaotic systems with level repulsion in the energy spectrum.} }
    \label{tab:comparison_props}
\end{table}

These quantities are often used to distinguish the chaotic versus integrable nature of quantum systems. However, as mentioned in the introduction, certain quantum systems that are classically integrable can nevertheless display features reminiscent of chaos. In saddle-dominated scrambling, an unstable fixed point gets smeared over a finite-measure region in phase space, leading to exponential
growth in the squared commutators \cite{rozenbaum2020early, Xu:2019lhc} and in Krylov complexity \cite{Bhattacharjee:2022vlt}, and a peak in spread complexity \cite{Huh:2023jxt}.
{In the following,  we confirm that these features also occur for the $\SU(2)$ version of the spin system \eqref{eq:hamds2}.} Even so, we argue that a careful analysis of these quantities still allows us to disentangle saddle-dominated effects from `true' quantum chaos in the sense of the presence of level repulsion in the energy spectrum.
{The results of our analysis are summarized in Tab.\,\ref{tab:comparison_props}}.

\subsection{Two-point functions}
We begin by calculating the normalized 2-point functions of spin operators $J_i$ defined as 
\begin{equation}
    G_i(t) = \frac{\tr(J_i(t) J_i)}{\tr(J^2_i)}~,~~ i = x,y,z\,.
\end{equation}
The time evolution of the correlation functions   $G_z(t)$ and $G_x(t)$ is shown in Figs.\,\ref{fig:2ptZfinal} and \ref{fig:2ptX}, respectively. There are several features of these plots that we can explain analytically: 

\begin{comment}
\begin{figure}
    \centering
    \includegraphics[width=0.6\linewidth]{figures/GZT_NU0.pdf}
    \caption{Plot of the normalized 2-point function $G_z(t)$ for $j=150$ (red), $j=200$ (brown) and $j=500$ with $\nu=0$. In all the cases, there is an identical initial decay up to a dip time, after which $G_z(t)$ grows again and start to differ for different values of $j$ (the early-time behavior is shown more clearly in the inset). At later times, the 2-point function oscillates around zero, with decreasing magnitude for larger values of $j$.} 
    \label{fig:2ptZ}
\end{figure}

\begin{figure}
    \centering
    \includegraphics[width=0.6\linewidth]{figures/GZT_NU4.pdf}
    \caption{Plot of the normalized 2-point function $G_z(t)$ for $j=100$ (red), $j=150$ (brown) and $j=200$ with $\nu=4$. In all the cases, there is an initial decay up to a dip, after which $G_z(t)$ grows again (the early-time behavior is shown more clearly in the inset). Compared to the plot for $\nu=0$ in Fig. \ref{fig:2ptZ}, the plots for different values of $j$ start to differ at early times after the initial decay and growth.  At later times, the 2-point function oscillates around zero, with decreasing magnitude for larger values of $j$.} 
    \label{fig:2ptZ2}
\end{figure}
\end{comment}

\begin{figure}[h!]
    \centering
  \begin{subfigure}{0.49\textwidth}
            \centering
            \begin{tikzpicture}
              \node[inner sep = 0pt] at (0,0) {\includegraphics[width=\textwidth]{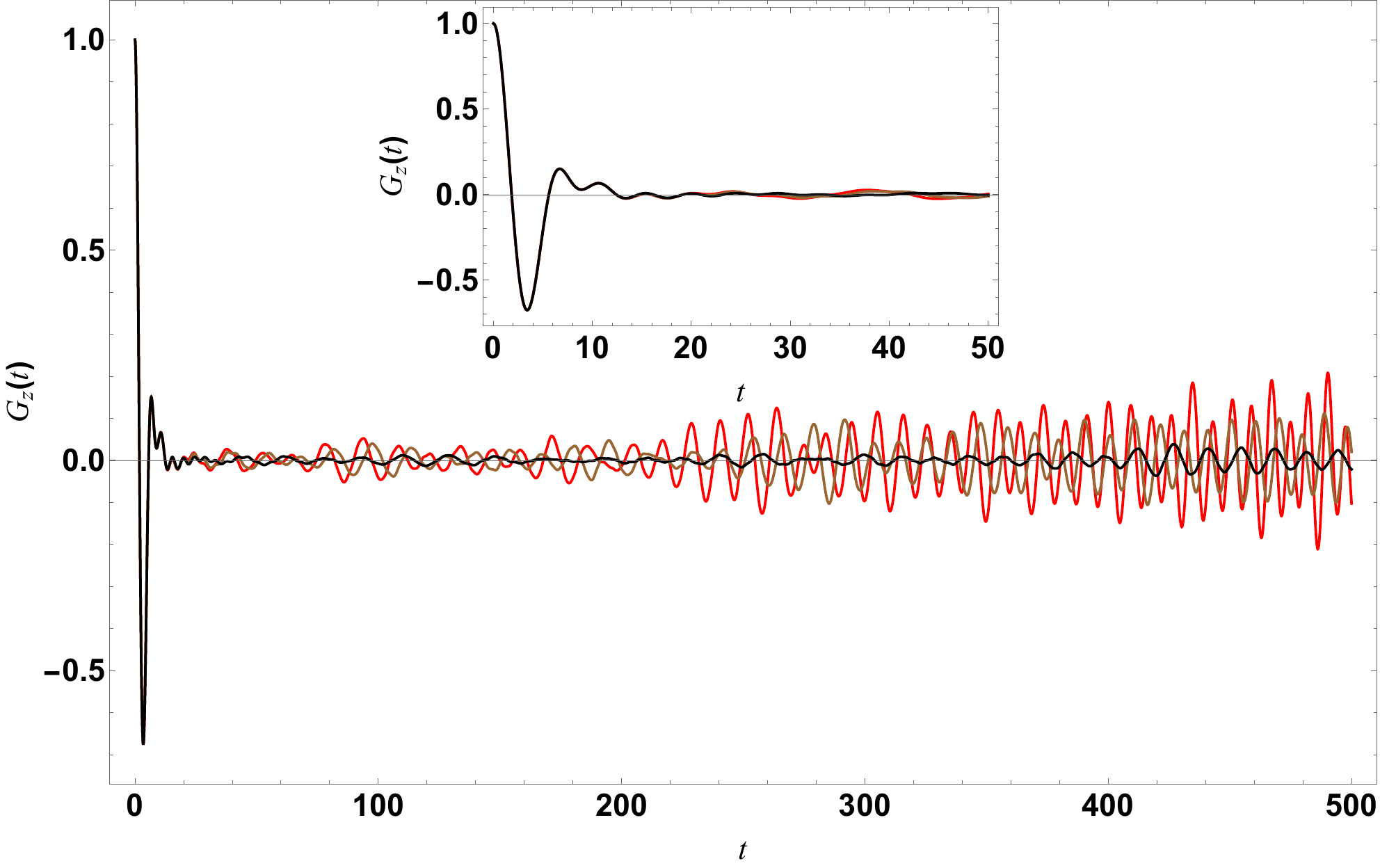}};
            \end{tikzpicture}
            \caption[]
            {\small $\nu=0$}
        \end{subfigure}
        \begin{subfigure}{0.49\textwidth}  
            \centering 
        \begin{tikzpicture}
              \node[inner sep = 0pt] at (0,0) {\includegraphics[width=\textwidth]{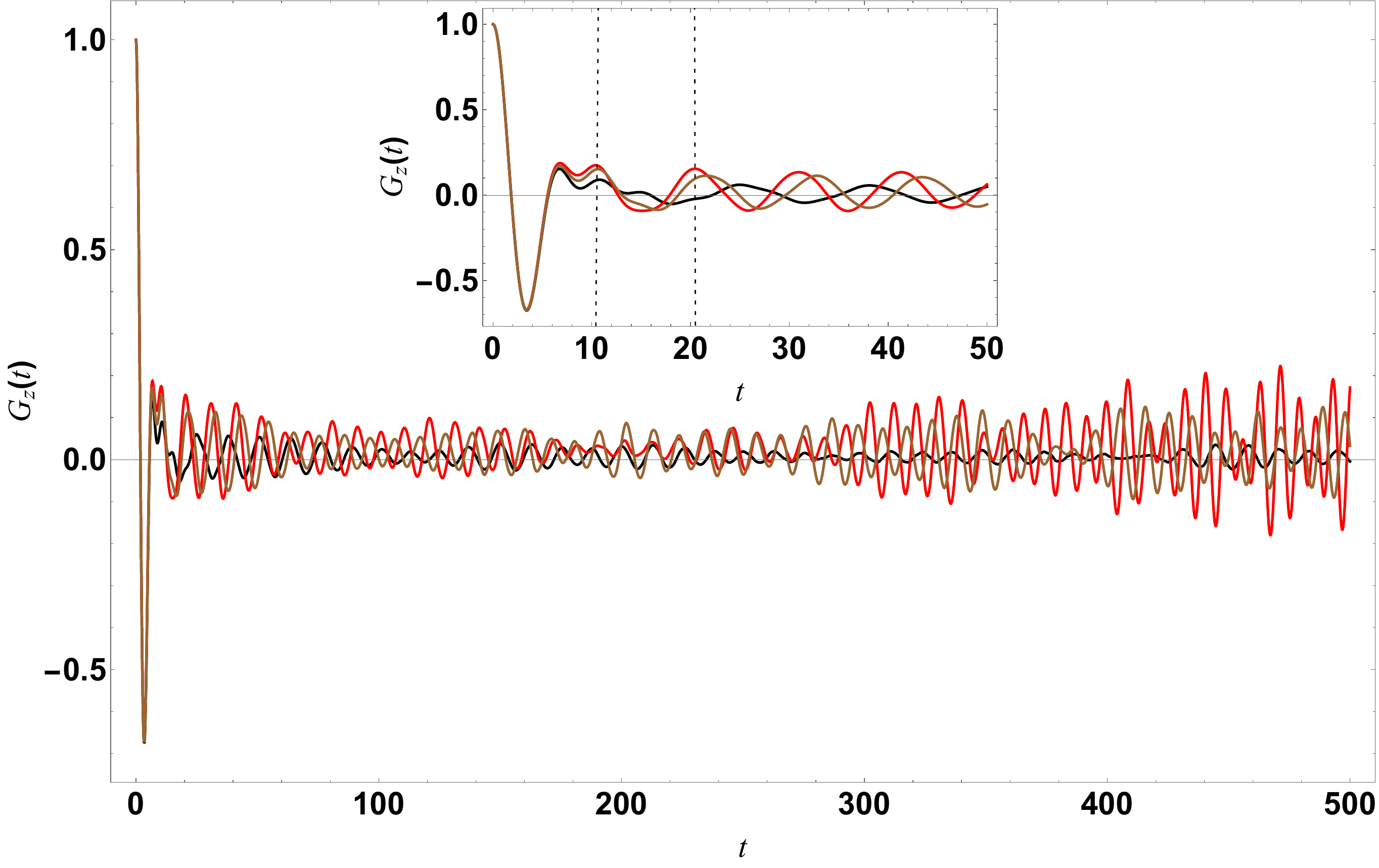}};
            \end{tikzpicture}
            \caption[]
            {\small $\nu=4$}
        \end{subfigure}
    \caption{Plots of the normalized 2-point function $G_z(t)$ obtained numerically for $j=150$ (red), $j=200$ (brown) and $j=500$ (black) with $\nu=0$ (panel (a)) and $\nu=4$ (panel (b)). All cases show an identical initial decay up to a dip time, after which $G_z(t)$ grows again and starts to differ for different values of $j$ (the early-time behavior is shown in the insets). For $\nu=4$, as compared to $\nu=0$, the 2-point function for different $j$ starts to differ at an early time.
    At later times, the 2-point function oscillates around zero, with decreasing magnitude for larger  $j$. {The two dashed lines in the inset of panel (b) indicate the time period of oscillations of the correlation function $G_z(t)$ for $j=150$ (red curve), which can be determined analytically from the formula in \eqref{t_period}. }}
    \label{fig:2ptZfinal}
\end{figure}

\begin{figure}[h!]
    \centering
  \begin{subfigure}{0.48\textwidth}
            \centering
            \begin{tikzpicture}
              \node[inner sep = 0pt] at (0,0) {\includegraphics[width=\textwidth]{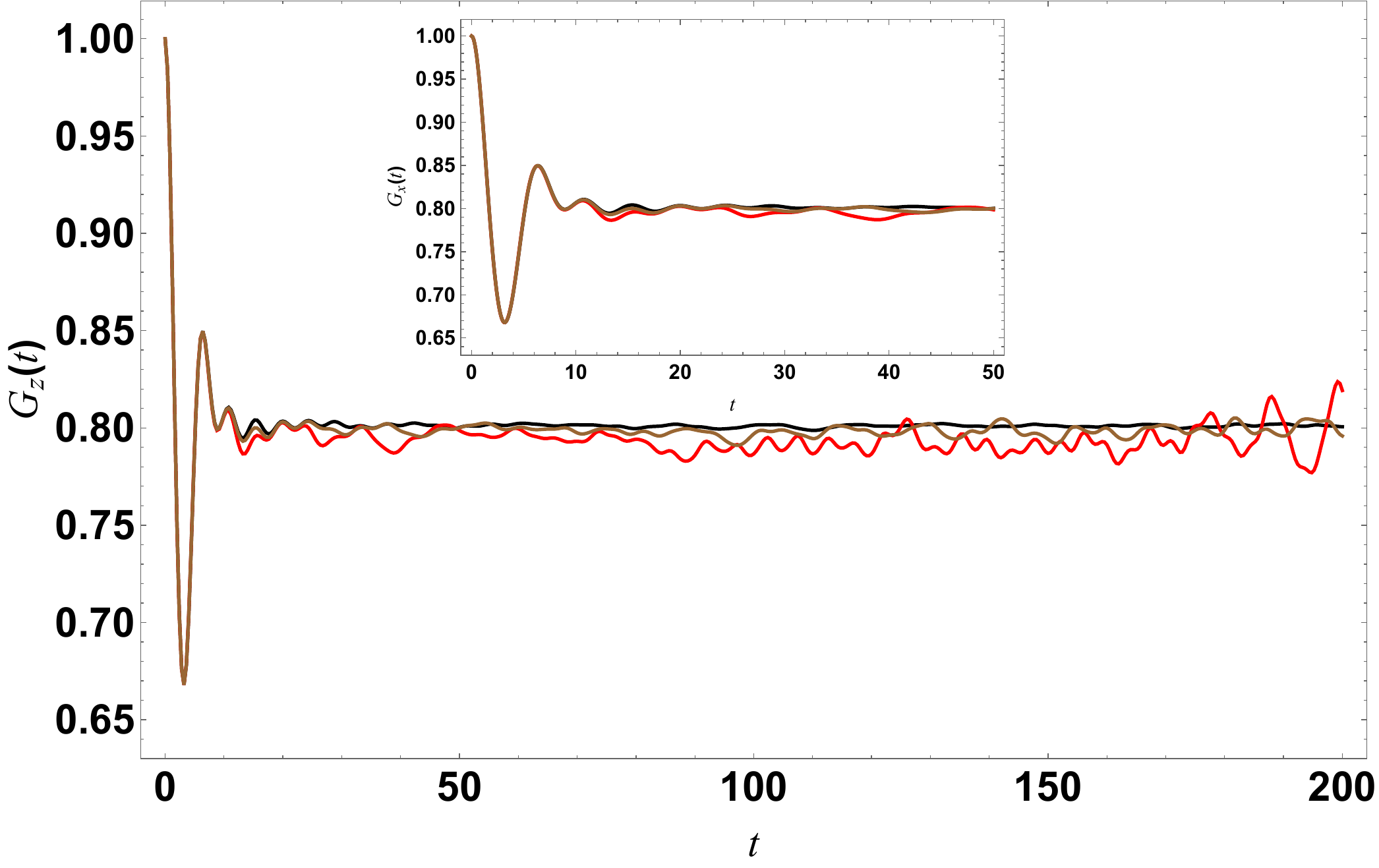}};
            \end{tikzpicture}
            \caption[]
            {\small $\nu=0$}
        \end{subfigure}
        \begin{subfigure}{0.48\textwidth}  
            \centering 
        \begin{tikzpicture}
              \node[inner sep = 0pt] at (0,0) {\includegraphics[width=\textwidth]{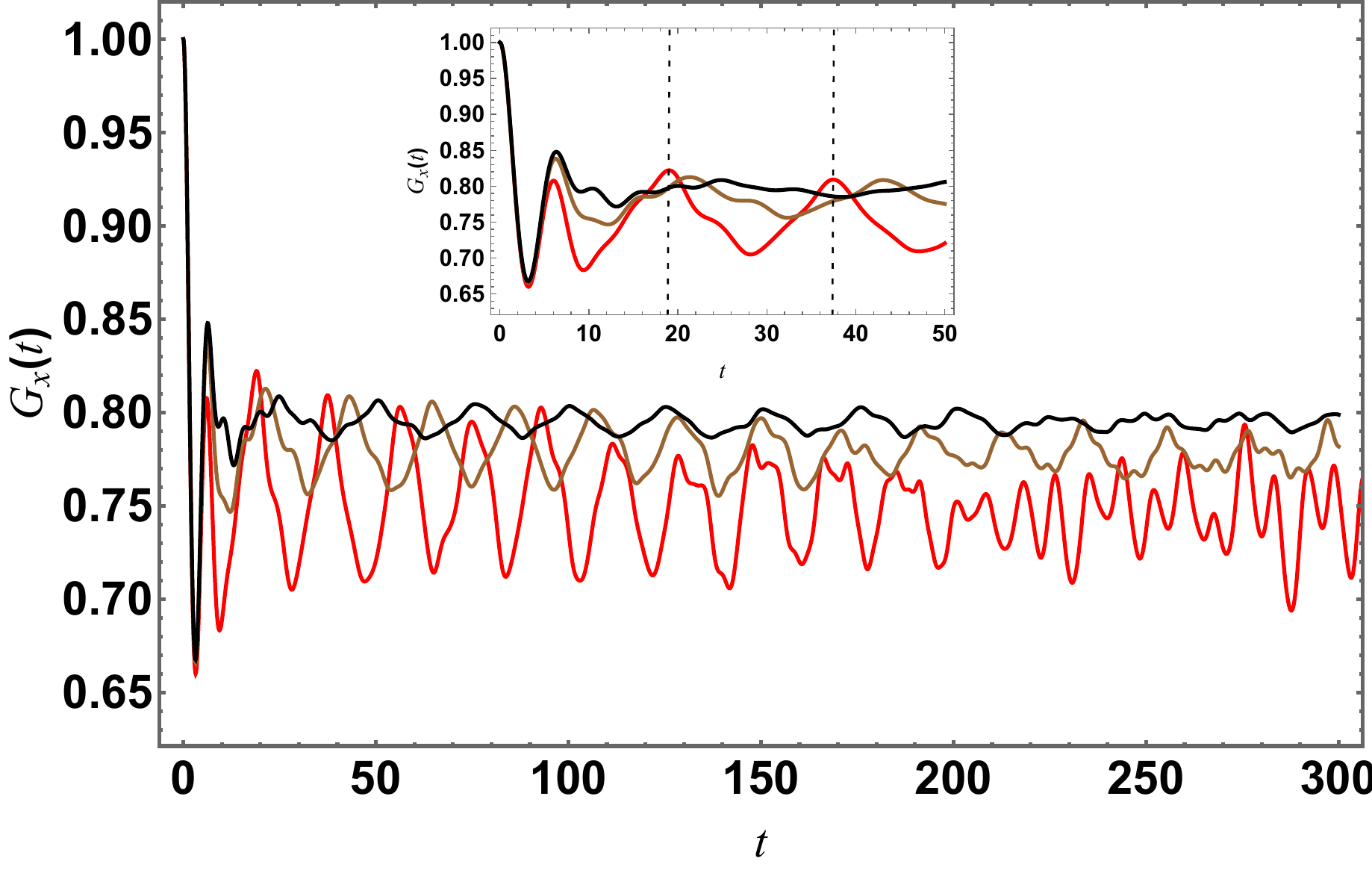}};
            \end{tikzpicture}
            \caption[]
            {\small $\nu=4$}
        \end{subfigure}
    \caption{{Plots of the normalized 2-point function $G_x(t)$ obtained numerically for $j=100$ (red), $j=200$ (brown), and $j=500$ (black)  for different values of $\nu$, $\nu=0$ (panel (a)) and $\nu=4$ (panel (b))}. Both cases are quite similar initially, up to the point where $G_x(t)$ reaches a minimum. At later times, it grows again before seemingly saturating at a non-zero value in the case of $\nu=0$ (in reality, at a very late time, the oscillations grow again). On the other hand, for non-zero $\nu$, there are large oscillations after the dip. However, these oscillations gradually decrease for higher values of $j$ (e.g., for $j=500$, the black curve shown in panel (b), the 2-point correlation function almost reaches a saturation value at the end of the time scale shown here. {The two dashed lines in the inset of panel (b) indicate the time period of oscillations of the correlation function $G_x(t)$ for $j=100$ (red curve).} }
    \label{fig:2ptX}
\end{figure}

\begin{itemize}
    \item The initial decay of the 2-point functions can be understood by explicitly calculating their second derivative  at the initial time, e.g., 
    \begin{equation}
        \ddot{G}_z(0) = \frac{\tr[H,J_z]^2}{\tr J^2_z} = -\frac{2\tr (J^4_x - J^2_x J^2_y)}{j^2 \tr J_z^2} = -\frac{3\tr J^4_x}{j^2\tr J^2_x}+\frac{j+1}{j} 
        = -\frac{4j^2+4j-3}{5j^2} \,.
        %= \frac{3-4j(j+1)}{5j^2}\,.
    \end{equation}
Here we have used the quadratic Casimir in the intermediate steps. Similarly 
    \begin{equation}
        \ddot{G}_x(0) = \frac{\tr[H,J_x]^2}{\tr J^2_x} = -\frac{ \tr (\rmi J_x J_y J_z + 2 J^2_x J^2_z)}{2j^2 \tr J_x^2}-\frac{\nu^2}{j^2} 
        = -\frac{4j^2+4j-3}{20j^2}-\frac{\nu^2}{j^2}  \,.
        %= \frac{3-4j(j+1)}{20j^2} -\frac{\nu^2}{j^2} \,.
    \end{equation}
    To find the final expression, we used $\tr(J_xJ_yJ_z) = \frac{\rmi}{2}\tr J^2_x$. 
    \item Calculating the exact early-time behavior of the correlators is harder. The explicit expressions of the orbits in terms of elliptic functions make it difficult to calculate the required phase space average explicitly. For instance, when $\nu=0$, one can solve for $Z(t)$ explicitly in terms of the Jacobi elliptic sine function $\mathsf{sn}$ and find, with $\alpha\equiv 2\omega/j$,
    \begin{equation}
        Z(t) = \sqrt{1-\alpha}\sin\bigg(\mathsf{sn}\big(\sqrt{1+\alpha}(t+t_0), \tfrac{1-\alpha}{1+\alpha}\big) \bigg)\,.
    \end{equation}
    It appears hard to now average $Z(0)Z(t)$ over phase space. 
   \item To get an estimate for the dip time, one could consider approximating the above expression for $Z(t)$ in terms of its period  $T(\omega)$, 
    \begin{equation}
        Z(t) \approx \sqrt{1-\alpha}\sin\bigg(\frac{2\pi (t+t_0)}{T(\omega)}\bigg)\, ,\quad T(\omega) = 4\mathsf{K}\bigg(\frac{1-\alpha}{1+\alpha}\bigg)/\sqrt{1+\alpha}\,.
    \end{equation}
    Here $\mathsf{K}$ represents the elliptic K-function. 
    Then, in this approximation,
    \begin{equation}
        G_z(t)\sim \int_0^{1} \rmd \alpha T(\alpha)(1-\alpha)\cos(2\pi t/T(\alpha))\,.
    \end{equation}
    To get a rough idea of the dip time, we estimate when $\dot{G}_z$ first vanishes. This will happens near some half-period $T(\omega_i)/2$ which can be estimated as:
    \begin{equation}\label{eq:Tdipest}
        T_\text{dip}\approx \frac{\int \rmd \alpha (1-\alpha)}{2\int \rmd \alpha (1-\alpha)/T(\alpha)}\approx 3.37\,.
    \end{equation}
    As can be seen from fig.\,\ref{fig:2ptZfinal}, this is rather close to the numerical value $T_{\text{dip}}=3.41$.
    \item For $G_z(t)$ the late time average is zero when $\nu=0$. On the other hand, for $G_x(t)$, it is strictly positive. Looking at the classical orbits in fig.\,\ref{fig:orbitds2} indeed shows that the average $Z(0)Z(t)$ vanishes for any given orbit, while the average $X(0)X(t)$ does not.  For $J_x$ the dip occurs around $3.2$ instead, which is also close to our estimate \eqref{eq:Tdipest}.
    \item Finally, when $\nu \neq 0$, both $G_x(t)$ and $G_z(t)$ show transient oscillations (after the dip but before the erratic oscillations at very late times) with periods that seem to satisfy $T_x = 2T_z$. When $\nu\neq 0$, there are zero-energy orbits that encircle a hyperbolic fixed point (as there are in the phase space of a pendulum, when the pendulum keeps going around), see fig.\,\ref{fig:orbitds2}(b). The period of these orbits is found to be
    \begin{equation}\label{t_period}
        T\Big(\frac{\nu}{j}\Big) = \frac{4j}{\nu}\mathsf{K}\Big(-j^2/\nu^2\Big)\,.
    \end{equation}
    This should be the period of the late-time oscillations in $G_x$, while that for $G_z$ would be half this value. Indeed, looking at fig.\,\ref{fig:orbitds2}(b), one notices that the orbits go up and down twice per rotation. Then, in fig.\,\ref{fig:2ptZfinal}(b) (inset) we have, for instance, that $\nu/j = 2/75$ for the red graph and therefore expect a period 10.02, which matches the numerical results quite well. Similarly, in fig.\,\ref{fig:2ptX}(b), we have $\nu/j = 1/25$ for the red graph and expect a period of 18.4, which is also close to the numerical results.
\end{itemize}

\subsection{Squared commutator and OTOCs}

In quantum systems that show signatures of chaos, the growth 
of operators under time evolution should be more pronounced 
compared to their integrable counterparts. The squared commutator is one particular 
quantity used to measure the growth of a given operator (say, $W$) under time evolution by computing its overlap with a probe operator (say $V$)\footnote{We assume both the operators $W$ and $V$ to be Hermitian.} \cite{Garcia-Mata:2022voo, Rozenbaum:2019nwn, Xu:2022vko}. It is usual in the literature to first define the following quantity 
\begin{equation}
    C(t)=- \big<[W(t), V]^2\big>_{\beta}~,
\end{equation}
which is essentially the thermal expectation value (at a finite inverse temperature $\beta$) of the squared-commutator between the time-evolved operator $W(t)$ and $V$.   Expanding the squared 
commutator $C(t)$, one can see that it contains correlation functions where operators appear in an out-of-time-ordered fashion. %i.e., terms proportional to $\tr(W(t)VW(t)V)$. 
We thus define the normalized OTOC as 
\begin{equation}\label{OTOC_def}
    \text{OTOC}(t)=\frac{\tr(W(t)VW(t)V)}{\tr(W(0)VW(0)V)}~.
\end{equation}

It has been proposed that the early-time exponential growth of a quantity like $C(t)$, defined for a suitable choice of operators, is a diagnostic for quantum chaos.\footnote{For chaotic systems, the exponential growth usually persists up to a time scale known as the scrambling time, after which it attains a saturation value \cite{rammensee2018many, garcia2018chaos}. For a bounded one-body system having a classical counterpart, the scrambling time is equal to the Ehrenfest time.} Nevertheless, there are quite a few counter-examples to this general expectation, indicating that early-time exponential growth of $C(t)$ is not necessarily a signature of the chaotic nature of the system \cite{Xu:2019lhc, pilatowsky, hashimoto2020exponential, rozenbaum2020early}. Specifically, it has been observed that the presence of a hyperbolic fixed point (saddle point) in the semi-classical phase space of integrable systems, can also result in exponential growth of $C(t)$. This is called saddle-dominated scrambling \cite{Xu:2019lhc, pilatowsky, kidd2021saddle}.

Below, we verify this behavior in the spin system \eqref{eq:hamds2}. However, we also note that, even though the early-time exponential growth of $C(t)$ can either be a genuine signature of quantum chaos or due to the presence of saddle points in the classical phase space, these two scenarios can be distinguished using the late-time behavior of $C(t)$. Namely, whereas quantum chaotic dynamics leads
to the saturation of $C(t)$, early-time growth due to saddle points leads to noticeable oscillations at late times \cite{kidd2021saddle}. 

%{Based on \cite{Rozenbaum:2019nwn}, expect OTOC to grow exponentially until a time $\log j/\lambda$. We confirm this here. See also \cite{Xu:2019lhc} who did this for LMG. They also mention \cite{pilatowsky} who start from state close to saddle, rather than trace. See also \cite{trunin} regarding work on how to avoid false positives.}

%%%%%%%%%%%%%%%%%%%%%%%%%%%%%%%%%%
\begin{figure}
    \centering
    \subfloat[Squared commutator $C_z(t)$]{
    \centering 
    \includegraphics[width=0.45\textwidth]{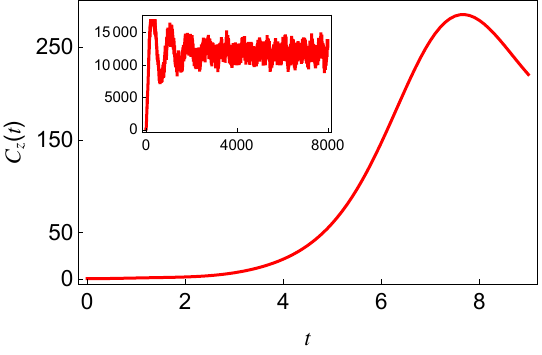}}
    \subfloat[Late-time normalized OTOC ($\nu=0$)]{
    \centering 
    \includegraphics[width=0.48\textwidth]{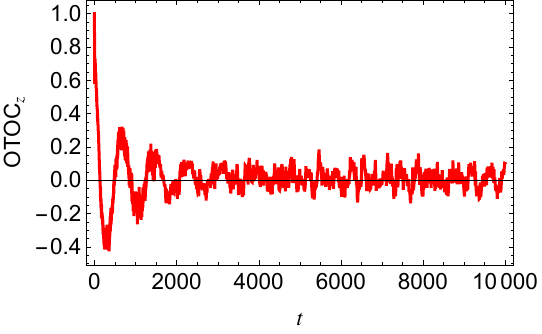}}
     \hfill
    \subfloat[Exponential growth of $C_z(t)$ ($\nu=0$)]{
    \centering 
    \includegraphics[width=0.45\textwidth]{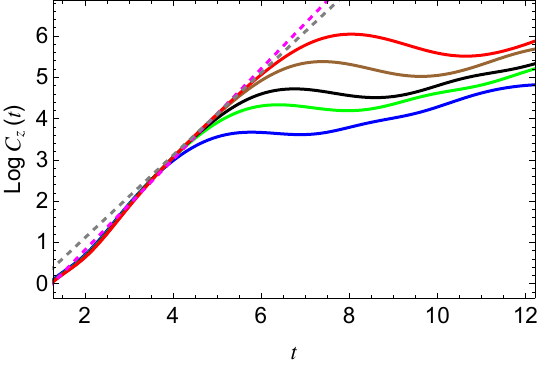}}\hspace{0.2cm}
    \subfloat[Exponential growth of $C_z(t)$ ($\nu=4$)]{
    \centering 
    \includegraphics[width=0.45\textwidth]{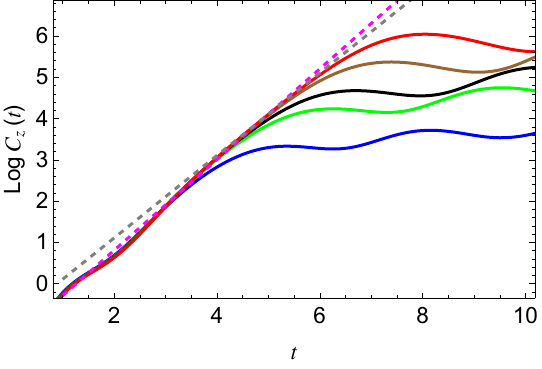}}
    \caption{Plots of the squared commutator $C_z(t)$ and OTOC of the $J_z$ operator obtained numerically with $j=200$, and $\nu=0$. Panel (a) shows the early-time exponential growth, while the inset shows the presence of large oscillations around a mean value at late times, indicating that the early-time growth is due to a saddle point. Panel (b) shows the late-time decay and oscillations of the OTOC. In panels (c) and (d), we show the exponential growth 
    of the squared commutator until the Ehrenfest time $1\leq t \lesssim \log j$, with $j=25$ (blue), $j=50$ (green), $j=75$ (black),  $j=150$ (brown), and $j=300$ (red) ($\nu=0$ and $\nu=4$, for panel (c) and (d) resp.). The {magenta} dashed line fits the early-time exponential growth of $C_z(t) \propto \exp (\Lambda_{\text{OTOC}} t)$. {The gray dashed line is the reference line from classical saddle point $\propto \exp (\lambda_{\mathrm{saddle}}t)$}. From the numerical fit {in time range $t\in[2.5,5.5]$}, we obtain $\Lambda_{\text{OTOC}} \approx 1.1$. Since the classical saddle has $\lambda=1$ (see app.\,\ref{app:saddle_derivation}), our results are consistent with the bound $\Lambda_{\text{OTOC}}\geq \lambda_{\text{saddle}}$ of \cite{Xu:2019lhc}.
    }
    \label{fig:OTOCZ}
\end{figure}
Here we are interested in calculating the trace of the following squared commutators (we consider infinite temperature $\beta=0$ from now on, which exhibits the maximal growth)
\begin{equation}\label{comm_squared}
    C_i(t) =  -\frac{1}{ (2j+1) j^2} \tr \big([J_i(t),J_i]^2\big)\,,~~~ i=x,y,z~,
\end{equation}
where we have included an appropriate normalization constant. This can be thought of as an \emph{infinite temperature thermal average} of $[J_i(t),J_i]^2$. 
%(with an additional normalization factor of $j^2$). 
Due to the saddle points,
we expect that in the spin system under consideration, the above 
squared commutators grow exponentially up to a time scale around $\log j/\lambda$ \cite{Rozenbaum:2019nwn}, where $\lambda$ is the Lyapunov exponent associated to the saddle. In our case $\lambda = 1$ in the large-spin limit.\footnote{We provide a brief classical analysis of this in app.\,\ref{app:saddle_derivation} where 
we obtain the location and stability properties of the stationary points of the Hamiltonian \eqref{eq:hamds2}. For $\nu>j$ none of the stationary points are unstable, while when $\nu<j$
there are two unstable saddle points that give rise to the `exponential growth' of the trajectories with classical Lyapunov exponent $\lambda=j^{-1}\sqrt{j^2-\nu^2}$.} Below, we will see that numerical computations confirm these expectations.\footnote{One difference between the model in \eqref{eq:hamds2} and the LMG model analyzed in the context of the saddle-dominated scrambling \cite{Xu:2019lhc} is that here there are two unstable saddle points in the classical phase space with energies $\pm\nu$, compared to a single one in \cite{Xu:2019lhc}. Only when $\nu=0$ two saddle points have the same energy. In that case, the classical Lyapunov exponent takes the maximum value of $\lambda=1$ irrespective of $j$.}

The early- and late-time behaviors of $C_z(t)$ and $C_{x+y/\sqrt{2}}(t)$ (as well as those of the corresponding OTOCs defined in \eqref{OTOC_def}) are shown 
in figs.\,\ref{fig:OTOCZ} and \ref{fig:OTOCX}, respectively. As one increases $j$ for fixed $\nu$, the time up to which the exponential behavior $\propto \exp(\Lambda_{\text{OTOC}}t)$ persists also increases, and the overall magnitude of $C_i(t)$ gets magnified as well (see fig.\,\ref{fig:OTOCZ}(c)). 

In all cases analyzed we find that $\Lambda_\text{OTOC} \geq \lambda_{\text{saddle}}=1$, consistent with the bound of \cite{Xu:2019lhc}. In the case of $J_z$ we find $\Lambda_{\text{OTOC}}\approx 1.1$.\footnote{Our $z$-direction takes on the same role with respect to the saddle as the $x$-direction does in the LMG model considered in \cite{Xu:2019lhc}. Analyzing the squared commutator $C_x$ in their system we also find that $\Lambda_{\text{OTOC}}$ is larger than $\lambda_{\text{saddle}}$ ($\sqrt{3}$ in their case) by a factor of roughly $1.1$.} In the case of $(J_x + J_y)/\sqrt{2}$ we find $\Lambda_{\text{OTOC}} = \lambda_{\text{saddle}}=1$. Essentially, the $(x+y)$-direction plays the same role in our saddle as the $z$-direction did in \cite{Xu:2019lhc}. The reason we do not consider $J_x$ by itself is because the $x$-direction is the direction of maximal growth near the saddle. To first order, the classical sensitivity $\{X(t),X(0)\}$ then vanishes at the saddle, and the argument of \cite{Xu:2019lhc} does not apply for this particular direction. Indeed, we still find an exponential growth, but it is much slower than $\lambda_{\text{saddle}}$ and moreover quite sensitive to the relative size of $\nu$ and $j$. 

As can also be seen in the plots, the early-time exponential growth of the squared commutator does not lead to  saturation at late times, where large-amplitude oscillations indicate that the early-time exponential growth of $C_i(t)$ is due to saddle-dominated scrambling.

%%%%%%%%%%%%%%%%%%%%%%%%%%%%%%%%%%
\begin{figure}[h!]
    \centering
    \subfloat[Commutator squared $C_{\frac{x+y}{\sqrt{2}}}(t)$]{
    \centering 
    \includegraphics[width=0.45\textwidth]{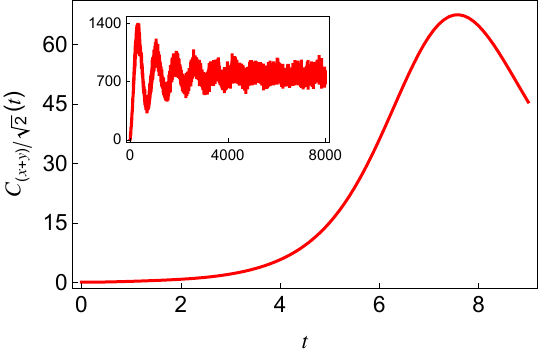}}
    \subfloat[Late-time normalized OTOC]{
    \centering 
    \includegraphics[width=0.48\textwidth]{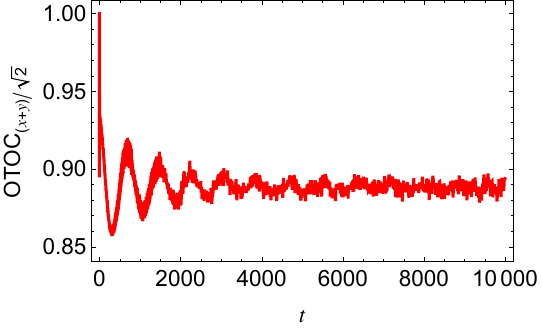}}
     \hfill
    \subfloat[Exponential growth of $C_{\frac{x+y}{\sqrt{2}}}(t)$ ($\nu=0$)]{
    \centering 
    \includegraphics[width=0.45\textwidth]{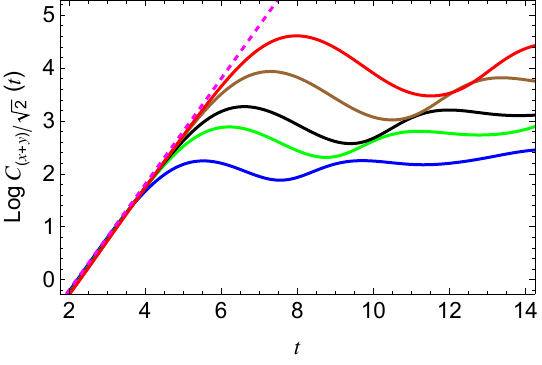}}\hspace{0.2cm}
    \subfloat[Exponential growth of $C_{\frac{x+y}{\sqrt{2}}}(t)$ ($\nu=4$)]{
    \centering 
    \includegraphics[width=0.45\textwidth]{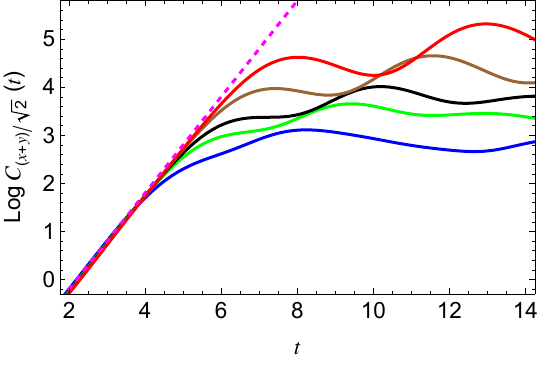}}
    \caption{Plots of the squared commutator $C_{(x+y)/\sqrt{2}}(t)$ and OTOC of the $(J_x+J_y)/\sqrt{2}$ operator obtained numerically  with $j=200$, and $\nu=0$. Panel (a) shows the early-time exponential growth, while the inset shows the presence of large oscillations around a mean value at late times. Panel (b) shows the late-time decay and oscillations of the OTOC. In panels (c) and (d), we show the exponential growth 
    of $C_{(x+y)/\sqrt{2}}(t)$ which persists for a time scale $1\leq t \lesssim \log j$, with $j=25$ (blue), $j=50$ (green), $j=75$ (black),  $j=150$ (brown), and $j=300$ (red) ($\nu=0$ and $\nu=4$, for panel (c) and (d) respectively). The dashed line indicates exponential growth of $C_x(t) \,\propto\; \exp (\Lambda_{\text{OTOC}} t)$. The linear fit is consistent with $\Lambda_{\text{OTOC}}= \lambda_{\text{saddle}}=1$.
    }
    \label{fig:OTOCX}
\end{figure}

Before moving on, let us make two further comments. First, we note that the early-time exponential growth of $C(t)$ in the presence of isolated saddle points is essentially due to the wrong order of averaging over the phase space, i.e., instead of obtaining the maximum Lyapunov exponent (which is the phase space average of the \textit{log} of sensitivity), as in the classical case, the quantity
used to quantify the exponential growth of the squared commutator 
$\Lambda_{\text{OTOC}}$ is calculated from \textit{log} of phase space average of the sensitivity squared. An alternative quantity that avoids this issue but also retains various useful properties of the square commutators has recently been proposed in \cite{trunin}, and it was shown that for the LMG model (which has an unstable saddle point) this new quantity does not have early-time exponential growth, even in the presence of saddle points. 

Second, as mentioned above, rather than the early-time exponential growth of the commutator squared, the vanishing of OTOC at late time is believed to be an indication of the chaotic nature of a quantum system. In this context, we note that the vanishing of not only the two-point function and OTOC, but also of all the higher-order mixed cumulants - a property of non-commutative free operators known as freeness, has recently been suggested as a unified notion of quantum chaos \cite{Camargo:2025zxr}. For finite-size systems, the OTOC should decay and saturate at a finite value of order $O(1/N)$ around the scrambling time $\sim\text{Log}(N/\hbar)/\lambda$ \cite{Cotler:2017myn}. For quantum systems, whose integrability is broken slightly, the freeness (i.e., when all the mixed cumulants go to zero) emerges at a rather long time scale \cite{Camargo:2025zxr}. Despite the appearance of early exponential growth for the commutator squared, we see in fig.\,\ref{fig:OTOCZ} and \ref{fig:OTOCX} that OTOCs do not completely saturate at late times. They decay but oscillations remain present for a very long time. This indicates that the system we studied is integrable rather than chaotic. From our observations above, we conclude that instead of the early exponential growth of the commutator squared, the vanishing or saturation of the OTOC around and after the scrambling time does seem to be a better indicator for chaos.

%%%%%%%%%%%%%
\subsection{Krylov operator complexity}
%{The initial exponential growth should be explained by the Lyapunov exponent of the classical dynamics (which also determines the quasinormal modes). \cite{Bhattacharjee:2022vlt} }

%{Linear growth rate of Lanczos coefficients in generic non-integrable systems, leads to exponential growth in expectation values and bounds Lyapunov coefficient \cite{parker}.}

In this section, we compute the Krylov operator complexity ($C_K$) in \eqref{eq:hamds2}.\footnote{It would be interesting to extend our study with the notion of microcanonical Krylov complexity developed in \cite{Bhattacharjee:2022vlt} for the LMG model (\ref{eq:more general LMG}) with $A=B$, since they argued there are other imprints of saddle dominated scrambling in its Lanczos coefficients.} This notion of quantum complexity measures how `deep' a local operator grows under Heisenberg evolution in a certain basis, known as the Krylov basis \cite{parker}. 
% This is in contrast with the Krylov state complexity that we will discuss in the next subsection, where the complexity of an initial state under time evolution is measured. 
To define the Krylov complexity for a Hermitian operator $\mathcal{O}$ in a quantum system described by the Hamiltonian $H$, one focuses on its time evolution governed by the Heisenberg equation:
\begin{equation}
    \partial_t \mathcal{O}(t)=\rmi[H, \mathcal{O}(t)] \equiv \rmi \mathcal{L} \mathcal{O}(t)~,
\end{equation}
where $\mathcal{L}\equiv [H,\cdot]$ denotes the \textit{Liouvillian} superoperator. Therefore, the time-evolved operator in the Heisenberg picture can be written as 
\begin{eqnarray}\label{O_time}
    \mathcal{O}(t)&=&\rme^{\rmi Ht} \mathcal{O} \rme^{-\rmi Ht}\equiv \rme^{\rmi\mathcal{L}t}\mathcal{O}\nonumber\\
    &=& \mathcal{O}+\rmi t \mathcal{L}\mathcal{O}+\frac{(\rmi t)^2}{2}\mathcal{L}^2 \mathcal{O}+\dots.~.
\end{eqnarray}
To construct a convenient orthonormal basis from a given operator $\mathcal{O}$, one can generate states $\{\vert \mathcal{O}_n)\}$ (Krylov basis) from the initial operator by the Gelfrand-Naimark-Segal (GNS) construction, which spans a Hilbert space $\mathcal{H_{\mathcal{O}}}$ (Krylov space). This projection process is performed through the Lanczos algorithm\footnote{This is essentially a Gram-Schmidt orthogonalization of the initial unnormalized states $\vert \tilde{\mathcal{O}}_n)=\mathcal{L}^n \mathcal{O}$.}, which consists of the following steps:
\begin{itemize}
    \item define the first basis state $\vert\mathcal{O}_0):=\frac{\vert \mathcal{O})}{\sqrt{(\mathcal{O}\vert \mathcal{O})}}$ and set $b_0$=0, $\vert \mathcal{O}_{-1})=0$.
    \item for $n\geq 1$, $\vert\mathcal{A}_n)=\mathcal{L} \vert \mathcal{O}_{n-1})-b_{n-1}\vert \mathcal{O}_{n-2})$, where  $b_{n}=\sqrt{(\mathcal{A}_n \vert \mathcal{A}_n)}$.
    \item if $b_n=0$, the algorithm stops; otherwise, add the normalized  $\vert\mathcal{O}_n) :=\frac{\vert\mathcal{A}_n)}{b_n}$ to the basis.
\end{itemize}

In the following, to perform this algorithm, we use the following inner product \footnote{Since in this section we consider the case of infinite temperature ($\beta=0$), the inner product is \eqref{eq:inner product inf temp}. For a general temperature, one usually considers the Wightman inner product to study operator growth \cite{parker}:
\begin{equation}
    (A\vert B)=\frac{1}{Z} \mathrm{Tr}\left( \rme^{-\beta H/2} A^\dagger \rme^{-\beta H/2} B\right),
\end{equation}
with $Z=\mathrm{Tr}\left( \rme^{-\beta H} \right)$ the partition function.}
\begin{equation}\label{eq:inner product inf temp}
    (A\vert B)=\frac{1}{D} \mathrm{Tr}\left(A^\dagger B\right),
\end{equation}
where $D$ is the Hilbert space dimension.

As established in \cite{Rabinovici:2020ryf}, the dimension of Krylov space is bounded by 
\begin{equation}
    D_{\mathcal{O}}=\mathrm{dim}(\mathcal{H}_{\mathcal{O}})\leq D^2-D+1.
\end{equation}
After obtaining the fully orthonormal Krylov basis, one can expand the time-evolved initial operator in this basis:
\begin{equation}
    \vert \mathcal{O}_t)=\sum_{n=0}^{ D_{\mathcal{O}}-1}\rmi^n \phi_n(t) \vert \mathcal{O}_n)~,
\end{equation}
Here the function $\phi_n(t)$ is called the \textit{operator wavefunction} or \textit{transition amplitude}, and satisfies the normalization condition 
\begin{equation}
    \sum_{n=0}^{ D_{\mathcal{O}}-1}|\phi_n(t)|^2=1~,
\end{equation}
as well as  the recursion relation following from the Heisenberg time evolution:
\begin{equation}
    \partial_t \phi_n(t)=b_n\phi_{n-1}(t)-b_{n+1}\phi_{n+1}(t)~.
\end{equation}
Finally, the Krylov operator complexity is defined as \cite{parker} 
\begin{equation}
    C_K(t)=\sum_{n=0}^{ D_{\mathcal{O}}-1} n\vert \phi_n(t)\vert^2~.
\end{equation}

With the Hamiltonian in \eqref{eq:hamds2}, and a given initial operator $J_z$, we perform the Lanczos algorithm to obtain the Lanczos coefficients $ b_n$. These are shown in fig.\,\ref{Fig: LanczosCoefficients} for different values of the parameters $\nu=0,4$ and $j=25,50,75$. Due to saddle-dominated scrambling, the $b_n$ sequence has an identical linear growth for small values of $n$, irrespective of the value of the spin $j$, in line with \cite{Bhattacharjee:2022vlt}. As shown a fit in fig.\,\ref{Fig: LanczosCoefficients}, the rate of this linear growth ($\alpha$) is given by
the saddle-point exponent $2\alpha=\lambda=j^{-1}\sqrt{j^2-\nu^2}$, which is $1$ when $\nu=0$ and slightly smaller when $\nu\neq 0$. After the initial linear growth, the $ b_n$ sequence approximately saturates at larger values of $n$, due to finite-size effects and finally reaches zero at the end of the Krylov basis as shown in fig.\,\ref{Fig: Full bn}.

%%%%%%%%%%%%%%%%%%%%%%%%%%%%%%%%%%
\begin{figure}[!ht]
    \centering
    \begin{subfigure}{0.48\textwidth}
            \centering
            \begin{tikzpicture}
             \node[inner sep = 0pt] at (0,0) {\includegraphics[width=\textwidth]{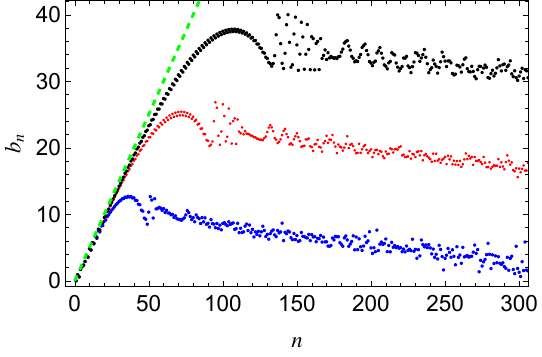}};
            \end{tikzpicture}
            \caption[]%
            {{\small Lanczos coefficients for $\nu=0$}}
        \end{subfigure}
        \hspace{0.2cm}
    \begin{subfigure}{0.48\textwidth}
            \centering
            \begin{tikzpicture}
             \node[inner sep = 0pt] at (0,0) {\includegraphics[width=\textwidth]{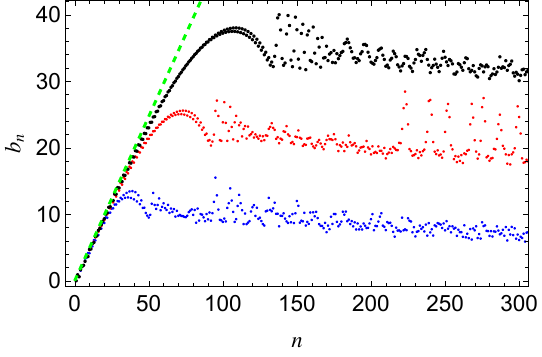}};
            \end{tikzpicture}
            \caption[]%
            {{\small Lanczos coefficients for $\nu=4$}}
        \end{subfigure}
        \caption{The Lanczos coefficients $b_n$ for the initial operator $J_z$ show linear growth at small values of $n$. We take $j=25~(\text{blue}),~50~(\text{red}),~75~(\text{black})$. The green dashed lines represent $b_n=0.5n$, corresponding to the growth rate $2\alpha=\lambda_\text{saddle}=1$ at fixed $\nu$ and large $j$.}
        \label{Fig: LanczosCoefficients}
\end{figure}
%%%%%%%%%%%%%%%%%%%%%%%%%%%%%%%%%%

As in \cite{Bhattacharjee:2022vlt} for the LMG model, we notice a bump in the Lanczos spectrum after the linear growth reaches its peak, which might be due to the fact that the system is integrable. Before the bump, the Lanczos coefficients have a small variance due to the unstable saddle points, which directly give rise to the exponential growth of complexity in early time. {After the bump, the underlying integrable nature of the systems is revealed, and the variance of the Lanczos sequence becomes larger \cite{Rabinovici:2022beu}, which will also be reflected in the late-time complexity.}

The initial linear growth of the Lanczos coefficients results in an early-time exponential growth of the Krylov operator complexity, shown in fig.\,\ref{Fig: Krylov Operator Complexity}. When increasing $\nu$ from zero but smaller than $j$, the growth rate of Lanczos coefficients gets smaller, resulting in a slower exponential growth of Krylov complexity. 

%%%%%%%%%%%%%%%%%%%%%%%%%%%%%%%%%%
\begin{figure}[!ht]
    \centering
    \begin{subfigure}{0.46\textwidth}
            \centering
            \begin{tikzpicture}
             \node[inner sep = 0pt] at (0,0) {\includegraphics[width=\textwidth]{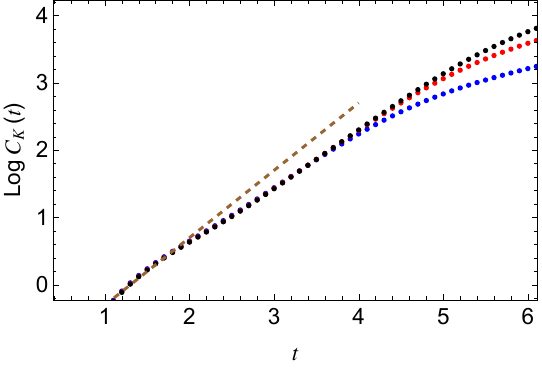}};
            \end{tikzpicture}
            \caption[]
            {\small Krylov operator complexity (log-plot) for $\nu=0$, and $j=25~(\text{blue}),~50~(\text{red}),~75~(\text{black})$}
        \end{subfigure}
        \hspace{0.5cm}
    \begin{subfigure}{0.46\textwidth}
            \centering
            \begin{tikzpicture}
             \node[inner sep = 0pt] at (0,0) {\includegraphics[width=\textwidth]{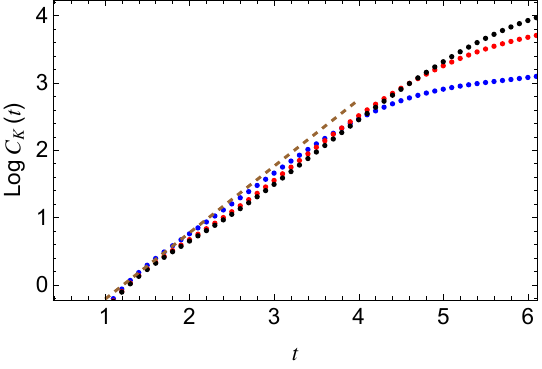}};
            \end{tikzpicture}
            \caption[]
            {\small Krylov operator complexity (log-plot) for $\nu=4$, and $j=25~(\text{blue}),~50~(\text{red}),~75~(\text{black})$}
        \end{subfigure}
        \caption{{Early-time evolution of the Krylov operator complexity for the initial operator $J_z$ obtained numerically.} The brown dashed lines are proportional to the function $\rme^{t}$. Here, we show the early-time behavior up to $t=6$. For the late-time behavior see fig.\,\ref{Fig: FullLanc and C}.
        }
        \label{Fig: Krylov Operator Complexity}
\end{figure}
%%%%%%%%%%%%%%%%%%%%%%%%%%%%%%%%%%

%%%%%%%%%%%%%%%%%%%%%%%%%%%%%%%%%%
\begin{figure}[h!]
    \centering
    \subfloat[Full set of Lanczos coefficients.]{
    \centering 
    \includegraphics[width=0.46\textwidth]{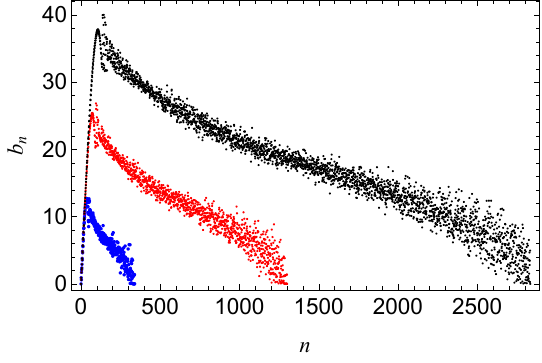}\label{Fig: Full bn}}
    \hfill
    \subfloat[Late-time Krylov operator complexity.]{
    \centering 
    \includegraphics[width=0.50\textwidth]{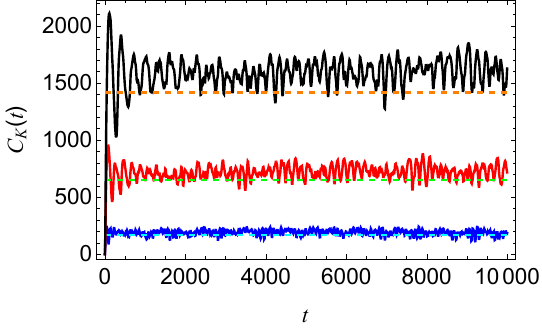} \label{Fig: Full C}}
    \caption{The full set of Lanczos coefficients and late-time Krylov complexity of the operator $J_z$ for $\nu=0$, and $j=25$ (blue), $j=50$ (red), and $j=75$ (black). The Lanczos coefficients (panel (a)) grow linearly, reach a peak, decay with oscillations, and eventually vanish due to the finite size effect. The Krylov complexity (b) grows exponentially at a very early time, reaches a peak, and then oscillates around a saturation value slightly above half of the Krylov operator space dimension $D_{\mathcal{O}}/2$,  as represented by the dashed lines.}
    \label{Fig: FullLanc and C}
\end{figure}
%%%%%%%%%%%%%%%%%%%%%%%%%%%%%%%%%%

The late-time behavior of Krylov operator complexity is shown in fig.\,\ref{Fig: Full C}. We find that $C_K(t)$ exhibits a pattern of exponential growth at early times, followed by large oscillations, and saturation at a finite value with smaller oscillations. The magnitude of the oscillations around the saturation value and the time at which $C_K(t)$ saturates increase with $j$. It was argued in \cite{Rabinovici:2021qqt, Rabinovici:2020ryf} that the saturation value is roughly (or a little below) half of the dimension of the Krylov operator space $D_{\mathcal{O}}/2$ for chaotic systems, and much lower than that in integrable systems. As seen in fig.\,\ref{Fig: Full C}, the saturation value actually seems to be slightly above $D_{\mathcal{O}}/2$. 
%This may indicate that the system is non-chaotic. 
The appearance of a slightly higher saturation value is due to a biased nature of the Krylov chain here. Namely, the time-averaged transition amplitudes $Q_{0n}$, defined as
\begin{equation}\label{eq:trans}
    Q_{0n}=\lim_{T\rightarrow \infty}\frac{1}{T} \int^T_0 \vert \phi_n(t)\vert^2 dt~,
\end{equation}
are bigger on the right side of the Krylov chain compared to those on the left side (see fig.\,\ref{Fig: biased chain}). This is similar to the case discussed in \cite{Rabinovici:2021qqt}, where the authors showed a right-sided biased Krylov chain from a phenomenological model by building a sequence of Lanczos coefficients. However, we observe that our spin model naturally exhibits this feature and is helpful in distinguishing it from chaotic systems.

%%%%%%%%%%%%%%%%%%%%%%%%%%%%%%%%%%
\begin{figure}[h!]
    \centering 
    \includegraphics[width=0.56\linewidth]{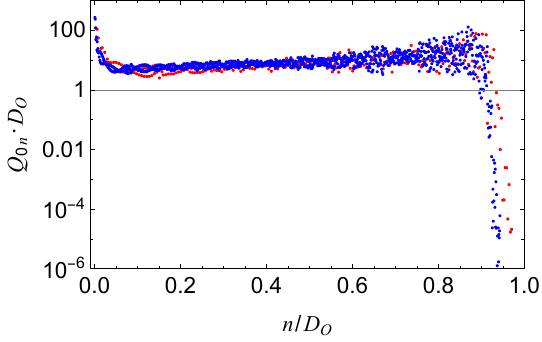}
    \caption{The right-sided biased $Q_{0n}\cdot D_O$ (eq. \eqref{eq:trans}) in the Krylov chain  for the operator $J_z$ with $j=25~\text{(red)},~50~\text{(blue)}$ and $\nu=0$.}
    \label{Fig: biased chain}
\end{figure}
%%%%%%%%%%%%%%%%%%%%%%%%%%%%%%%%%%

%%%%%%%%%%%%%%%%%
\subsection{Krylov state complexity}\label{spread_complexity}
Finally, we consider Krylov state or spread complexity \cite{Balasubramanian:2022tpr} in the system with Hamiltonian \eqref{eq:hamds2}.\footnote{The spread complexity of a related LMG model was studied in \cite{Huh:2023jxt}, where the authors showed the existence of a peak in spread complexity for non-chaotic systems, due to a saddle point. Our system has the additional benefit of having a geometric interpretation. In particular, as discussed in the previous sections, this toy model reproduces the QNM spectrum of a particle in de Sitter. The spread complexity of our model can be interpreted in terms of wavefunction delocalization of a single particle, as in \cite{Bhattacharya:2024hto}. } Compared to Krylov operator complexity, one considers the time-evolution of a given initial state $\ket{\psi_0}$ instead of an initial operator. The Krylov subspace associated with the time-evolved state
is generated by acting with the Hamiltonian $H$ on this state. The action of $H$ on the $n$th element of the Krylov basis can be written as  
\begin{equation}\label{eq:Hamiltonian Krylov}
    H\ket{K_n}=a_n\ket{K_n}+b_n\ket{K_{n-1}}+b_{n+1}\ket{K_{n+1}}~,
\end{equation}
with $\bra{K_m}\ket{K_n}=\delta_{nm}$ and $\ket{K_0}=\ket{\psi_0}$.
Thus, in this approach, the unitary time-evolution is mapped to the hopping motion of a fictitious particle on a one-dimensional discrete chain, known as the Krylov chain. At the initial
time, the motion starts from the left-most site on the chain, and the Krylov chain ends when the final element of the Krylov subspace is reached, i.e. when $b_n=0$ for some non-zero $n$ and the Lanczos algorithm ends \cite{Balasubramanian:2022tpr,Nandy:2024htc}. 

The dynamics is then quantified by calculating the expectation value of the position of this fictitious particle on the Krylov chain. This quantity, known as the Krylov state complexity or the spread complexity of the time-evolved state,\footnote{For a recent set of works on various aspects of spread complexity, see \cite{Balasubramanian:2024ghv,Fu:2024fdm, Afrasiar:2022efk, Medina-Guerra:2025rwa, Camargo:2024deu, Camargo:2024rrj, Baggioli:2024wbz, Takahashi:2025mdt, Chakrabarti:2025hsb, Chattopadhyay:2024pdj, Hu:2025zvv, Gautam:2023bcm, Jeong:2024oao, Huh:2023jxt,Bhattacharjee:2024yxj,Bhattacharjee:2022qjw,Nandy:2023brt,Aguilar-Gutierrez:2023nyk,Aguilar-Gutierrez:2024nau,Aguilar-Gutierrez:2025pqp,Aguilar-Gutierrez:2025kmw,Das:2024tnw}. See \cite{Baiguera:2025dkc,Nandy:2024htc} for comprehensive reviews of the topic.} measures the spreading of an initial reference state $\ket{\psi_0}$ in the Krylov subspace under the time-evolution generated by $H$
\begin{equation}\label{eq:spread complexity}
    \mathcal{C}_{\rm S}(t)=\expval{\hat{n}}=\sum_nn\abs{\bra{K_n}\ket{\psi(t)}}^2~,\quad \hat{n}\equiv\sum_nn\ket{K_n}\bra{K_n}~,
\end{equation}
We now proceed to computing $\mathcal{C}_{\rm S}$ for the Hamiltonian \eqref{eq:hamds2}, taking the thermofield double (TFD) state as initial state.\footnote{Behavior of spread complexity for other initial states is briefly discussed in app.\,\ref{SC_lowest}.} It has been reported in the literature that for this state, the spread complexity shows a pronounced peak when the Hamiltonian is chaotic (in the sense that it has level repulsion in the spectrum, and the level spacing statistics follows the Wigner-Dyson distribution) \cite{Balasubramanian:2022dnj, Erdmenger:2023wjg, Camargo:2024deu}, whereas for other states, such as the domain wall state in spin chains, the peak is either absent or suppressed compared to that of the infinite-temperature TFD state \cite{Gautam:2023bcm, Camargo:2024rrj}.  
Specifically, the initial state considered here is this infinite-temperature version of the usual TFD state. It is defined in a double copy Hilbert space constructed by taking a tensor product of the original Hilbert space with itself, i.e., 
\begin{equation}\label{TFD_inft}
    \ket{\TFD}_{\infty}=\frac{1}{\sqrt{N}} \sum_n \ket{E_n} \otimes \ket{E_n}~, 
\end{equation}
where $\ket{E_n}$ and $N$ denote the eigenstates of the Hamiltonian \eqref{eq:hamds2}, and the dimension of the energy eigenbasis respectively. We then consider the time evolution of this state with a Hamiltonian of the form $H \otimes \mathbb{I}$
and evaluate the spread complexity of the time-evolved state. 

\begin{figure}[ht]
    \centering
    \includegraphics[width=0.56\linewidth]{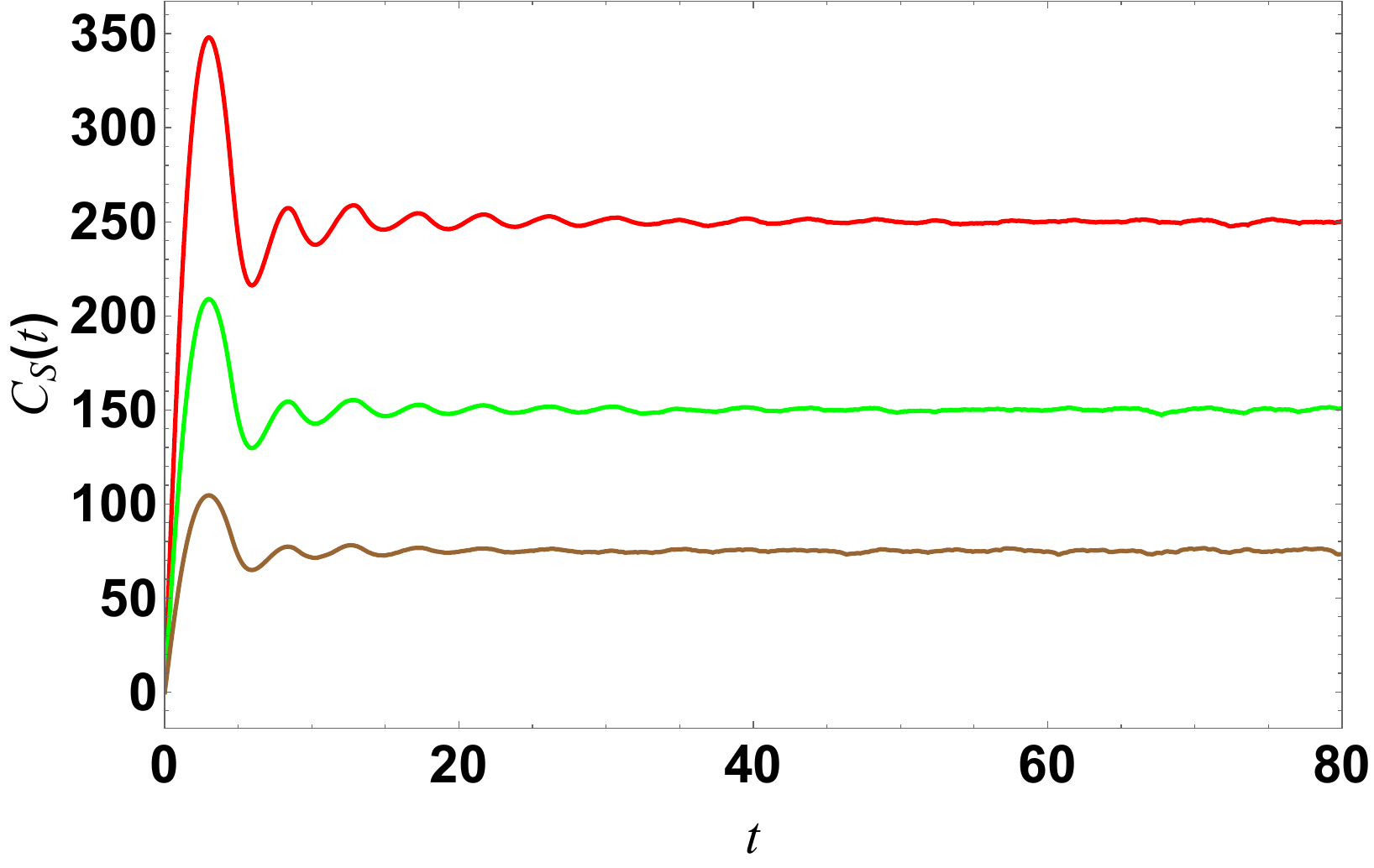}
    \caption{{Numerically obtained time evolution pattern of the spread complexity when the infinite temperature TFD state is taken as the initial state.} Here $j=301/2$ (brown), $j=601/2$ (green) and  $j=1001/2$ (red) with $\nu=0$. The complexity attains a peak before reaching a saturation value (see the discussion on the saturation value of complexity in the main text). }
    %The dashed line indicates the late time saturation value of $\mathcal{C}_{\rm S}(t)/N$, which is equal to $0.3$ for the TFD initial state. The plot in the inset shows the early time growth of the spread complexity.}
    \label{fig:C_TFD_DEVEN}
\end{figure}

In fig.\,\ref{fig:C_TFD_DEVEN}, we show the spread complexity 
of the time-evolved \eqref{TFD_inft}. We have set $\nu=0$, and chose $j$ such that $2j+1$ is even. For these choices, the energy spectrum has two-fold degeneracy at each energy
level, and the dimension $D_K$ of the Krylov subspace is $(2j+1)/2$. 
To numerically compute the spread complexity, among the two degenerate levels, we removed one, and constructed an infinite temperature TFD state from the resulting Hamiltonian having dimension $(2j+1)/2$.
The resulting complexity
has a clear peak at early times, followed by a series of increasingly damped oscillations. Finally, the complexity reaches a saturation value of $D_K/2$, as expected when the infinite-temperature TFD state is taken as the initial state \cite{Erdmenger:2023wjg}.

Fig.\,\ref{fig:C_TFD_DODD} shows spread complexity with TFD initial state, when  $\nu$ is non-zero, while $j$ is chosen in such a way that $2j+1$ is odd. The complexity again follows a general pattern of rise, peak, (damped) oscillations, and saturation. However, note that the saturation value of the complexity is significantly lower than the expected saturation value $D_K/2$ for infinite temperature TFD state and $D_K=2j+1$. In fact, the value of $\mathcal{C}_S(t)$ at the peak is smaller than the expected saturation value. Indeed, even though for these parameter choices, there are no exact degeneracies in the energy spectrum, there is an approximate two-fold degeneracy in each eigenvalue. Apart from a small region in the middle of the spectrum, the eigenvalues for the two invariant subsectors of even and odd spin states lie very close to each other. This reduces the effective dimension of the Krylov subspace below $2j+1$.

Hence, even though there is an early-time peak, it is smaller even than the saturation value when $2j+1$ is odd or $\nu \neq 0$. Only at very late times, the tiny differences (around $\mathcal{O}(10^{-12})$ for $j=1000$) between approximately degenerate eigenvalues are resolved. Therefore, if one considers very late times scales, greater than $\sim (\Delta E_{\text{deg}})^{-1}$, the spread complexity reaches its expected saturation value $D_K/2$ from below (see fig.\,\ref{fig:C_TFD_DODD}b where we have shown this behavior) and oscillates around this value.
This fact helps us to distinguish saddle-dominated scrambling from chaos, since in the latter case, the peak value of spread complexity is larger than the saturation value.  A schematic picture of the time evolution of the spread complexity 
up to its saturation is shown in fig.\,\ref{fig:C_TFD_DODD}c.

{Moreover, we note that the peak in the spread complexity appear at a time scale of $\mathcal{O}(1)$%\footnote{The peak time scale does not have the same $\lambda^{-1}\log j$ behavior as the peak in Krylov operator complexity that we considered in the previous section.}
, for both the cases of $\nu=0$ and $2j+1$ is even (spectrum with exact two-fold degeneracy) and $\nu\neq0$.} For comparison, for a random matrix Hamiltonian, drawn, e.g., from the Gaussian unitary ensemble, the peak in spread complexity for the TFD state typically appears at a time scale $\mathcal{O}(N)$, where $N$ is the size of the matrix. In our case, though there is a peak in the spread complexity, it appears much earlier compared to chaotic Hamiltonians of similar size. {This fact helps us to distinguish the saddle-dominated scrambling from genuine quantum chaos, specifically for the $\nu=0$ and $2j+1= \text{even}$ case (Fig.\,\ref{fig:C_TFD_DEVEN}).}  

\begin{figure}[h!]
    \centering
    \subfloat[Early time]{
    \centering
    \includegraphics[width=0.33\textwidth]{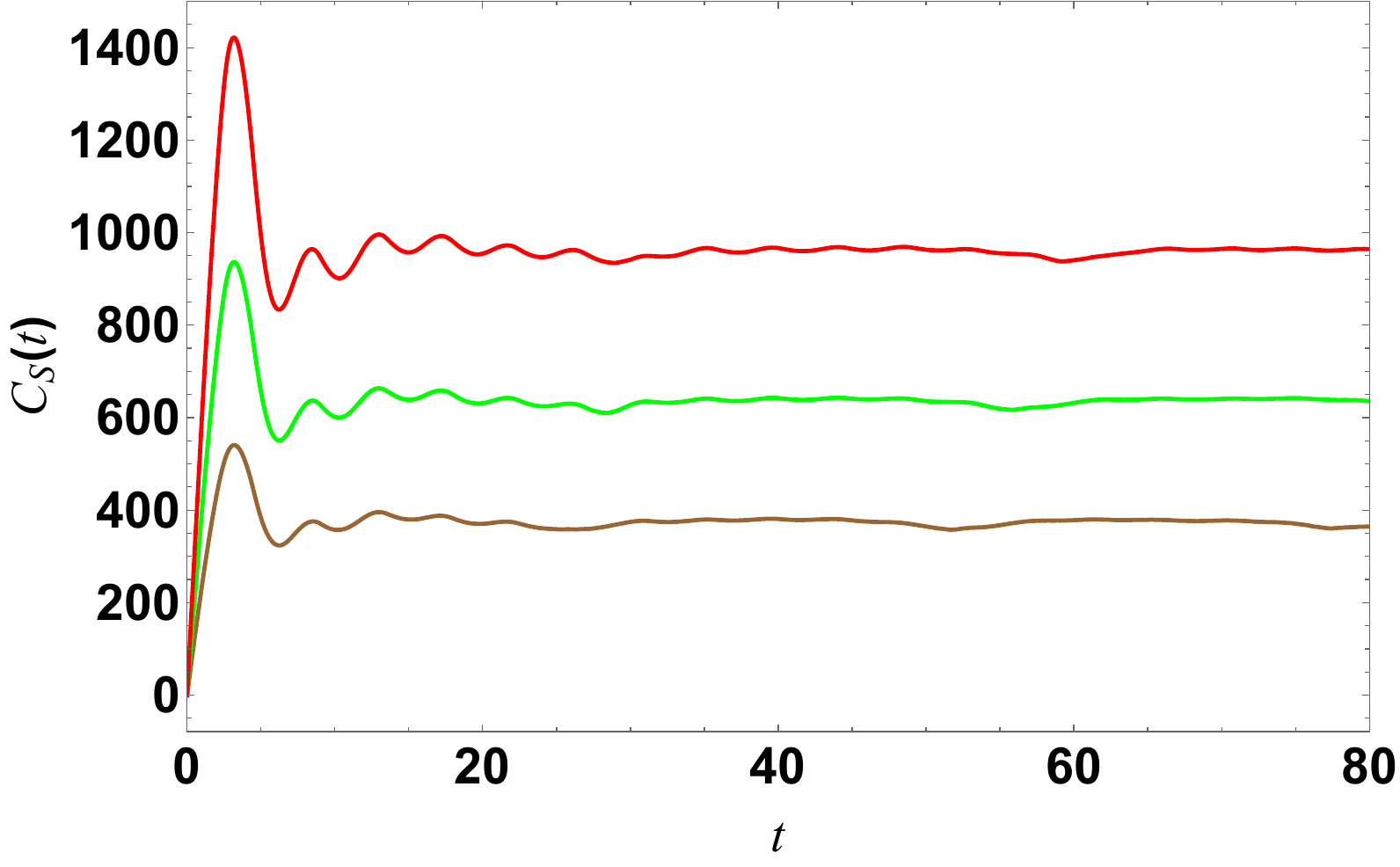}}
    \subfloat[Late time]{
    \centering 
    \includegraphics[width=0.33\textwidth]{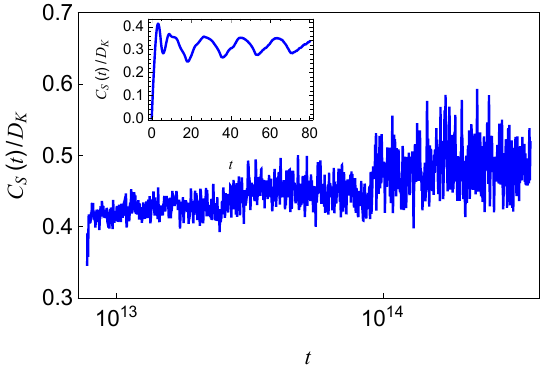}}
    \subfloat[Schematic picture]{
    \centering 
    \includegraphics[width=0.33\textwidth]{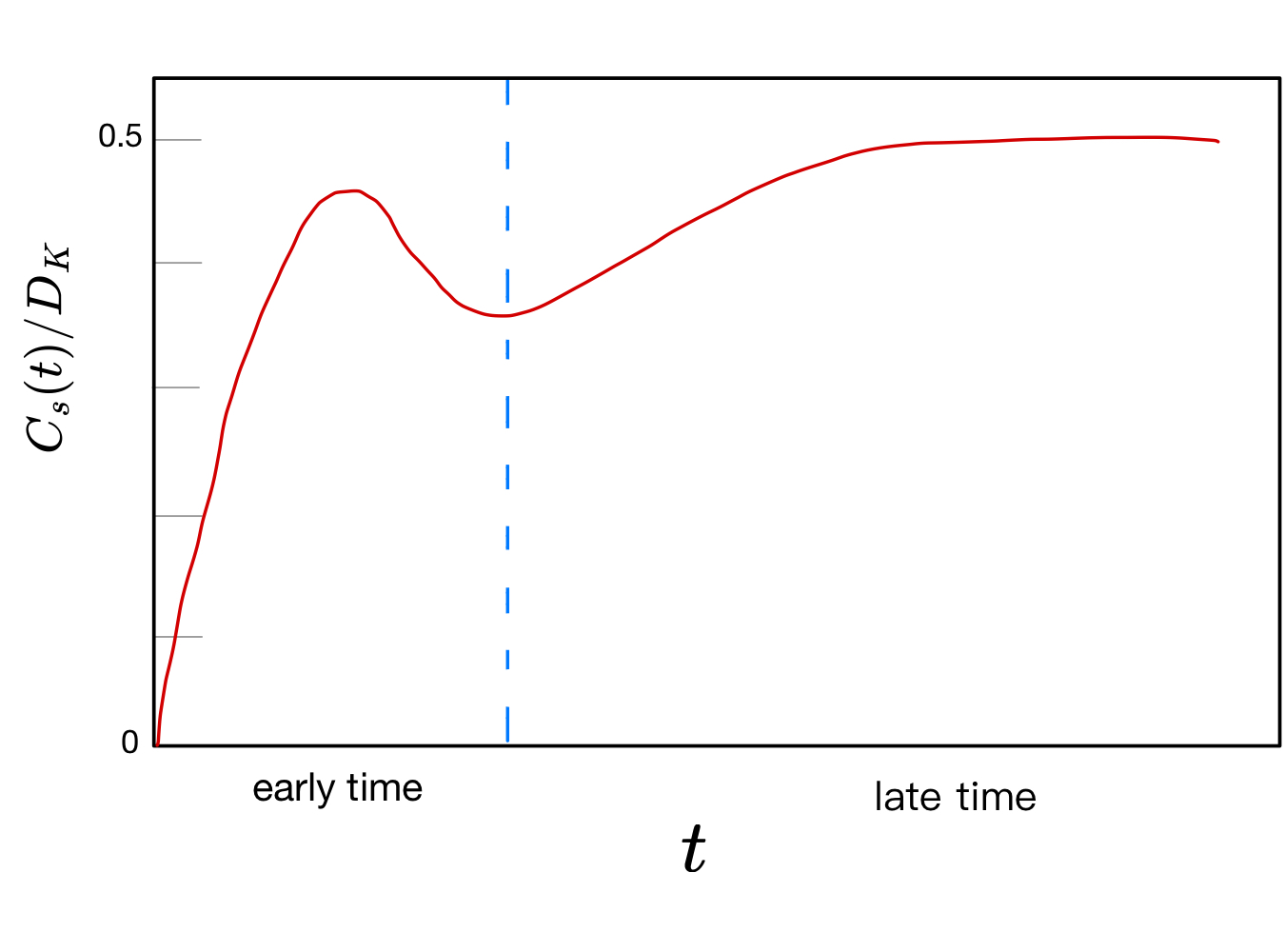}}
    \caption{Plot (a) shows the time evolution of the spread complexity for the infinite temperature TFD state as the initial state. Here $j=600$ (brown),  $j=1000$ (green), and $j=1500$ (red) with $\nu=4$. The complexity attains a peak before reaching an approximate saturation value, which, for the relatively early time scale shown here, is lower than the expected saturation value $D_K/2$ for the TFD initial state \eqref{TFD_inft} due to an approximate two-fold degeneracy in most of the energy spectrum (see the discussion in the main text). Panel (b) shows the late-time spread complexity for the infinite-temperature TFD initial state, with $j=200$ and $\nu=10$. Panel (c) is a schematic picture of spread complexity for the infinite-temperature TFD state with non-zero $\nu$. The saturation to $D_K/2$ from below is reached at a very late time around a time scale $\sim 1/\Delta E_{\text{deg}}$, where $\Delta E_{\text{deg}}$ is the difference of two approximately degenerate energies close to the edge of the spectrum.}
    %The dashed line indicates the late time saturation value of $\mathcal{C}_{\rm S}(t)/N$, which is equal to $0.3$ for the TFD initial state. The plot in the inset shows the early time growth of the spread complexity.}
    \label{fig:C_TFD_DODD}
\end{figure}

%%%%%%%%%%%%%%%%%%%%%%%%%%%%%%%%%%%%%%%%%%%%%%%%%%%
\section{Complementary series and PT-symmetry breaking }\label{sec:complementary series}
%%%%%%%%%%%%%%%%%%%%%%%%%%%%%%%%%%%%%%%%%%%%%%%%%%%

Before concluding, let us take a moment to discuss the complementary series in $\dS_2$. This is the representation in which light scalars ($2m\ell_{\dS} < 1$) transform.  In this case $\nu \in \rmi (-\frac12,\frac{1}{2})$, such that the scaling dimension $\Delta \in (0,1)$\,. Is there a spin Hamiltonian similar to \eqref{eq:hamds2} which has emergent complementary series QNMs in the large-spin limit?

One could imagine analytically continuing $\nu$. Formally, the proof of convergence given in \cite{Parmentier:2023axg} and reviewed in sec\,.\,\ref{sec:SU(N)} remains the same, yielding the complementary series character at large $j$. Unfortunately, Hamiltonian $H_j$ in \eqref{eq:hamds2} is no longer Hermitian when $\nu$ is imaginary. It is, however, PT-symmetric. In what follows, we drop the subscript $j$ to reduce clutter.

Let us first define an anti-Hermitian operator $\mathrm{T}$, acting simply by complex conjugation. It maps the eigensystem $(\omega, \psi)$ of $H$ to the eigensystem $(\omega^*, \psi^*)$ of $H^*$. For the LMG Hamiltonian with  $\nu\in\rmi\IR$: $\mathrm{T}H\mathrm{T}^{-1} = H^* = -H$. Defining also $\mathrm{P}=\rme^{\rmi \pi J_x}$, we find that since $\mathrm{P}H\mathrm{P}^{-1} = -H$, the combined action of $\mathrm{PT}$ leaves the Hamiltonian invariant.\footnote{\label{footPT}We are thinking of a particle living on a lattice with position $J_z$ and momentum $J_y$. Then indeed $\mathrm{T}$ is supposed to flip $J_y$, while $\mathrm{P}$ flips both $J_z$ and $J_y$. This definition makes the spectral analysis more parallel to what is usually done for so-called $\mathrm{PT}$-symmetric Hamiltonians \cite{Bender:1998ke}. Of course, strictly speaking, the physical time-reversal is supposed to flip all spin operators, and would correspond to what we defined as $\mathrm{PT}$ here. }

If the eigenfunctions of a PT-symmetric $H$ are themselves invariant under PT, the corresponding eigenvalues satisfy $\omega = \omega^*$ and must be real. This scenario occurs for several non-Hermitian Hamiltonians, which typically interpolate between PT-broken and unbroken phases. In the unbroken phase, the eigenvalues are all real and despite non-Hermiticity, the Hamiltonian still defines a healthy quantum mechanical system \cite{Bender:1998ke, Bender:2023cem}. 

We might thus hope to still find real eigenvalues when $|\nu| \leq |\nu_c|$ for some critical $\nu_c$. In the large-spin limit one would guess that $|\nu_c|\to \frac12$, i.e. the unitarity bound  for the $\SO(1,2)$ complementary series. We will discuss how this comes about, working in holomorphic polarization to study the spectrum. After introducing the Hamiltonian in holomorphic polarization, we study the spectrum at $\nu=0$, before analytically continuing $\nu \in \rmi \IR$.

%%%%%%%%%%%%%%%%%%%%%%%%%%%%%%%%%%%%%%%%%%%%%%%%%%%%%
\subsection{Spin-model spectrum and Heun polynomials}
%%%%%%%%%%%%%%%%%%%%%%%%%%%%%%%%%%%%%%%%%%%%%%%%%%%%%

The $2j+1$ eigenvalues of the spin Hamiltonian $H_j$ in \eqref{eq:hamds2} can, in principle, be found as roots of its characteristic polynomial. In our case, it does not take on a particularly pleasant form, so we proceed differently. In holomorphic polarization -- where the spin operators act as differential operators \eqref{eq:homspin} --  $H_j$ acts on states $\psi(z)$ as: 
\begin{equation}\label{eq:holoHnu}
    H_j\, \psi(z) = \Big(\, \tfrac{\rmi}{4j }\big(\,(1-z^4)\partial_z^2 +(4j-2 )z^3\partial_z + (2j-4j^2)z^2 \,\big) +\tfrac{\nu}{j}( z\partial_z-j)\, \Big) \psi(z)\,.
\end{equation}
 Solutions of the finite-$j$ eigenvalue equation are then polynomials $p(z)$ satisfying 
\begin{equation}\label{eq:eigeqn}
    H_j\, p(z) = \lambda\, p(z)\,,\qquad \text{deg}(p)\leq 2j\,.
\end{equation}
In fact, deg$(p)$ must be either $2j$ or $2j-1$; otherwise, one finds from \eqref{eq:eigeqn} that $p(z)$ vanishes. Not restricting to polynomials, one finds eigenfunctions for every $\lambda$:
\begin{equation}\begin{split}
    f_1(z) &= \mathsf{H}\ell(-1,\;\rmi j (\nu+\lambda),\; \frac{1}{2}-j,\; -j,\;\frac12,\;\frac{1}{2}-j+\rmi \nu,\;z^2)\,,\\
    f_2(z) &= z\;\mathsf{H}\ell(-1,\rmi j (\nu+\lambda)-\rmi\nu,\; \frac{1}{2}-j,\;1- j,\; \frac{3}{2},\; \frac{1}{2}-j+\rmi \nu,\; z^2)\,.
    \end{split}
\end{equation}
In the above, $\mathsf{H}\ell$ is the Heun function. The finite-$j$ eigenvalues are then those $\lambda$ for which one of these functions reduces to a polynomial of degree $2j$ or $2j-1$, corresponding to the even- and odd-spin invariant subspaces, respectively. 

\subsection{Spin-model spectrum at \texorpdfstring{$\nu=0$}{}}
Let us now discuss the Hamiltonian at $\nu=0$, before continuing $\nu$ along the imaginary axis. First of all, the parity of $2j$ plays an important role. When $2\mid 2j$, there is an odd number of states in total. The eigenvalue pairs $\pm \omega$ correspond to states $\psi$ and $\mathrm{T}\psi = \mathrm{P}\psi =\psi^*$. The unpaired eigenstate must have zero energy and is an eigenstate of both $\mathrm{P}$ and $\mathrm{T}$. On the other hand, when $2 \mid 2j+1$, there is an even number of states in total. Both $\mathrm{P}$ and $\mathrm{T}$ flip $\omega\to-\omega$. However, in this case, the energy eigenfunctions are not eigenfunctions of $\mathrm{PT}$. They cannot be, since under $\mathrm{P}$, $z^n\to z^{2j-n}$ changes parity, whereas $\mathrm{T}$ leaves parity as it is. Consequently, the spectrum is exactly doubly degenerate, in agreement with Kramer's theorem, see also footnote\,\ref{footPT}. Moreover, one finds that for $2j+1 = 0\; \text{mod}\; 4$ there are no states with $\omega=0$, while for $2j+1 = 2\; \text{mod}\; 4$ there are two.

To demonstrate these last claims, we explicitly solve for the $\omega=0$ (ground) states. From \eqref{eq:holoHnu}, the zero-energy eigenvalue equation at $\nu=0$ becomes:
\begin{equation}\label{eq:holoH}
   \big((1-z^4)\partial^2_z + (4j-2)z^2(z \partial_z -j)\big) p(z)=0\,,
\end{equation}
which has two independent solutions 
\begin{equation}
    p_1(z)= {}_2\mathsf{F}_1\Big(\frac14 - \frac{j}{2}, -\frac{j}{2},\frac34, z^4\Big) ,\quad p_2(z) = z\,{}_2\mathsf{F}_1\Big(\frac14 - \frac{j}{2},\frac{1}{2} -\frac{j}{2},\frac54, z^4\Big)\,.
\end{equation}
First, consider $2\mid 2j$. When $j$ is even, $p_1$ truncates to a polynomial, while for odd $j$ it is $p_2$ which does. Now take $2\mid 2j+1$. When $2j = 1\; \text{mod}\; 4$, both truncate, while for $2j= -1\; \text{mod}\; 4$ there is no polynomial solution.

\subsection{Continuing \texorpdfstring{$\nu\in\rmi\IR$}{} and  PT-symmetry breaking}

Now, consider moving $\nu$ along the imaginary axis.

Let us first discuss the case $2\mid2j+1$, where $\mathrm{PT}$-symmetry is broken at the level of the wave functions at $\nu=0$. When $2j+1 = 0\; \text{mod}\; 4$, the eigensystem splits into quadruples $\pm \omega, \pm \omega^*$ corresponding to acting with $\one, \mathrm{P}, \mathrm{T}, \mathrm{PT}$\,. With $2j+1 = 2\; \text{mod}\; 4$, the 2 zero-energy states are special. Their eigenvalues must continue to $\pm \omega = \mp \omega^*$, and therefore along the imaginary axis. $\mathrm{P}$ maps one state to the other, while $\mathrm{T}$ leaves the state invariant. For a generic $\nu\in\rmi\IR$, all eigenvalues are complex.

When $2\mid 2j$, we start out in the PT-unbroken phase at $\nu=0$, and the situation is more interesting. The eigenvalues remain real until a critical $\nu_c$. When $j$ is odd, this value turns out to be precisely $\nu_c = \rmi/2$, at which point the two smallest eigenvalues -- with eigenfunctions having even powers of $z$ -- have moved to $\omega=0$. At this point, $H$ is no longer diagonalizable. There is the original ground state (with odd powers of $z$) and then a Jordan block with $H\psi_0=0$ and $H\psi_1=\psi_0$. These two states then move past the ground state as $\nu$ is increased, and get purely imaginary eigenvalues. Below the critical value, the large-spin density of states is that of a light scalar in $\dS_2$, see fig.\,\ref{fig:doscomp}a.

When $j$ is even, the ground state has even powers of $z$, and $\nu = \rmi/2$, turns out to be the value at which the second smallest (closest to zero) eigenvalues (whose $\psi(z)$ has even powers of $z$) cross the smallest ones (odd powers of $z$). Since they are even/odd, they move past each other without going into the complex plane. At a later value of $\nu_c$, the even ones arrive at $\omega=0$ before becoming purely imaginary eigenvalues. From a fit, see fig.\,\ref{fig:doscomp}b, we find 
\begin{equation}\label{eq:fitnuc}
    |\nu_c|\approx \frac12 + \frac{1}{1+\log{2j}}\quad \text{as } j\to\infty\,,\quad j\in 2\mathbb{N}+1\,,
\end{equation}
so that at large $j$ the critical value becomes $\frac12$. Further increasing $\nu$ to $3\rmi/2$, we now get that at $\omega=0$ there are three states again. One is the ground state with even powers of $z$, and the others form a Jordan block with states of odd powers of $z$. 

\begin{figure}
    \centering
  \begin{subfigure}{0.49\textwidth}
            \centering
            \begin{tikzpicture}
              \node[inner sep = 0pt] at (0,0) {\includegraphics[width=\textwidth]{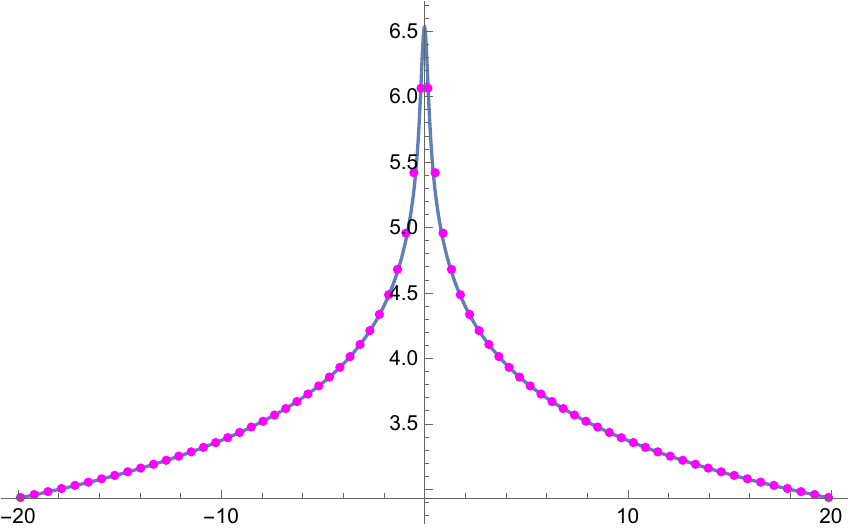}};
              \node at (0.7,2) {\text{$\rho(\omega)$}};
              \node at (3.9,-1.9) {\text{$\omega$}};
            \end{tikzpicture}
            \caption[]
            {\small complementary series}
        \end{subfigure}
        \hspace{0.3cm}
        \begin{subfigure}{0.45\textwidth}  
            \centering 
        \begin{tikzpicture}
              \node[inner sep = 0pt] at (0,0) {\includegraphics[width=\textwidth]{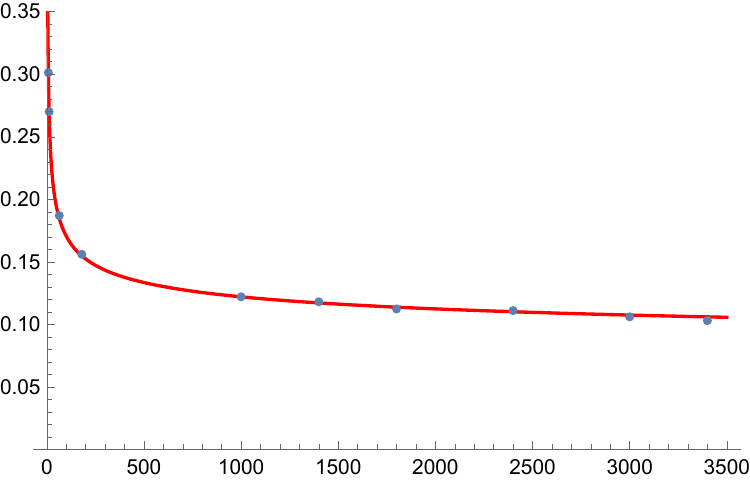}};
              \node at (-2.3,2.) {\text{$\nu_c-\frac12$}};
              \node at (3.4,-1.7) {\text{$j$}};
            \end{tikzpicture}
            \caption[]
            {\small approach to $\nu_c=\frac12$}
        \end{subfigure}       
    \caption{In (a) we see the density of states $\rho(\omega)$ for \eqref{eq:hamds2} at $j=501$ and the complementary series value $\nu=\frac{\rmi}{4}$, compared to the analytical $\dS_2$ result \eqref{eq:exactrho}. When $j \in 2\mathbb{N}+1$, the critical $\nu_c\in \rmi\mathbb{R}$ at which the first imaginary eigenvalue appears is $|\nu_c|=1/2$. In (b), we show the behavior of the critical $\nu_c$ for different values of $j\in 2\mathbb{N}$, which is consistent with \eqref{eq:fitnuc}.}
    \label{fig:doscomp}
\end{figure}
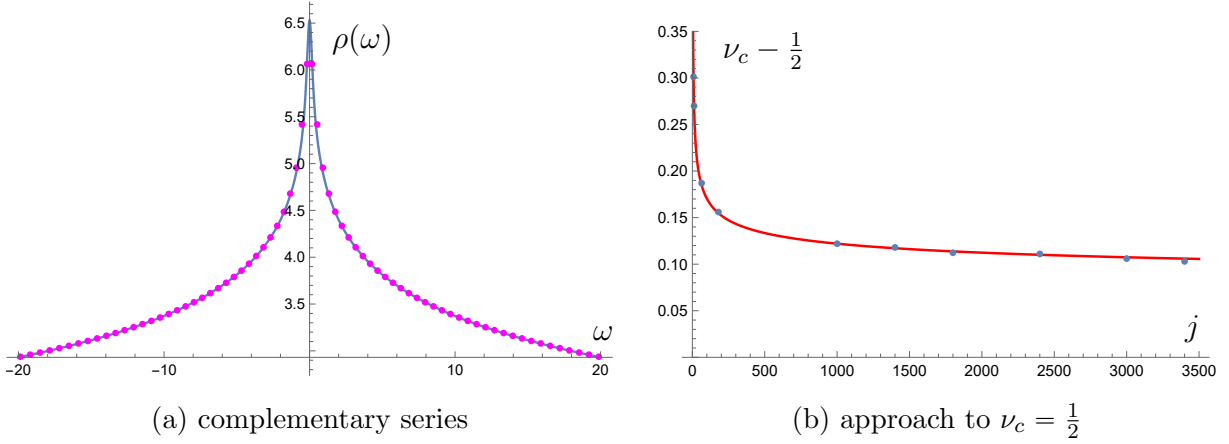

We can prove this property of the spectrum, namely that for even $j$, there is a double degeneracy at $\omega=0$ when $\nu\in \rmi (\frac32+2\mathbb{N})$, while for odd $j$ this happens when $\nu\in\rmi(\frac12 + 2\mathbb{N})$. (In all cases $\nu \to -\nu$ states are found by mapping $z\to \rmi z$.) What makes the values $\nu=\rmi(\frac12+m)$ special is that the Hamiltonian \eqref{eq:hamds2} then takes the form:
\begin{equation}
    H\, \propto \; J_x J_y + m[J_x,J_y]\,.
\end{equation}
First, we look for zero-energy states with even powers of $z$ by taking an Ansatz
\begin{equation}
    \psi^{(0)}= \sum_{i=0}^{j}b_i (1-z^2)^{j-i}(1+z^2)^i\,.
\end{equation}
Using the action of $H$ in holomorphic polarization \eqref{eq:holoHnu}, we find the conditions 
\begin{equation}
 (i+2)(j+m-i-1)  b_{i+2}= b_i (i-m)(j-i)  ,\quad b_1=b_{j-1}=0\,.
\end{equation}
Now, for even $j$, we can indeed consistently put $b_1=b_{j-1}=0$. Then, we find an even polynomial in $z$, which is the ground state at any value of $m$. On the other hand, when $j$ is odd, we can take $b_{1}=0$, but since $b_{j-1}$ would be generated from $b_0$, the series must truncate at an even value of $i$. Note that starting with $b_0\neq 0$, the series can indeed truncate at $i=m$. Hence, the requirement that $m\in 2\mathbb{N}$ in this case, and we have found the excited state which reaches $\omega=0$ at this particular $\nu$. Note that for any $j$, when $m=0$ and hence $\nu=\rmi/2$, the series truncates at the first term and we find $\psi^{(0)}=(1-z^2)^j$.  

Second, we look for zero-energy states with odd powers of $z$ with the Ansatz
\begin{equation}
    \psi^{(1)}= \sum_{i=0}^{j-1}c_i z(1-z^2)^{j-i-1}(1+z^2)^i\,,
\end{equation}
so that this time, we obtain,
\begin{equation}
     (i+2)(i+1-j-m)  c_{i+2}= c_{i} (j-i-1)(m-i-1) ~,\quad c_1=c_{j-2}=0\,.
\end{equation}
The situation is now reversed. For odd $j$ we can always consistently put $c_1=c_{j-2}=0$ and find a ground state solution with odd powers of $z$. On the other hand, for even $j$ we only find a solution, starting from $c_{0}=1$, when the series truncates at $i=m-1$. This requires odd $m$, in which case we have found the excited state that reaches $\omega=0$ at this particular value of $\nu$. Note that in both cases, when $\nu=3\rmi/2$, the series truncates at the first term, and we find a solution $\psi^{(1)}= z(1-z^2)^{j-1}$.

\subsection{Summary of the analysis}
{Due to the technical nature of this section, we conclude with a brief summary that addresses affirmatively the question posed at the start of this section, and in sec.\,\ref{sec:intro}. Namely, we have shown that by allowing the mass parameter ($\nu$) in the SU$(2)_j$ LMG Hamiltonian \eqref{eq:hamds2} to take imaginary values we recover emergent complementary series QNMs in dS$_2$ space in the large-spin limit. Crucially, although the resulting Hamiltonian is non-Hermitian in this case, we have shown through holomorphic polarization methods that the Hamiltonian with $\nu\in\rmi\mathbb{R}$ is PT-symmetric as long as $\nu$ remains below a critical value (which in the large spin limit becomes the unitarity bound for SO$(d,1)$). This implies that the energy eigenvalues remain real and the resulting theory still shares similar well-behavior as if it were Hermitian, as it is argued in different parts of extensive literature on PT symmetric systems (e.g. \cite{Bender:1998ke, Bender:2023cem}). We believe it will be interesting to continue the study of these complexified LMG systems, which we initiated in the above subsections.}

\section{Discussion}\label{sec:discussion}
Below, we briefly summarize our results and list a few suggestions for future work.
\paragraph{Summary}
We have analyzed various extensions of the SU$(2)$ LMG system \eqref{eq:hamds2} whose classical dynamics is characterized by having saddle points in the classical phase space, see sec.\,\ref{sec:SU(N)}. Notably, we initiated in sec.\,\ref{sec:complementary series} the study of a complexified but PT-symmetric version of \eqref{eq:hamds2} which reproduces the density of states associated to a light particle in dS$_2$, transforming in the complementary series representation. This density has poles at the $\dS_2$ QNMs. In the spin system these are emergent, as they do not exist at finite $j$. Yet, these exponentially damped modes do govern the large-$j$ dynamics of wave packets near the saddle points in phase space. We also considered an extension to an SU$(3)$ spin model \eqref{eq:spds3} for which we wrote down the analytic large-$j$ density of states $\rho(\omega)$ in sec.\,\ref{sec:spectra}. It has $\Re(\omega)>0$ poles at the QNMs for a massive particle in dS$_3$, while its $\Re(\omega)<0$ poles correspond to QNMs in $\dS_2$ instead. Besides the density of states, in sec.\,\ref{sec:spectra} we also studied other measures of the spectrum, namely the unfolded level spacing statistics and spectral form factor:
\begin{itemize}
\item The unfolded level spacing statistics $P(s)$ in the SU$(2)$ shows $\delta$-spikes at $s=1$, due to the regularity of the spectrum, indicating the theory is integrable. In the SU$(3)$ case the same is true in a fixed angular momentum sector. Considering the combination of such sectors does yield a standard Poisson distribution for the level spacings.
\item The SFF for the $\SU(2)$ and $\SU(3)$ systems at fixed angular momentum are also quite similar; both display a slope-dip-plateau structure, with essentially no ramp. The sudden way in which the plateau is reached after the dip again indicates the integrable nature of the system. In the $\SU(3)$ case, summing over angular momentum $L$ essentially modifies the time scale for the transition slope to plateau and does introduce a ramp-like feature, due to the fact that the spacing between the same eigenvalue in different $L$-sectors is much smaller than that of different eigenvalues with the same $L$.
\end{itemize}
In sec.\,\ref{sec:spread complexity} we analyzed various other dynamical probes of the system. Focusing on the SU$(2)$ case, we studied in particular the imprint left by saddle-dominated scrambling on these probes, and how to differentiate it from quantum chaos. Our main observations are:
\begin{itemize}
\item \emph{Two-point correlators}: These display a decay up to a dip time and subsequently oscillate around a given late-time value. We estimated the period of some of these transient oscillations by analyzing the classical phase space orbits.
\item \emph{Squared commutator and OTOCs}: We evaluated $\tr [J_i(t),J_i]^2$ for various spin operators. The presence of unstable saddle points, with classical Lyapunov exponent $\lambda_{\text{saddle}}$, leads to an early-time exponential growth of the squared commutator with $\Lambda_{\text{OTOC}}\geq \lambda_{\text{saddle}}$, consistent with \cite{Xu:2019lhc}. While this behavior is similar to chaotic systems, the integrable nature of the system reveals itself in large oscillations at later times. 

\item \emph{Krylov operator complexity}: We found a linear growth in the Lanczos coefficients determined by  $\lambda_{\text{saddle}}$ and correspondingly an early-time exponential growth in the Krylov operator complexity. At late times, the complexity saturates at a value over half of the dimension of the Krylov space with oscillations due to the right-sided biased time-averaged transition amplitudes in the Krylov chain. It would be interesting to study to what extent this feature holds more generally in integrable systems with saddle points.
%which shall indicate that the system is non-chaotic.

\item \emph{Spread complexity}: Considering the infinite-temperature TFD state as the initial state, we found that the spread complexity follows a general pattern of with a peak followed by (damped) oscillations, and saturation -- quite analogous to that of the chaotic systems having level repulsion in the spectrum. However, we observed that due to (approximate) degeneracies in the spectrum, the intermediate saturation value of the complexity can be significantly lower than the expected one. At very late times (of order of inverse of the level spacings of the approximately-degenerate energies), when the final saturation is eventually reached, it is reached from below, as opposed to what happens in quantum chaotic systems.% where the level repulsion in the spectrum leads to a peak value in the spread complexity larger than the saturation value (leading to a late time saturation value that is approached from above).
\end{itemize}

\paragraph{Outlook}
We conclude with a few speculative comments and suggestions for future work. 
\begin{itemize}
   \item \textit{Quenches and phase transitions}: Performing quenches in other versions of the LMG model leads to phase transitions \cite{Bento:2023bjn}. It would be worth exploring this in our model as well, and perhaps to find an interpretation in terms of de Sitter space and quasinormal modes, as we have conveyed in most of the present work.

    \item \textit{Geometry and complexity}: The systems we studied appear to mimic the spectrum and dynamics of particles in de Sitter. It would be interesting to generalize this: can similar saddle-dominated spin systems describe particles in other spacetimes? Perhaps one can use the relation between spacetime geometry and Lanczos coefficients for Krylov operator complexity as discussed in \cite{Dodelson:2025rng}. In particular, the exponential growth of Krylov operator complexity reveals information about the QNMs. See also sec.\,4.2 in \cite{Parmentier:2023axg} for a different way in which \eqref{eq:hamds2} discretizes the dynamics of a particle in $\dS_2$.
    \item \textit{Effective temperature}: We have studied spin systems using probes evaluated at infinite temperature. The classical Lyapunov exponent at the saddle point does set a particular time scale. Equivalently, the spacing between de Sitter QNM frequencies, which show up as poles in the analytic large-$j$ limit of the density of states, corresponds to an effective (de Sitter) temperature. Its scale is controlled by the overall scaling of the spin Hamiltonians \eqref{eq:hamds2} and \eqref{eq:spds3} respectively. Relatedly, it is known that linear growth of the Lanczos coefficients and $\exp(-\beta \omega/2)$ fall-off in the Fourier transform of the OTOC are equivalent \cite{Avdoshkin:2019trj, Dodelson:2025rng}. Both have the interpretation of an inverse temperature. In the de Sitter context, this temperature is reminiscent of the `fake' temperature of \cite{Susskind:2021esx}. 
    \item \textit{Complexified LMG and discrete series}: It would be interesting to continue the study of the complexified LMG system and its number of complex eigenvalues at various values of $\nu \in \rmi\mathbb{R}$. We initiated this in sec.\,\ref{sec:complementary series} and expect that the large literature on exact results \cite{ORTIZ2005421, LERMAH2013421} may be helpful in proving more general statements. 
    %Also interesting is the relation between Heun functions (which appear as eigenfunctions of the spin Hamiltonian in holomorphic polarization \eqref{eq:holoHnu}), Fuchsian equations and Stieltjes polynomials, see \href{https://dlmf.nist.gov/31}{DLMF}. 
    Particularly intriguing is 
   the feature is that for any $j$, at  $\nu \in \rmi \mathbb{N}$, \eqref{eq:hamds2} appears to have precisely $2|\nu|$ imaginary modes. Since these values of $\nu$ correspond to the $\SO(1,2)$ fermionic discrete series, this observation is somewhat reminiscent of the number of discrete series modes that have to be gauged or rotated for the quantum system to make sense \cite{Anninos:2023lin, Letsios:2025pqo}. Perhaps the complexified but PT-symmetric LMG system also allows us to make the link to QNMs more direct. In PT-symmetric systems, the inner product is constructed using CPT conjugation, reminiscent of the inner product between QNMs \cite{Jafferis:2013qia, Parmentier:2023axg}.   
   \item \textit{SU(3) SFF and brick-wall systems}: Our results for the SFF in fig.\,\ref{fig:spectral_form_factor} essentially do not display ramp at fixed angular momentum, only after summing of angular momentum sectors does a ramp-like feature appear. This is similar to what was found recently in various brick-wall systems \cite{Das:2022evy, Jeong:2024oao, Ageev:2024gem}, where the appearance of a ramp can also be traced back to spectral rigidity between different angular momentum sectors. The brick-wall systems are discretized toy models for quantum particles in a spacetime with horizons. Similarly, our spin systems reproduce features of particles in de Sitter space. It would be interesting to see if there is a further connection.  Based on \cite{Law:2022zdq} one would think that the brick wall density of states would be a sum of a universal Rindler density and a `renormalized' density of states $\rho(\omega)$ associated with the specific spacetime, $\dS$ in our case. It is the latter which contains the information about the QNMs, and which was reproduced by the large-$j$ inverse level spacing of our spin systems. Our current toy models are good at describing the QNMs, but particles bounce back fast, after a time $\log j$. Can one think of this as the time needed to reach the brick wall? One could imagine combining the saddle-dominated properties of our spin systems with a truly chaotic system accounting for horizon degrees of freedom. In the case of the $\SU(3)$ system, it is somewhat tempting to imagine the sphere at infinity of zero-energy fixed points as horizon states and to modify their dynamics. 
\end{itemize}

\subsection*{Acknowledgements}
We thank Hugo A. Camargo, Viktor Jahnke,  Hyun-Sik Jeong, Jeremy Mann, Pratik Nandy, and Kunal Pal for valuable discussions and correspondence. %Add support XXXX. 
We thank the organizers of the workshop ``New Avenues in Quantum Many-body Physics and Holography'' in APCTP, which helped part of this collaboration to start. SEAG thanks Don Marolf in UC Santa Barbara for hospitality during the development of this work, and the QISS consortium for travel support. SEAG is supported by the Okinawa Institute of Science and Technology Graduate University. This project/publication was also made possible through the support of the ID\#62312 grant from the John Templeton Foundation, as part of the ‘The Quantum Information Structure of Spacetime’ Project (QISS), as well as Grant ID\# 62423 from the John Templeton Foundation. The opinions expressed in this project/publication are those of the author(s) and do not necessarily reflect the views of the John Templeton Foundation. Kl.\,P was supported by the US DOE DESC0011941. 

\appendix
% \section*{Accronyms}
% \setlist{nolistsep}
% \begin{itemize}
% \item (A)dS : (Anti-)de Sitter
% \item IHO: Inverted harmonic oscillator
% \item GOE: Gaussian orthogonal ensemble
% \item SFF: Spectral from factor
% \item RMT: Random matrix theory
% \item QNMs : Quasi-normal modes
% \item PT : Parity–time 
% \item OTOC: Out-of-time-order correlator
% \item LMG : Lipkin-Meshkov-Glick
% \item TFD : Thermofield double
% \end{itemize}

\newpage\section{de Sitter density of states }\label{ds_dos}
Let us introduce the Harish-Chandra character $ \chi(t)= \tr \rme^{-\rmi t H}$ associated with the static patch Hamiltonian $H$ \cite{Anninos:2020hfj}. It is related to the dS density of states by
\begin{eqnarray}
    \rho_{\dS}(\omega)=\int^{\infty}_{\Lambda^{-1}} \frac{\rmd t}{2\pi} \,\big(\rme^{\rmi \omega t}+\rme^{-\rmi \omega t}\big)\, \chi(t)~.
\end{eqnarray}
For instance, the spectrum of a massive particle in $\dS_2$, with principal series scaling dimension $\frac12 
+\rmi \nu$, lies encoded in the character $\chi(t)$, which can be found by summing over QNMs:
\begin{equation}\label{eq:chards2}
    \chi(t)= \sum_{n} \rme^{-nt}(\rme^{-\Delta t}+\rme^{-\bar\Delta t})=  \frac{\rme^{-\Delta t}+\rme^{-\bar\Delta t}}{|1-\rme^{-t}|}~,
\end{equation}and equivalently, the $\dS_2$ density of states \cite{Anninos:2020hfj, Law:2022zdq}:
\begin{equation}\label{eq:exactrho}\begin{split}
    \rho_{\dS_2}(\omega) &= \frac{2}{\pi}\log\big(\rme^{-\gamma}\Lambda\big) - \frac{1}{2\pi}\sum_{\pm,\pm}\psi\big(\tfrac12 \pm \rmi \nu \pm \rmi \omega\big)\,,
    \end{split}
\end{equation}
where $\psi(x) = \Gamma'(x)/\Gamma(x)$ is the digamma function and $\gamma$ the Euler-Mascheroni constant. The UV-regulator $\Lambda$ shifts $\rho$ without affecting its shape; it is a constant to be matched\footnote{In the spin model, for instance, it scales like $\Lambda \,\propto\; j$.}. The poles of $\rho$ are the QNM frequencies. In $\dS$ these are the same as those of two IHOs \cite{Parmentier:2023axg}. It will also be useful to recall that
\begin{equation}\label{eq:psisum}
    \psi(z) = -\gamma + \sum^{\infty}_{n=0}\Big(\frac{1}{n+1}-\frac{1}{n+z}\Big)\, ,
\end{equation}
allowing to split each $\psi$ in \eqref{eq:exactrho} into contributions coming from even and odd resonances:
\begin{equation}\label{eq:even/odd res}
    \psi\big(\tfrac12 \pm \rmi \nu \pm \rmi \omega\big) = \frac12\,\Big(\,\psi\big(\tfrac14 \pm  \tfrac{\rmi\nu}{2} \pm  \tfrac{\rmi\omega}{2}\big) + \psi\big(\tfrac34 \pm  \tfrac{\rmi\nu}{2} \pm  \tfrac{\rmi\omega}{2}\big)\,\Big)\,.
\end{equation}
Similarly, the $\dS_3$ character, with $\Delta = 1+\rmi \nu$, is given by
\begin{equation}
    \chi(t) = \sum_{n_1,n_2} \rme^{-(n_1+n_2)t}(\rme^{-\Delta t}+\rme^{-\bar\Delta t}) = \frac{\rme^{-\Delta t}+\rme^{-\bar\Delta t}}{|1-\rme^{-t}|^2}\,.
\end{equation}
It can be decomposed into fixed angular momentum sectors $L$
\begin{equation}
    \chi(t) = \sum_{l} \rme^{-(2 n +|L|)t}(\rme^{-\Delta t}+\rme^{-\bar\Delta t}) = \rme^{-|L|t}\frac{\rme^{-\Delta t}+\rme^{-\bar\Delta t}}{|1-\rme^{-2t}|}\,.
\end{equation}
Note that each takes the form of a $\dS_2$ character evaluated at $2t$ and $\Delta = \frac12 + \frac{|L|}{2}+\rmi\frac{\nu}{2}$. Using this, it follows from \eqref{eq:exactrho} that (up to a logarithmic divergence):
\begin{equation}\label{eq:rhods3}
    \rho_{\dS_3}(\omega,L) = -\frac{1}{4\pi} \sum_{\pm\pm} \psi\Big(\frac12 + \frac{|L|}{2}\pm \rmi \frac{\nu}{2}\pm\rmi\frac{\omega}{2}\Big)\,.
\end{equation}

%\newpage
\section{\texorpdfstring{$\SU(N)$}{} coherent states }\label{app:cohstates}
In this appendix, we review some aspects of coherent states that will be useful in the main text. 
Coherent states can be defined for general Lie groups \cite{Perelomov:1971bd}. This is based on the correspondence between irreps and coadjoint orbits. The latter are symplectic manifolds and can be thought of as phase spaces of classical dynamical systems. Coherent states are then quantum states labeled by points on the coadjoint orbit. They satisfy a minimal uncertainty property. In this sense, they are the most classical states possible \cite{kirillov1976elements, Perelomov_1977}. In what follows, we will focus on the concrete case of $\SU(N)$ coherent states, for which several useful formulas can also be found in \cite{Gitman_1993, Nemoto_2000}.  

\subsection{Coherent states}
We will begin by introducing the $\SU(N)$ irreps which arise by considering fixed-energy states of $N$ coupled harmonic oscillators. These will lead to the coherent states of interest. Defining the standard raising and lowering operators
\begin{equation}
    [a_I, a^\dagger_J] = \delta_{IJ},\qquad I=1,\dots N\,,
\end{equation}
one may impose the fixed-energy constraint
\begin{equation}\label{eq:con}
    a^\dagger \cdot a = 2j\,.
\end{equation}
The Fock states satisfying this constraint transform in the degenerate level $j$ irrep of $\SU(N)$. It is the one labeled by $\Box\hspace{-0.03cm}\Box \cdots \Box$ ($2j$ boxes), and has dimension $\tfrac{(2j+1)_{N-1}}{(N-1)!}$. 
A general state can be represented a homogeneous polynomial $\psi(Z^I)$, on which the operators act as
\begin{equation}
    a_I = \partial_{Z^I}\,, \qquad a^\dagger_{I} = Z^I\,.
\end{equation}
The inner product is then
\begin{equation}\label{eq:in1}
    \langle \Phi | \Psi\rangle = \frac{1}{\pi^N} \int \rmd^{2N} Z\, \rme^{-Z\cdot\bar{Z}}\, \bar{\Phi}(\bar Z)\Psi(Z)\, ,
\end{equation}
corresponding to K\"ahler quantization with K\"ahler potential $Z\cdot\bar{Z}$. An orthonormal basis is provided by products of monomials of the form $(Z^I)^n/\sqrt{n!}$.

Now we want to impose the constraint in \eqref{eq:con}. This means we restrict to homogeneous polynomials of degree $2j$, on which $\SU(N)$ acts through
\begin{equation}
    M_{IJ} = a^\dagger_Ia_J = Z^I\partial_{Z_J}\, ,\qquad [M_{IJ},M_{KL}] = \delta_{JK} M_{IL} - \delta_{IL}M_{KJ}\,.
\end{equation}
We can factor out the coordinate $Z^N$ whenever it is non-vanishing and, in doing so, define new wavefunctions $\psi$ and coordinates $z^i$ (with convenient normalization)
\begin{equation}
    \Psi(Z) = \frac{(Z^N)^{2j}}{(2j)!} \psi(z)\,, \qquad z^i Z^N = Z^i\,.
\end{equation}
Integrating out $Z^n\equiv U$ in \eqref{eq:in1}, we find:
\begin{equation}\label{eq:innerCoh}\begin{split}
   \langle \phi | \psi \rangle = \langle \Phi | \Psi\rangle &= \frac{1}{(2j)!\pi^N} \int \rmd^{2(N-1)} Z\,\rmd^2U\,(U\bar U)^{2j+N} \rme^{-U\bar U(1+z\cdot\bar{z})}\, \bar{\phi}(\bar z)\psi(z)\\
    &= \frac{\Gamma(2j+N)}{(2j)!\pi^{N-1}} \int \frac{\rmd^{2(N-1)} z}{(1+z\cdot\bar z)^{2j+N}} \, \bar{\phi}(\bar z)\psi(z)\,.
    \end{split}
\end{equation}
One can check explicitly that the following basis of spin eigenstates is orthonormal:
\begin{equation}\label{eq:basis}
   (z |n_1, n_2, \dots, n_{N-1}\rangle = \frac{\sqrt{(2j)!} (z_1)^{n_1} (z_2)^{n_2}\cdots (z_{N-1})^{n_{N-1}}}{\sqrt{n_1 !n_2! \cdots n_N!}}\,, \quad\sum^N_{I=1} n_I =2j\,.  
\end{equation}
After transforming to the inhomogeneous coordinates $z^i$, the spin operators become
\begin{equation}\label{eq:homspin}
    M_{Ni} = \partial_{z^i}\, , \quad M_{iN}=z^i(2j -\sum z^i\partial_{z^i})\, ,\quad  M_{ij} = z^i\partial_{z^j} \,.
\end{equation}
The inner product can be read as a completeness relation for coherent states, namely
\begin{equation}
    \frac{\Gamma(2j+N)}{(2j)!\pi^{N-1}} \int \frac{\rmd^{2(N-1)} z}{(1+z\cdot\bar z)^{2j+N}} \, |\bar z)(z| =\one\,.
\end{equation}
The overlap of coherent states is easy to find by inserting a complete set of states \eqref{eq:basis}
\begin{equation}
    (\bar{w}|z) = \sum_{n_i}\,(\bar w|n_i\rangle\langle n_i|z) = (1+\bar{w}\cdot z)^{2j}\,.
\end{equation}

\subsection{Roots, weights, and irreps}
More abstractly, we can discuss irreps in terms of roots $\alpha_i$ and weights $\mu_i$. The Dynkin diagram for $\SU(N)$ is
\begin{equation}
   \hspace{-2cm} A_{N-1} \qquad \begin{tikzpicture}[inner sep=0pt, scale=0.7, xshift=1.cm, yshift=-0.35cm, overlay]
        	   \draw[pink, line width=2] (0,0.5) circle (0.3);\draw[pink, line width=2] (1.2,0.5) circle (0.3);\draw[pink, line width=2] (4,0.5) circle (0.3); \draw[-, color=blue!30,line width=2] (0.33,0.5)  -- (0.87,0.5);\draw[-, color=blue!30,line width=2] (1.53,0.5)  -- (2.07,0.5);\draw[-, color=blue!30,line width=2] (3.13,0.5)  -- (3.67,0.5);\end{tikzpicture}{\hspace{2.27cm}\cdots}
\end{equation}
Let us focus on $\SU(3)$, since the discussion is straightforward to generalize. The Lie algebra has rank two, and we can take as normalized ($2\tr T_aT_b = \delta_{ab}$) Cartan generators 
\begin{equation}
    H_1 = \frac{1}{2}(U\partial_U-V\partial_V)\, , \qquad H_2 = \frac{1}{2\sqrt{3}}(U\partial_U+V\partial_V-2W\partial_W)\,.
\end{equation}
These act as $3\times 3$ matrices in the fundamental representation, which has three states $U,V,W$. The raising operators for the simple roots can be taken as
\begin{equation}
    J_{\alpha_1} = U\partial_V\, ,\quad J_{\alpha_2} = V\partial_W\,.
\end{equation}
One checks that 
\begin{equation}
   [H_a,J_{\alpha_i}]=(\alpha_i)_a J_{\alpha_i} \,,\qquad \alpha_1 = (1,0)\,, \quad\alpha_2 = (-\tfrac12,\tfrac{\sqrt{3}}{2})\,,
\end{equation}
so that the roots are normalized to $\alpha^2= 1$ and have inner product $\alpha_1\cdot\alpha_2= -\frac12$, in line with the Dynkin diagram notation. We can label representations by the Dynkin vector $2\mu\cdot( \alpha_1,\alpha_2)$, which are essentially the coordinates in a fundamental weight basis  ($\tilde{\mu}_i$ with $\tilde{\mu}_i\cdot \alpha_j=\delta_{ij}$). The representations of immediate interest are degenerate in the sense that they are labeled by $(2j,0)$ (and for general $\SU(N)$ by $(2j,0,\dots,0)$). 

\subsection{The classical limit}
The space of $\SU(N)$ coherent states is labeled by points in $\IC \rP^{N-1}$. This is the group $\SU(N)$ divided by the stabilizer (isotropic subgroup) of the highest weight state, which is $\SU(N-1)\times \U(1)$. In the large $j$ limit, the overlap between coherent states $|w)$ and $|z)$ becomes increasingly sharply peaked near $z\approx w$. This is a classical limit in which operators\footnote{At least those operators composed of a finite number of elementary spin operators, or in any case a number which grows more slowly than $\sqrt{j}$. If not, the number of commutators can compete with the decreasing size of each individual one.} can be replaced by their symbols, i.e. their coherent state expectation values: 
\begin{equation}
    A(w, \bar z) = \frac{(w|A|\bar z)}{(w|\bar z)}\,.
\end{equation}
A star product of symbols is defined as the symbol of the product operator:
\begin{equation}
     A\star B = (AB)(w,\bar z)\,.
\end{equation}
In the large spin limit, star commutators become Poisson brackets \cite{Berezin:1974du}
\begin{equation}
    A\star B - B\star A \to \{A,B\} = \Omega^{ij}\partial_iA\partial_j B\,,\quad \Omega = 2 \rmi j  \,\Big(\frac{\rmd z_i \wedge \rmd \bar z^i}{1+ z\cdot \bar{z}} - \frac{(\bar z \cdot \rmd z)\wedge(z\cdot\rmd \bar z)}{(1+ z\cdot \bar{z})^2}\Big)\,.
\end{equation}
This is a prime example of Berezin quantization of a compact K\"ahler manifold, where the symplectic form is related to the K\"ahler potential
\begin{equation}\label{eq:symp}
    \Omega_{ij} = \rmi \partial_{i}\partial_j K\, , \qquad K = 2 j \log (1+z\cdot \bar z)\,.
\end{equation}
In this case, we get the familiar Fubini-Study metric on $\IC \rP^{N-1}$. The role of $\hbar$ is played by $1/j$. In fact, if we had kept $\hbar$ explicitly, then the classical dynamics at fixed phase space volume becomes exact in the limit $\hbar \to 0,\; j\to \infty$ with $j\hbar $ fixed.  

%\newpage
\section{Saddle point analysis of the \texorpdfstring{$\SU(2)$ spin model}{}}\label{app:saddle_derivation}
In this appendix, we briefly present a classical analysis to determine the saddle points in the classical phase space dynamics \eqref{H_classical}.  The classical variables $X,Y,Z$ satisfy the Poisson brackets $\{X,Y\}=Z/j$ and cyclic permutations thereof. Denoting $X_i=\{X,Y,Z\}$, the Hamilton equations of motion, $\dot{X}_i(t) = \{X_i,H\}$ are then given by 
\begin{equation}\label{classical_eom}
   \dot{X}(t) =  X Z - \tfrac{\nu}{j} Y~, ~~\dot{Y}(t)  =
    - YZ + \tfrac{\nu}{j} X~,~~\text{and}~~ \dot{Z}(t)  =Y^2-X^2~.
\end{equation}
Fixed points satisfy the equations $\dot{X}_i(t) =0$. Using \eqref{classical_eom}, along with the constraint $X^2+Y^2+Z^2=1$, we determine the locations of the fixed points to be 
\begin{equation}\label{saddle_points}
    \begin{split}
        X=Y=0,~~ Z=\pm 1~~\text{and}\\
        X=\pm\frac{\sqrt{j^2-\nu^2}}{\sqrt{2}j}~, ~Y= \mp \frac{\sqrt{j^2-\nu^2}}{\sqrt{2}j}~,~~ Z=~ -\frac{\nu}{j}~~\text{and}\\
        X=Y=\pm\frac{\sqrt{j^2-\nu^2}}{\sqrt{2}j}~, ~~ Z=\frac{\nu}{j}~.
    \end{split}
\end{equation}
Thus, there are six of these points, and we refer to them as $S_i$, $i=1,2, \cdots6$.
Note that when $j>\nu$, the last four points do not represent physical solutions, whereas the first two are always stationary points regardless of the numerical values $j$ and $\nu$. 

To determine their stability properties, we construct the Jacobian matrix between $\dot{X}_i$ and $X_i$, given by
\begin{equation} \label{Jacobian}
		\mathbf{J} = \left(
		\begin{array}{ccc}
			Z & -\nu/j &  X   \\
			\nu/j & -  Z & - Y \\
              - 2  X & 2 Y & 0
		\end{array}
		\right)~.
\end{equation}
A stationary point is unstable if one of the eigenvalues of this matrix has a positive real part. 
%However, the exact analytical expressions for the eigenvalues at these saddle points are cumbersome, and we do not provide them here. 
The expressions for eigenvalues of $\mathbf{J}$ at these points, depending on whether $\nu>j$ or not, are given as follows. For $\nu<j$, 
\begin{equation}
    \begin{split}
e=\big(0,\pm\lambda\big)~~&\text{for}~~S=S_1, S_2~\\
        e=\big(0, \pm \rmi \sqrt{2} \lambda \big)~~&\text{for}~~S=S_3, S_4, S_5, S_6\,,
    \end{split}
\end{equation}
where $\lambda=j^{-1}\sqrt{j^2-\nu^2}$ is real in this case. On the other hand, for $\nu>j$, we have 
\begin{equation}
    e=\big(0, \pm   \lambda\big)~~\text{for}~~S=S_1, S_2~.
\end{equation}
Note that in this case $\lambda$ is purely imaginary.  

From the eigenvalues listed above, we see that when $\nu<j$, among the six fixed points, only two are unstable ($S_1$ and $S_2$). When $\nu=0$, these saddle points both have zero energy and $\lambda=1$. For $\nu>j$ there are only two fixed points and both are stable (elliptic).

%\sa{maybe}
We remind the reader that saddle-dominated behavior refers to the appearance of normal modes in \eqref{classical_eom} due to the saddle points ($S_1$, $S_2$ in this case), i.e.
\begin{equation}
    \dv{X_i}{t}=\pm\lambda X_i~.
\end{equation}
This leads to the exponential growth of the distance of classical orbits with respect to the origin, captured by the classical Lyapunov exponent ($\lambda=\sqrt{1-(\nu/j)^2}$) of this system.

\section{Spread complexity for other initial states}\label{SC_lowest}

In sec.\,\ref{spread_complexity}, we studied spread complexity 
with the TFD state as initial state, since for this state the signature of chaos is argued to be more prominent compared to other initial states, which have overlap with only a finite number of energy eigenstates. Here, for completeness, we briefly discuss the evolution 
of the spread complexity for some other relevant initial states. We note that at finite temperature, the system is more sensitive to special states like the one with the highest/lowest weight (indeed, at finite $\beta$, inverted harmonic oscillator also shows exponential growth in spread complexity, see \cite{Balasubramanian:2022tpr}).

\paragraph{Lowest weight state as the initial state.}

A relevant initial state is the lowest weight state: $\ket{j,-j}$, for which the Krylov basis elements are directly related to the states $\ket{j,n-j}$, and it simplifies the evaluation of the spread complexity.

To start with, we notice that (\ref{eq:Hamiltonian spin model}) is reminiscent of the action of a  generic Hamiltonian on the Krylov basis elements $\ket{K_n}$, given in \eqref{eq:Hamiltonian Krylov}.  Indeed, one can perform a redefinition $n=2m$ and a state renormalization $\ket{j,n-j}\xrightarrow{}\rmi^m\ket{j,2m-j}$, such that (\ref{eq:Hamiltonian spin model}) is equivalently:
\begin{equation}\label{eq:Hamiltonian j Krylov}
    H_j\ket{j,2m-j}=\nu\qty(\frac{2m}{j}-1)\ket{j,2m-j}+c_{2m}\ket{j,2m-j-2}+c_{2m+2}\ket{j,2m-j+2}~.
\end{equation}
We can then identify the Krylov basis and Lanczos coefficients for the $\ket{K_0}=\ket{j,-j}$ initial state:
\begin{equation}
    \ket{K_n}=\rmi^n\ket{j,2n-j}~,\quad a_n=\nu\bigg(\frac{2n}{j}-1\bigg)~,\quad b_n=c_{2n}~,\quad n\in \mathbb{R}~.
\end{equation}
Note that the relation above also applies when $j$ is a half-integer, but we must have $0\leq n\leq [j]$ (i.e. the Krylov algorithm finishes before reaching the maximal spin state due to the fact that the $(J_+)^2$ operator in $H_j$ only hops in jumps of two). The time evolved states is given by, $\ket{\psi(t)}=\rme^{-\rmi H t}\ket{j,-j}$.
%\begin{equation}\label{eq:phi(t)}
    %\ket{\psi(t)}=\rme^{-\rmi H t}\ket{j,-j}~.
%\end{equation}
% We will interpret $t$ as the proper time of the particle. 
By finding the eigenvalues of \eqref{eq:Hamiltonian j Krylov}, we can numerically evaluate \eqref{eq:spread complexity} to obtain the complexity, which in this case is essentially the spin expectation value.
%{\cp Add plot showing both the spread complexity (average spin) for a maximal spin initial state. both late times and initial times in log scale.}
Fig.\,\ref{fig:C_lowest_DODD} is a typical plot for the spread complexity with the  lowest weight state as initial state. As expected, it shows large oscillations without reaching a saturation value.

\begin{figure}
    \centering
    \includegraphics[width=0.55\linewidth]{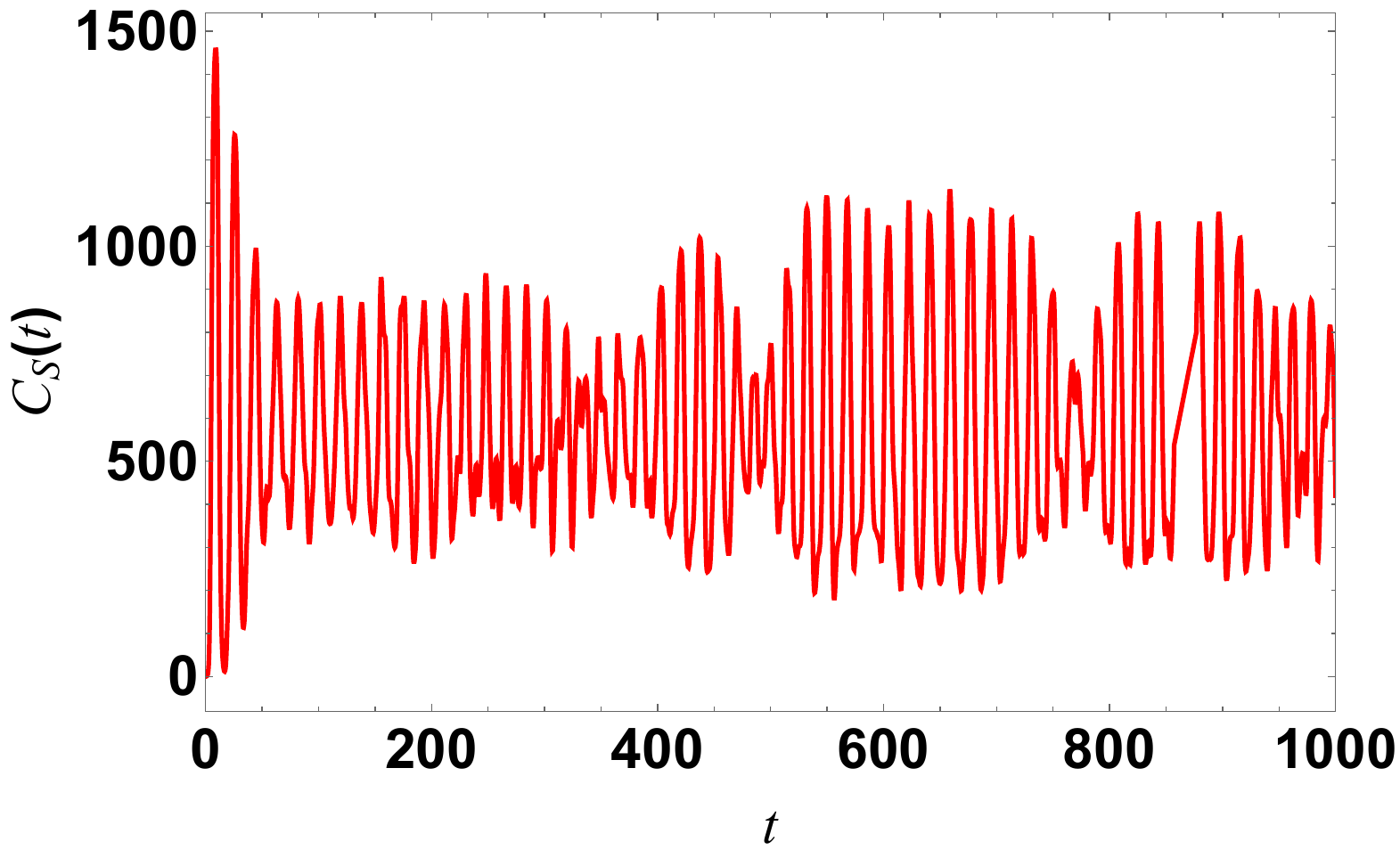}
    \caption{Time evolution under \eqref{eq:hamds2} of the spread complexity with the lowest weight state as the initial state. Here $j=1500$ and $\nu=4$. }
    \label{fig:C_lowest_DODD}
\end{figure}

\paragraph{Non-maximal initial state}
Meanwhile, if we had started from an arbitrary initial state $\ket{K_0}=\ket{j,n-j}$, we would have to find the new Krylov basis. In this case, the spread complexity is not the same as the average spin value. Indeed, for generic states, the late-time oscillations in the average spin are quite small (the wave function appears very delocalized in spin basis), whereas the spread complexity still has rather larger oscillations.

\newpage
{\small
\bibliographystyle{JHEP.bst}
\bibliography{references}}

\providecommand{\href}[2]{#2}\begingroup\raggedright\begin{thebibliography}{10}

\bibitem{Rozenbaum:2019nwn}
E.B.~Rozenbaum, L.A.~Bunimovich and V.~Galitski, \emph{{Early-Time Exponential Instabilities in Nonchaotic Quantum Systems}}, \href{https://doi.org/10.1103/PhysRevLett.125.014101}{\emph{Phys. Rev. Lett.} {\bfseries 125} (2020) 014101} [\href{https://arxiv.org/abs/1902.05466}{{\ttfamily 1902.05466}}].

\bibitem{Xu:2019lhc}
T.~Xu, T.~Scaffidi and X.~Cao, \emph{{Does scrambling equal chaos?}}, \href{https://doi.org/10.1103/PhysRevLett.124.140602}{\emph{Phys. Rev. Lett.} {\bfseries 124} (2020) 140602} [\href{https://arxiv.org/abs/1912.11063}{{\ttfamily 1912.11063}}].

\bibitem{Nandy:2024zcd}
P.~Nandy, \emph{{Tridiagonal Hamiltonians modeling the density of states of the Double-Scaled SYK model}},  \href{https://arxiv.org/abs/2410.07847}{{\ttfamily 2410.07847}}.

\bibitem{Erdmenger:2023wjg}
J.~Erdmenger, S.-K.~Jian and Z.-Y.~Xian, \emph{{Universal chaotic dynamics from Krylov space}}, \href{https://doi.org/10.1007/JHEP08(2023)176}{\emph{JHEP} {\bfseries 08} (2023) 176} [\href{https://arxiv.org/abs/2303.12151}{{\ttfamily 2303.12151}}].

\bibitem{Balasubramanian:2022dnj}
V.~Balasubramanian, J.M.~Magan and Q.~Wu, \emph{{Tridiagonalizing random matrices}}, \href{https://doi.org/10.1103/PhysRevD.107.126001}{\emph{Phys. Rev. D} {\bfseries 107} (2023) 126001} [\href{https://arxiv.org/abs/2208.08452}{{\ttfamily 2208.08452}}].

\bibitem{Balasubramanian:2022tpr}
V.~Balasubramanian, P.~Caputa, J.M.~Magan and Q.~Wu, \emph{{Quantum chaos and the complexity of spread of states}}, \href{https://doi.org/10.1103/PhysRevD.106.046007}{\emph{Phys. Rev. D} {\bfseries 106} (2022) 046007} [\href{https://arxiv.org/abs/2202.06957}{{\ttfamily 2202.06957}}].

\bibitem{Balasubramanian:2023kwd}
V.~Balasubramanian, J.M.~Magan and Q.~Wu, \emph{{Quantum chaos, integrability, and late times in the Krylov basis}},  \href{https://arxiv.org/abs/2312.03848}{{\ttfamily 2312.03848}}.

\bibitem{Baggioli:2024wbz}
M.~Baggioli, K.-B.~Huh, H.-S.~Jeong, K.-Y.~Kim and J.F.~Pedraza, \emph{{Krylov complexity as an order parameter for quantum chaotic-integrable transitions}},  \href{https://arxiv.org/abs/2407.17054}{{\ttfamily 2407.17054}}.

\bibitem{Alishahiha:2024vbf}
M.~Alishahiha, S.~Banerjee and M.J.~Vasli, \emph{{Krylov Complexity as a Probe for Chaos}},  \href{https://arxiv.org/abs/2408.10194}{{\ttfamily 2408.10194}}.

\bibitem{Jha:2024nbl}
R.G.~Jha and R.~Roy, \emph{{Sparsity dependence of Krylov state complexity in the SYK model}},  \href{https://arxiv.org/abs/2407.20569}{{\ttfamily 2407.20569}}.

\bibitem{Camargo:2024deu}
H.A.~Camargo, K.-B.~Huh, V.~Jahnke, H.-S.~Jeong, K.-Y.~Kim and M.~Nishida, \emph{{Spread and spectral complexity in quantum spin chains: from integrability to chaos}}, \href{https://doi.org/10.1007/JHEP08(2024)241}{\emph{JHEP} {\bfseries 08} (2024) 241} [\href{https://arxiv.org/abs/2405.11254}{{\ttfamily 2405.11254}}].

\bibitem{Bhattacharjee:2022vlt}
B.~Bhattacharjee, X.~Cao, P.~Nandy and T.~Pathak, \emph{{Krylov complexity in saddle-dominated scrambling}}, \href{https://doi.org/10.1007/JHEP05(2022)174}{\emph{JHEP} {\bfseries 05} (2022) 174} [\href{https://arxiv.org/abs/2203.03534}{{\ttfamily 2203.03534}}].

\bibitem{Huh:2023jxt}
K.-B.~Huh, H.-S.~Jeong and J.F.~Pedraza, \emph{{Spread complexity in saddle-dominated scrambling}}, \href{https://doi.org/10.1007/JHEP05(2024)137}{\emph{JHEP} {\bfseries 05} (2024) 137} [\href{https://arxiv.org/abs/2312.12593}{{\ttfamily 2312.12593}}].

\bibitem{pilatowsky}
S.~Pilatowsky-Cameo, J.~Ch\'avez-Carlos, M.A.~Bastarrachea-Magnani, P.~Str\'ansk\'y, S.~Lerma-Hern\'andez, L.F.~Santos et~al., \emph{Positive quantum lyapunov exponents in experimental systems with a regular classical limit}, \href{https://doi.org/10.1103/PhysRevE.101.010202}{\emph{Phys. Rev. E} {\bfseries 101} (2020) 010202}.

\bibitem{kidd2021saddle}
R.~Kidd, A.~Safavi-Naini and J.~Corney, \emph{Saddle-point scrambling without thermalization}, {\emph{Physical Review A} {\bfseries 103} (2021) 033304}.

\bibitem{Lipkin:1964yk}
H.J.~Lipkin, N.~Neshkov and A.J.~Glick, \emph{{Validity of many-body approximation methods for a solvable model. 1. Exact solutions and perturbation theory}}, \href{https://doi.org/10.1016/0029-5582(65)90862-X}{\emph{Nucl. Phys.} {\bfseries 62} (1965) 188}.

\bibitem{Meshkov:1965btx}
N.~Meshkov, A.J.~Glick and H.J.~Lipkin, \emph{{Validity of many-body approximation methods for a solvable model}: {(II). Linearization procedures}}, \href{https://doi.org/10.1016/0029-5582(65)90863-1}{\emph{Nucl. Phys.} {\bfseries 62} (1965) 199}.

\bibitem{Glick:2002fef}
A.J.~Glick, H.J.~Lipkin and N.~Meshkov, \emph{{Validity of many-body approximation methods for a solvable model}: {(III). Diagram summations}}, \href{https://doi.org/10.1016/0029-5582(65)90864-3}{\emph{Nucl. Phys.} {\bfseries 62} (1965) 211}.

\bibitem{Parmentier:2023axg}
K.~Parmentier, \emph{{Coherent spin states and emergent de Sitter quasinormal modes}},  \href{https://arxiv.org/abs/2312.08430}{{\ttfamily 2312.08430}}.

\bibitem{Aalsma:2020aib}
L.~Aalsma and G.~Shiu, \emph{{Chaos and complementarity in de Sitter space}}, \href{https://doi.org/10.1007/JHEP05(2020)152}{\emph{JHEP} {\bfseries 05} (2020) 152} [\href{https://arxiv.org/abs/2002.01326}{{\ttfamily 2002.01326}}].

\bibitem{Susskind:2021esx}
L.~Susskind, \emph{{Entanglement and Chaos in De Sitter Space Holography: An SYK Example}}, \href{https://doi.org/10.22128/jhap.2021.455.1005}{\emph{JHAP} {\bfseries 1} (2021) 1} [\href{https://arxiv.org/abs/2109.14104}{{\ttfamily 2109.14104}}].

\bibitem{Berezin:1974du}
F.A.~Berezin, \emph{{General Concept of Quantization}}, \href{https://doi.org/10.1007/BF01609397}{\emph{Commun. Math. Phys.} {\bfseries 40} (1975) 153}.

\bibitem{Perelomov_1977}
A.M.~Perelomov, \emph{Generalized coherent states and some of their applications}, \href{https://doi.org/10.1070/PU1977v020n09ABEH005459}{\emph{Soviet Physics Uspekhi} {\bfseries 20} (1977) 703}.

\bibitem{Gitman_1993}
D.M.~Gitman and A.L.~Shelepin, \emph{Coherent states of su(n) groups}, \href{https://doi.org/10.1088/0305-4470/26/2/018}{\emph{Journal of Physics A: Mathematical and General} {\bfseries 26} (1993) 313}.

\bibitem{PhysRevE.78.021106}
P.~Ribeiro, J.~Vidal and R.~Mosseri, \emph{Exact spectrum of the lipkin-meshkov-glick model in the thermodynamic limit and finite-size corrections}, \href{https://doi.org/10.1103/PhysRevE.78.021106}{\emph{Phys. Rev. E} {\bfseries 78} (2008) 021106}.

\bibitem{poincare1886courbes}
H.~Poincar{\'e}, \emph{Sur les courbes d{\'e}finies par les {\'e}quations diff{\'e}rentielles (quatri{\`e}me partie)}, {\emph{Journal de math{\'e}matiques pures et appliqu{\'e}es} {\bfseries 2} (1886) 151}.

\bibitem{bendixson1901courbes}
I.~Bendixson, \emph{Sur les courbes d{\'e}finies par des {\'e}quations diff{\'e}rentielles}, .

\bibitem{berry77}
M.V.~Berry and M.~Tabor, \emph{{Level clustering in the regular spectrum}}, \href{https://doi.org/10.1098/rspa.1977.0140}{\emph{Proc. R. Soc. Lond. A} {\bfseries 356} (1977) 375}.

\bibitem{Jeong:2024oao}
H.-S.~Jeong, A.~Kundu and J.F.~Pedraza, \emph{{Brickwall One-Loop Determinant: Spectral Statistics \& Krylov Complexity}},  \href{https://arxiv.org/abs/2412.12301}{{\ttfamily 2412.12301}}.

\bibitem{brezin1997spectral}
E.~Br{\'e}zin and S.~Hikami, \emph{Spectral form factor in a random matrix theory}, {\emph{Physical Review E} {\bfseries 55} (1997) 4067}.

\bibitem{Liu:2018hlr}
J.~Liu, \emph{{Spectral form factors and late time quantum chaos}}, \href{https://doi.org/10.1103/PhysRevD.98.086026}{\emph{Phys. Rev. D} {\bfseries 98} (2018) 086026} [\href{https://arxiv.org/abs/1806.05316}{{\ttfamily 1806.05316}}].

\bibitem{Gaikwad:2017odv}
A.~Gaikwad and R.~Sinha, \emph{{Spectral Form Factor in Non-Gaussian Random Matrix Theories}}, \href{https://doi.org/10.1103/PhysRevD.100.026017}{\emph{Phys. Rev. D} {\bfseries 100} (2019) 026017} [\href{https://arxiv.org/abs/1706.07439}{{\ttfamily 1706.07439}}].

\bibitem{Cotler:2016fpe}
J.S.~Cotler, G.~Gur-Ari, M.~Hanada, J.~Polchinski, P.~Saad, S.H.~Shenker et~al., \emph{{Black Holes and Random Matrices}}, \href{https://doi.org/10.1007/JHEP05(2017)118}{\emph{JHEP} {\bfseries 05} (2017) 118} [\href{https://arxiv.org/abs/1611.04650}{{\ttfamily 1611.04650}}].

\bibitem{berry85}
M.V.~Berry, \emph{{Semiclassical theory of spectral rigidity}}, \href{https://doi.org/10.1098/rspa.1985.0078}{\emph{Proc. R. Soc. Lond. A} {\bfseries 400} (1985) 229}.

\bibitem{Das:2022evy}
S.~Das, C.~Krishnan, A.P.~Kumar and A.~Kundu, \emph{{Synthetic fuzzballs: a linear ramp from black hole normal modes}}, \href{https://doi.org/10.1007/JHEP01(2023)153}{\emph{JHEP} {\bfseries 01} (2023) 153} [\href{https://arxiv.org/abs/2208.14744}{{\ttfamily 2208.14744}}].

\bibitem{Ageev:2024gem}
D.S.~Ageev, V.V.~Pushkarev and A.N.~Zueva, \emph{{Spectral form factors for curved spacetimes with horizon}},  \href{https://arxiv.org/abs/2412.19672}{{\ttfamily 2412.19672}}.

\bibitem{rozenbaum2020early}
E.B.~Rozenbaum, L.A.~Bunimovich and V.~Galitski, \emph{Early-time exponential instabilities in nonchaotic quantum systems}, {\emph{Physical Review Letters} {\bfseries 125} (2020) 014101}.

\bibitem{Garcia-Mata:2022voo}
I.~Garc\'\i{}a-Mata, R.A.~Jalabert and D.A.~Wisniacki, \emph{{Out-of-time-order correlators and quantum chaos}}, \href{https://doi.org/10.4249/scholarpedia.55237}{\emph{Scholarpedia} {\bfseries 18} (2023) 55237} [\href{https://arxiv.org/abs/2209.07965}{{\ttfamily 2209.07965}}].

\bibitem{Xu:2022vko}
S.~Xu and B.~Swingle, \emph{{Scrambling Dynamics and Out-of-Time-Ordered Correlators in Quantum Many-Body Systems}}, \href{https://doi.org/10.1103/PRXQuantum.5.010201}{\emph{PRX Quantum} {\bfseries 5} (2024) 010201} [\href{https://arxiv.org/abs/2202.07060}{{\ttfamily 2202.07060}}].

\bibitem{rammensee2018many}
J.~Rammensee, J.D.~Urbina and K.~Richter, \emph{Many-body quantum interference and the saturation of out-of-time-order correlators}, {\emph{Physical Review Letters} {\bfseries 121} (2018) 124101}.

\bibitem{garcia2018chaos}
I.~Garc{\'\i}a-Mata, M.~Saraceno, R.A.~Jalabert, A.J.~Roncaglia and D.A.~Wisniacki, \emph{Chaos signatures in the short and long time behavior of the out-of-time ordered correlator}, {\emph{Physical review letters} {\bfseries 121} (2018) 210601}.

\bibitem{hashimoto2020exponential}
K.~Hashimoto, K.-B.~Huh, K.-Y.~Kim and R.~Watanabe, \emph{Exponential growth of out-of-time-order correlator without chaos: inverted harmonic oscillator}, {\emph{Journal of High Energy Physics} {\bfseries 2020} (2020) 1}.

\bibitem{trunin}
D.A.~Trunin, \emph{Quantum chaos without false positives}, \href{https://doi.org/10.1103/PhysRevD.108.L101703}{\emph{Phys. Rev. D} {\bfseries 108} (2023) L101703}.

\bibitem{Camargo:2025zxr}
H.A.~Camargo, Y.~Fu, V.~Jahnke, K.~Pal and K.-Y.~Kim, \emph{{Quantum Signatures of Chaos from Free Probability}},  \href{https://arxiv.org/abs/2503.20338}{{\ttfamily 2503.20338}}.

\bibitem{Cotler:2017myn}
J.S.~Cotler, D.~Ding and G.R.~Penington, \emph{{Out-of-time-order Operators and the Butterfly Effect}}, \href{https://doi.org/10.1016/j.aop.2018.07.020}{\emph{Annals Phys.} {\bfseries 396} (2018) 318} [\href{https://arxiv.org/abs/1704.02979}{{\ttfamily 1704.02979}}].

\bibitem{parker}
D.E.~Parker, X.~Cao, A.~Avdoshkin, T.~Scaffidi and E.~Altman, \emph{A universal operator growth hypothesis}, \href{https://doi.org/10.1103/PhysRevX.9.041017}{\emph{Phys. Rev. X} {\bfseries 9} (2019) 041017}.

\bibitem{Rabinovici:2020ryf}
E.~Rabinovici, A.~S\'anchez-Garrido, R.~Shir and J.~Sonner, \emph{{Operator complexity: a journey to the edge of Krylov space}}, \href{https://doi.org/10.1007/JHEP06(2021)062}{\emph{JHEP} {\bfseries 06} (2021) 062} [\href{https://arxiv.org/abs/2009.01862}{{\ttfamily 2009.01862}}].

\bibitem{Rabinovici:2022beu}
E.~Rabinovici, A.~S{\'a}nchez-Garrido, R.~Shir and J.~Sonner, \emph{{Krylov complexity from integrability to chaos}}, \href{https://doi.org/10.1007/JHEP07(2022)151}{\emph{JHEP} {\bfseries 07} (2022) 151} [\href{https://arxiv.org/abs/2207.07701}{{\ttfamily 2207.07701}}].

\bibitem{Rabinovici:2021qqt}
E.~Rabinovici, A.~S\'anchez-Garrido, R.~Shir and J.~Sonner, \emph{{Krylov localization and suppression of complexity}}, \href{https://doi.org/10.1007/JHEP03(2022)211}{\emph{JHEP} {\bfseries 03} (2022) 211} [\href{https://arxiv.org/abs/2112.12128}{{\ttfamily 2112.12128}}].

\bibitem{Bhattacharya:2024hto}
A.~Bhattacharya, R.N.~Das, B.~Dey and J.~Erdmenger, \emph{{Spread complexity and localization in $\mathcal{PT}$-symmetric systems}},  \href{https://arxiv.org/abs/2406.03524}{{\ttfamily 2406.03524}}.

\bibitem{Nandy:2024htc}
P.~Nandy, A.S.~Matsoukas-Roubeas, P.~Mart\'\i{}nez-Azcona, A.~Dymarsky and A.~del Campo, \emph{{Quantum Dynamics in Krylov Space: Methods and Applications}},  \href{https://arxiv.org/abs/2405.09628}{{\ttfamily 2405.09628}}.

\bibitem{Balasubramanian:2024ghv}
V.~Balasubramanian, R.N.~Das, J.~Erdmenger and Z.-Y.~Xian, \emph{{Chaos and integrability in triangular billiards}}, \href{https://doi.org/10.1088/1742-5468/adba41}{\emph{J. Stat. Mech.} {\bfseries 2025} (2025) 033202} [\href{https://arxiv.org/abs/2407.11114}{{\ttfamily 2407.11114}}].

\bibitem{Fu:2024fdm}
Y.~Fu, K.-Y.~Kim, K.~Pal and K.~Pal, \emph{{Statistics and Complexity of Wavefunction Spreading in Quantum Dynamical Systems}},  \href{https://arxiv.org/abs/2411.09390}{{\ttfamily 2411.09390}}.

\bibitem{Afrasiar:2022efk}
M.~Afrasiar, J.~Kumar~Basak, B.~Dey, K.~Pal and K.~Pal, \emph{{Time evolution of spread complexity in quenched Lipkin\textendash{}Meshkov\textendash{}Glick model}}, \href{https://doi.org/10.1088/1742-5468/ad0032}{\emph{J. Stat. Mech.} {\bfseries 2310} (2023) 103101} [\href{https://arxiv.org/abs/2208.10520}{{\ttfamily 2208.10520}}].

\bibitem{Medina-Guerra:2025rwa}
E.~Medina-Guerra, I.V.~Gornyi and Y.~Gefen, \emph{{Correlations and Krylov spread for a non-Hermitian Hamiltonian: Ising chain with a complex-valued transverse magnetic field}},  \href{https://arxiv.org/abs/2502.07775}{{\ttfamily 2502.07775}}.

\bibitem{Camargo:2024rrj}
H.A.~Camargo, Y.~Fu, V.~Jahnke, K.-Y.~Kim and K.~Pal, \emph{{Higher-Order Krylov State Complexity in Random Matrix Quenches}},  \href{https://arxiv.org/abs/2412.16472}{{\ttfamily 2412.16472}}.

\bibitem{Takahashi:2025mdt}
K.~Takahashi, \emph{{Dynamical quantum phase transition, metastable state, and dimensionality reduction: Krylov analysis of fully-connected spin models}},  \href{https://arxiv.org/abs/2504.07474}{{\ttfamily 2504.07474}}.

\bibitem{Chakrabarti:2025hsb}
N.~Chakrabarti, N.~Nirbhan and A.~Bhattacharyya, \emph{{Dynamics of monitored SSH Model in Krylov Space: From Complexity to Quantum Fisher Information}},  \href{https://arxiv.org/abs/2502.03434}{{\ttfamily 2502.03434}}.

\bibitem{Chattopadhyay:2024pdj}
A.~Chattopadhyay, V.~Malvimat and A.~Mitra, \emph{{Krylov complexity of deformed conformal field theories}}, \href{https://doi.org/10.1007/JHEP08(2024)053}{\emph{JHEP} {\bfseries 08} (2024) 053} [\href{https://arxiv.org/abs/2405.03630}{{\ttfamily 2405.03630}}].

\bibitem{Hu:2025zvv}
Q.~Hu, W.-Y.~Zhang, Y.~Han and W.-L.~You, \emph{{Krylov complexity in quantum many-body scars of spin-1 models}}, \href{https://doi.org/10.1103/PhysRevB.111.165106}{\emph{Phys. Rev. B} {\bfseries 111} (2025) 165106} [\href{https://arxiv.org/abs/2503.24073}{{\ttfamily 2503.24073}}].

\bibitem{Gautam:2023bcm}
M.~Gautam, K.~Pal, K.~Pal, A.~Gill, N.~Jaiswal and T.~Sarkar, \emph{{Spread complexity evolution in quenched interacting quantum systems}}, \href{https://doi.org/10.1103/PhysRevB.109.014312}{\emph{Phys. Rev. B} {\bfseries 109} (2024) 014312} [\href{https://arxiv.org/abs/2308.00636}{{\ttfamily 2308.00636}}].

\bibitem{Bhattacharjee:2024yxj}
B.~Bhattacharjee and P.~Nandy, \emph{{Krylov fractality and complexity in generic random matrix ensembles}}, \href{https://doi.org/10.1103/PhysRevB.111.L060202}{\emph{Phys. Rev. B} {\bfseries 111} (2025) L060202} [\href{https://arxiv.org/abs/2407.07399}{{\ttfamily 2407.07399}}].

\bibitem{Bhattacharjee:2022qjw}
B.~Bhattacharjee, S.~Sur and P.~Nandy, \emph{{Probing quantum scars and weak ergodicity breaking through quantum complexity}}, \href{https://doi.org/10.1103/PhysRevB.106.205150}{\emph{Phys. Rev. B} {\bfseries 106} (2022) 205150} [\href{https://arxiv.org/abs/2208.05503}{{\ttfamily 2208.05503}}].

\bibitem{Nandy:2023brt}
S.~Nandy, B.~Mukherjee, A.~Bhattacharyya and A.~Banerjee, \emph{{Quantum state complexity meets many-body scars}}, \href{https://doi.org/10.1088/1361-648X/ad1a7b}{\emph{J. Phys. Condens. Matter} {\bfseries 36} (2024) 155601} [\href{https://arxiv.org/abs/2305.13322}{{\ttfamily 2305.13322}}].

\bibitem{Aguilar-Gutierrez:2023nyk}
S.E.~Aguilar-Gutierrez and A.~Rolph, \emph{{Krylov complexity is not a measure of distance between states or operators}}, \href{https://doi.org/10.1103/PhysRevD.109.L081701}{\emph{Phys. Rev. D} {\bfseries 109} (2024) L081701} [\href{https://arxiv.org/abs/2311.04093}{{\ttfamily 2311.04093}}].

\bibitem{Aguilar-Gutierrez:2024nau}
S.E.~Aguilar-Gutierrez, \emph{{Towards complexity in de Sitter space from the doubled-scaled Sachdev-Ye-Kitaev model}}, \href{https://doi.org/10.1007/JHEP10(2024)107}{\emph{JHEP} {\bfseries 10} (2024) 107} [\href{https://arxiv.org/abs/2403.13186}{{\ttfamily 2403.13186}}].

\bibitem{Aguilar-Gutierrez:2025pqp}
S.E.~Aguilar-Gutierrez, \emph{{From chords to dynamical wormholes with matter: Towards a bulk double-scaled (SYK) algebra}},  \href{https://arxiv.org/abs/2505.22716}{{\ttfamily 2505.22716}}.

\bibitem{Aguilar-Gutierrez:2025kmw}
S.E.~Aguilar-Gutierrez, H.A.~Camargo, V.~Jahnke, K.-Y.~Kim and M.~Nishida, \emph{{Krylov operator complexity in holographic CFTs: Smeared boundary reconstruction and the dual proper radial momentum}},  \href{https://arxiv.org/abs/2506.03273}{{\ttfamily 2506.03273}}.

\bibitem{Das:2024tnw}
R.N.~Das, S.~Demulder, J.~Erdmenger and C.~Northe, \emph{{Spread complexity for the planar limit of holography}},  \href{https://arxiv.org/abs/2412.09673}{{\ttfamily 2412.09673}}.

\bibitem{Baiguera:2025dkc}
S.~Baiguera, V.~Balasubramanian, P.~Caputa, S.~Chapman, J.~Haferkamp, M.P.~Heller et~al., \emph{{Quantum complexity in gravity, quantum field theory, and quantum information science}},  \href{https://arxiv.org/abs/2503.10753}{{\ttfamily 2503.10753}}.

\bibitem{Bender:1998ke}
C.M.~Bender and S.~Boettcher, \emph{{Real spectra in nonHermitian Hamiltonians having PT symmetry}}, \href{https://doi.org/10.1103/PhysRevLett.80.5243}{\emph{Phys. Rev. Lett.} {\bfseries 80} (1998) 5243} [\href{https://arxiv.org/abs/physics/9712001}{{\ttfamily physics/9712001}}].

\bibitem{Bender:2023cem}
C.M.~Bender and D.W.~Hook, \emph{{PT-symmetric quantum mechanics}},  \href{https://arxiv.org/abs/2312.17386}{{\ttfamily 2312.17386}}.

\bibitem{Bento:2023bjn}
P.H.S.~Bento, A.~del Campo and L.C.~C\'eleri, \emph{{Krylov complexity and dynamical phase transition in the quenched Lipkin-Meshkov-Glick model}}, \href{https://doi.org/10.1103/PhysRevB.109.224304}{\emph{Phys. Rev. B} {\bfseries 109} (2024) 224304} [\href{https://arxiv.org/abs/2312.05321}{{\ttfamily 2312.05321}}].

\bibitem{Dodelson:2025rng}
M.~Dodelson, \emph{{Black holes from chaos}},  \href{https://arxiv.org/abs/2501.06170}{{\ttfamily 2501.06170}}.

\bibitem{Avdoshkin:2019trj}
A.~Avdoshkin and A.~Dymarsky, \emph{{Euclidean operator growth and quantum chaos}}, \href{https://doi.org/10.1103/PhysRevResearch.2.043234}{\emph{Phys. Rev. Res.} {\bfseries 2} (2020) 043234} [\href{https://arxiv.org/abs/1911.09672}{{\ttfamily 1911.09672}}].

\bibitem{ORTIZ2005421}
G.~Ortiz, R.~Somma, J.~Dukelsky and S.~Rombouts, \emph{Exactly-solvable models derived from a generalized gaudin algebra}, \href{https://doi.org/https://doi.org/10.1016/j.nuclphysb.2004.11.008}{\emph{Nuclear Physics B} {\bfseries 707} (2005) 421}.

\bibitem{LERMAH2013421}
S.~{Lerma H.} and J.~Dukelsky, \emph{The lipkin–meshkov–glick model as a particular limit of the su(1,1) richardson–gaudin integrable models}, \href{https://doi.org/https://doi.org/10.1016/j.nuclphysb.2013.01.019}{\emph{Nuclear Physics B} {\bfseries 870} (2013) 421}.

\bibitem{Anninos:2023lin}
D.~Anninos, T.~Anous, B.~Pethybridge and G.~\c{S}eng\"or, \emph{{The discreet charm of the discrete series in dS$_{2}$}}, \href{https://doi.org/10.1088/1751-8121/ad14ad}{\emph{J. Phys. A} {\bfseries 57} (2024) 025401} [\href{https://arxiv.org/abs/2307.15832}{{\ttfamily 2307.15832}}].

\bibitem{Letsios:2025pqo}
V.A.~Letsios, B.~Pethybridge and A.~Rios~Fukelman, \emph{{Quite Discrete for a fermion}},  \href{https://arxiv.org/abs/2501.03724}{{\ttfamily 2501.03724}}.

\bibitem{Jafferis:2013qia}
D.L.~Jafferis, A.~Lupsasca, V.~Lysov, G.S.~Ng and A.~Strominger, \emph{{Quasinormal quantization in de Sitter spacetime}}, \href{https://doi.org/10.1007/JHEP01(2015)004}{\emph{JHEP} {\bfseries 01} (2015) 004} [\href{https://arxiv.org/abs/1305.5523}{{\ttfamily 1305.5523}}].

\bibitem{Law:2022zdq}
Y.T.A.~Law and K.~Parmentier, \emph{{Black hole scattering and partition functions}}, \href{https://doi.org/10.1007/JHEP10(2022)039}{\emph{JHEP} {\bfseries 10} (2022) 039} [\href{https://arxiv.org/abs/2207.07024}{{\ttfamily 2207.07024}}].

\bibitem{Anninos:2020hfj}
D.~Anninos, F.~Denef, Y.T.A.~Law and Z.~Sun, \emph{{Quantum de Sitter horizon entropy from quasicanonical bulk, edge, sphere and topological string partition functions}}, \href{https://doi.org/10.1007/JHEP01(2022)088}{\emph{JHEP} {\bfseries 01} (2022) 088} [\href{https://arxiv.org/abs/2009.12464}{{\ttfamily 2009.12464}}].

\bibitem{Perelomov:1971bd}
A.M.~Perelomov, \emph{{Coherent states for arbitrary lie groups}}, \href{https://doi.org/10.1007/BF01645091}{\emph{Commun. Math. Phys.} {\bfseries 26} (1972) 222}.

\bibitem{kirillov1976elements}
A.~Kirillov, \emph{Elements of the Theory of Representations}, Grundlehren der mathematischen Wissenschaften, Springer-Verlag (1976).

\bibitem{Nemoto_2000}
K.~Nemoto, \emph{Generalized coherent states for su(n) systems}, \href{https://doi.org/10.1088/0305-4470/33/17/307}{\emph{Journal of Physics A: Mathematical and General} {\bfseries 33} (2000) 3493}.

\end{thebibliography}\endgroup

\end{document}